\documentclass[sigconf]{acmart}

\usepackage{lipsum}
\usepackage{subfig}
\usepackage{amsmath}
\usepackage{wrapfig}
\usepackage{dblfloatfix}
\usepackage{lipsum,multicol}

\AtBeginDocument{%
  \providecommand\BibTeX{{%
    \normalfont B\kern-0.5em{\scshape i\kern-0.25em b}\kern-0.8em\TeX}}}

\ifodd 2
\newcommand{\rev}[1]{{\color{blue}#1}} %
\else
\newcommand{\rev}[1]{#1}
\fi

\copyrightyear{2022}
\acmYear{2022}
\setcopyright{acmcopyright}\acmConference[ACM MobiCom'21]{The 27th Annual International Conference on Mobile Computing and Networking}{January 31-February 4, 2022}{New Orleans, LA, USA}
\acmBooktitle{The 27th Annual International Conference on Mobile Computing and Networking (ACM MobiCom'21), January 31-February 4, 2022, New Orleans, LA, USA}
\acmPrice{15.00}
\acmDOI{10.1145/3447993.3483258}
\acmISBN{978-1-4503-8342-4/22/01}

\usepackage{ifpdf}
\ifpdf
\else
  
\fi

\begin{document}

\newcommand{\sysname}{SiWa}

\title{\sysname: See into Walls via Deep UWB Radar}

\author{Tianyue Zheng$^{1*}$,\hspace{3pt} Zhe Chen$^{1*}$,\hspace{3pt} Jun Luo$^1$,\hspace{3pt} Lin Ke$^2$,\hspace{3pt} Chaoyang Zhao$^3$,\hspace{3pt} Yaowen Yang$^3$}\thanks{*~Both authors contributed equally to this research.}

\affiliation{
    \institution{$^1$School of Computer Science and Engineering, Nanyang Technological University, Singapore  \\
    $^2$ Institute of Materials Research and Engineering, A*STAR, Singapore \\
    $^3$ School of Civil and Environmental Engineering, Nanyang Technological University, Singapore}
    \country{Email:~\{tianyue002, junluo\}@ntu.edu.sg, chenz@ssijri.com, karen-kl@imre.a-star.edu.sg, \{cy.zhao, cywyang\}@ntu.edu.sg}
}
\renewcommand{\authors}{T. Zheng, Z. Chen, J. Luo, L. Ke, C. Zhao, and Y. Yang}
\renewcommand{\shortauthors}{T. Zheng, Z. Chen, J. Luo, L. Ke, C. Zhao, and Y. Yang}

\begin{abstract}
Being able to \textit{see into walls} is crucial for diagnostics of building health; it enables inspections of wall structure without undermining the structural integrity. However, existing sensing devices do not seem to offer a full capability in mapping the in-wall structure while identifying their status (e.g., seepage and corrosion). In this paper, we design and implement \sysname\ as a low-cost and portable system for wall inspections. Built upon a customized IR-UWB radar, \sysname\ scans a wall as a user swipes its probe along the wall surface; it then analyzes the reflected signals to synthesize an image and also to identify the material status. Although conventional schemes exist to handle these problems individually, they require troublesome calibrations that largely prevent them from practical adoptions. To this end, we equip \sysname\ with a deep learning pipeline to parse the rich sensory data. With \rev{innovative construction and training,} the deep learning modules perform structural imaging and the subsequent analysis on material status, without the need for repetitive parameter tuning and calibrations. We build \sysname\ as a prototype and evaluate its performance via extensive experiments and field studies; results evidently confirm that \sysname\ accurately maps in-wall structures, identifies their materials, and detects possible defects, suggesting a promising solution for diagnosing building health with minimal effort and cost.
\end{abstract}




\maketitle

\section{Introduction}\label{sec:intro}
Walls in modern architectures are becoming increasingly complex due to functional~\cite{han2017smart, song2008smart}, structural~\cite{li2005analysis, zhang2020self}, and aesthetic~\cite{hopkins2011living,sutton2014aesthetics} advancements. However, the intricate and complex structures inside these walls may become deteriorated and even damaged as buildings age. Whereas breaking and reconstructing a damaged wall (or part) is certainly the last resort, a better maintenance method is to detect in-wall conditions non-destructively~\cite{helal2015non, hobbs2007non,maierhofer2010non}. Presumably, it would be most helpful to open a ``virtual window'' into walls to expose the structure inside. Aided by such ``visual'' cues, we may diagnose the healthiness of the wall structure without unnecessary destruction that can be dangerous and costly. In addition, we may detect leakage pathways and rebar (reinforcing bar) corrosion at their early stage, hence repairing them with a less destructive method~\cite{panasyuk2013injection,safan2019evaluation}.

One promising approach to non-destructive structural inspection is \textit{radio frequency} (\textit{RF-})\textit{sensing}. In fact, the wireless sensing community has witnessed a rapid growth of sensing capability enabled by RF signals in the past decade. By far, several systems have been implemented to ``see'' the location, shape, and motion of objects~\cite{jiang2020towards,adib2013see,lu2020see,yang2015see, zhu2015reusing,WiFiThruW,ThruFog,movifi,v2ifi}, 
as well as their material compositions~\cite{zhang2019feasibility,lu2020see}. By leveraging the penetration and diffraction of RF signals, \rev{some of these proposals deliver an ``X-ray''-like capability to see \textit{through} certain types of occlusions (e.g., drywall or fog)~\cite{adib2013see,yang2015see,amin,WiFiThruW,ThruFog}.
However, they lack specific capabilities of seeing \textit{into} walls (especially those built by concrete and bricks), which 
require processing schemes distinct from those commonly applied to RF signals propagating mainly through air.}

Designing a fully functional system to see into walls is very challenging. First of all, low-cost yet high-performance hardware simply does not exist. While cheap entry-level Stud Finders~\cite{bosch} provide only rudimentary functionalities,
professional Ground-Penetrating Radars (GPRs)~\cite{proceq, solla2019assessing} deliver imaging results at a cost over 10k~\!USD, yet they penetrate only asphalt and offer no material identification ability.
Second, though new hardware (e.g., the IR-UWB radar adopted by us~\cite{xethru}) have emerged in recent years, leveraging signals propagating in walls for diagnosis purpose is highly non-trivial, as RF-related research has focused mainly on \rev{air-propagating signals~\cite{ilocscan,deepmeet,ThruFog,zhang2019feasibility,lu2020see,polardraw,zhu2015reusing,ding2020rf}},
leaving us with only a handful of experimental algorithms~\cite{cafforio1991sar,  dorney2001material, rocca1989synthetic,prabhakara2020osprey}.
Last but not least, experimental algorithms often fail when facing realistic situations.
For example, excessive reflection at the rough air-wall boundary can submerge useful signals, while the heterogeneity of walls (e.g., material and depth) may require algorithms to be substantially calibrated for every new situation.

\begin{figure}[t]
    	\centering
	\includegraphics[width=.96\linewidth]{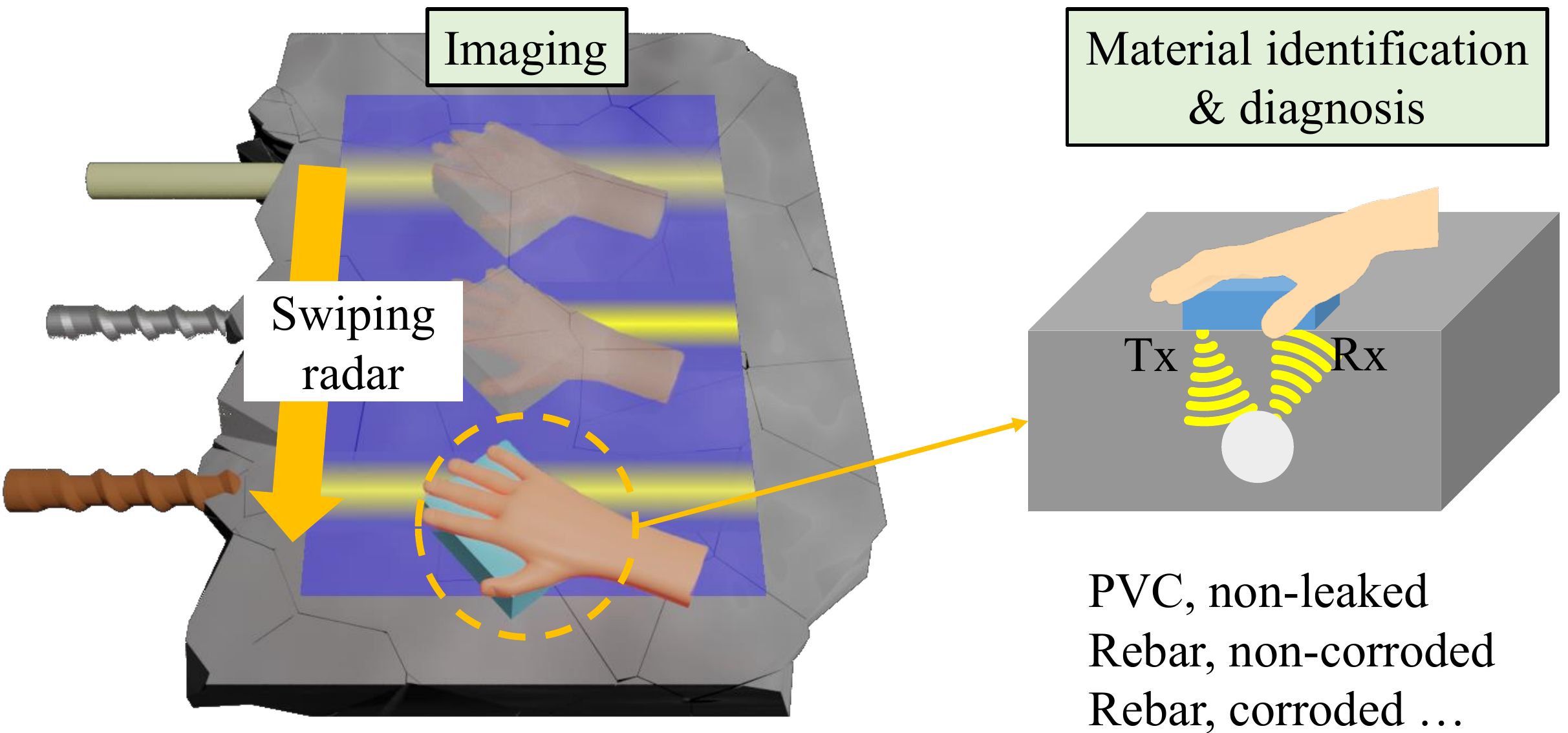}
	\caption{%
	\sysname\ collects RF reflections as a user swipes its probe across a wall surface, then it intelligently processes the signals for in-wall structural imaging and material identification \rev{(e.g, rebar layout and if a rebar is corroded or not)}.}
	\label{fig:SiWa}
	\vspace{-2ex}
\end{figure}

To tackle these challenges, we propose \textbf{\sysname} for \textbf{S}eeing \textbf{i}nto \textbf{Wa}lls. Constructed based on a commodity-grade IR-UWB radar~\cite{xethru}, \sysname\ can be held in hand to swipe across a wall surface. It processes the reflected signals by emulating \textit{synthetic aperture radar} (SAR) and \textit{spectrometer} for in-wall structural imaging and material identification, as illustrated in Figure~\ref{fig:SiWa}. \rev{In particular, \sysname\ aims to i) detect in-wall structures (e.g., rebar) and defects (e.g., cracks) via imaging, and ii) recognize their status (e.g., whether a rebar is corroded or a crack contains water seepage) via material identification.}
In order to avoid the complicated calibration of model-based approaches, \sysname\ exploits an end-to-end deep learning pipeline to process signals in a self-adaptive manner. 
Our major contributions in designing and implementing \sysname\ are summarized as follows:
\begin{itemize}%
    \item We design a low-cost, easy-to-carry prototype with a competitive \textit{see into wall} performance. Leveraging an IR-UWB radar to achieve both cost-effectiveness and compactness, we exploit the radar's wide bandwidth and extend it with dual-mode polarization antennas to obtain high imaging resolution and classification accuracy. 
    \item We conduct detailed investigations on existing model-based techniques for RF-based structural imaging and material identification. Whereas these two functionalities have always been treated separately, our study indicates that combining them into a single RF-sensing device is possible, but model-based approaches fail to handle realistic sensing tasks.
    \item We equip \sysname\ with a deep learning pipeline to extract useful information from the rich sensory data delivered by \sysname\ hardware. This pipeline is designed to perform in-wall structural imaging and material identification in a self-adaptive manner.
    Essentially, \sysname\ exploits novel encoder-decoder and adversarial learning to fulfill its tasks robustly when facing diverse environments, which, to the best of our knowledge, have never been achieved in a non-destructive way.
    \item We thoroughly evaluate \sysname\ with extensive experiments and field studies.
    The promising results show that \sysname\ can indeed see into walls, so as to generate high-quality structural imaging, identify materials, and diagnose potential problems with high accuracy. 
\end{itemize}
The overall idea of \sysname\ is to showcase the potential of RF-sensing in realizing an important application. Therefore, \sysname\ (albeit being a fully functional prototype) is by no means close to a final product, as its many design aspects are still tentative and experimental. 
The rest of the paper is organized as follows. Section~\ref{sec:background} motivates our design by introducing the background for in-wall structural imaging and material identification. Section~\ref{sec:design} presents the system design of \sysname, with detailed implementation discussed in Section~\ref{sec:implementation}. Section~\ref{sec:evaluation} reports the evaluation results. Technical limitations are discussed in Section~\ref{sec:related}.  Finally, Section~\ref{sec:conclusion} concludes this paper and points out future directions.

\section{Background and Motivations}\label{sec:background}
In this section, we provide a brief background on in-wall imaging and material identification. Detection of subsurface targets is first studied by geologists~\cite{clark2003seeing}, physicists~\cite{halabe2007detection}, and chemists~\cite{pickering2003important}. They mainly employ different physical and (electro-)chemical properties for the detection of targets. However, these methods only partially solve the problem of seeing into walls, so we hereafter focus on schemes that do have certain practical adoptions.

\subsection{Destructive and Intrusive Methods}
In order to know the structures inside a specimen, destructive
\setlength{\columnsep}{18pt}%
\begin{wrapfigure}{l}{0.4\linewidth}
  \begin{center}
  \vspace{-10pt}
    \includegraphics[width = \linewidth]{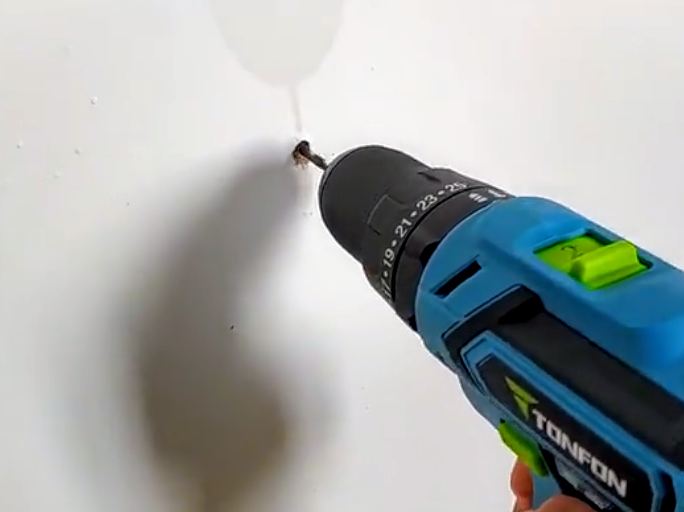}
  \end{center}
  \vspace{-6pt}
  \caption{Drilling for sensor embedding.}
  \vspace{-8pt}
\end{wrapfigure}
methods are intuitive, and the results of them are easier to interpret~\cite{zhang2013destructive, halsey1996destructive}. However, since these methods require the specimen to be destroyed, they are only suitable for objects that are mass-produced and not currently in use. For large building structures still in use, inspection by destruction is clearly not a sensible approach. As a more reasonable method, intrusive testing encompasses embedding sensors in a wall~\cite{duffo2009development, donelli2017remote} and sticking electrodes on a wall~\cite{subbiah2018novel, hoshi2019non}. However, embedding sensors into the wall and applying electrodes with conductive fluids can be costly and also degrade the structural integrity of the wall~\cite{cleland2008structural}.

\subsection{Wireless Sensing Methods}
To avoid the destructive and intrusive nature of the aforementioned methods, people have been seeking alternative methods to see into walls for a long time. Among them, various wireless methods employing acoustic and electromagnetic waves are most worthy of investigation. In the following, we discuss the pros and cons of these methods and also perform a comparative study.

\subsubsection{Acoustic Sensing}
Although proposals on non-destructive acoustic imaging~\cite{le2016plane, mao2018aim} and a small number of industrial products are also available~\cite{acs}, this technology is not suitable for our purpose because of the following reasons. First of all, acoustic signals are easily corrupted by interference, thus inapt for construction sites commonly filled with interference. Second, 
generating acoustic waves that propagate transversely requires sophisticated antenna arrays, rendering it very hard to support polarimetry for material identification as described in~\cite{zhang2019feasibility}. Third, most commercial acoustic transceivers are narrowband, hence the reflected signals barely carry any dispersion information pertinent to characterizing materials. Last but not least, generating wideband acoustic signals with commercial transceivers can be impractical since impedance mismatch~\cite{li2017acoustic} caused by a high frequency largely prevents the waves to be delivered into concrete walls.

\subsubsection{Ground Penetrating Radar (GPR)}
As discussed in Section~\ref{sec:intro}, professional GPR systems with imaging capabilities are rather expensive and mainly used for penetrating asphalt pavement~\cite{proceq}. 
\setlength{\columnsep}{18pt}%
\begin{wrapfigure}{r}{0.4\linewidth}
  \begin{center}
  \vspace{-10pt}
    \includegraphics[width = \linewidth]{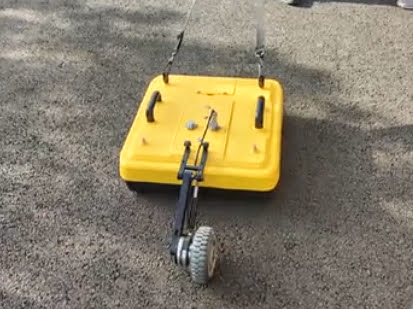}
  \end{center}
  \vspace{-10pt}
  \caption{GPR.}
  \vspace{-8pt}
\end{wrapfigure}
Moreover, these devices are closed source and provide no Application Programming Interface (API), so it is nearly impossible to extend them to achieve material identification and structural diagnosis, which are indispensable capabilities for actually inspecting the status of in-wall structures.

\begin{figure}[b]
    \setlength\abovecaptionskip{8pt}    
    \vspace{-2.5ex}
    \centering
	   \captionsetup[subfigure]{justification=centering}
		\centering
		\subfloat[\rev{Wi-Fi: 2.4~\!GHz, 20~\!MHz.}]{
		    \begin{minipage}[b]{0.48\linewidth}
		        \centering
			    \includegraphics[width = 0.96\textwidth]{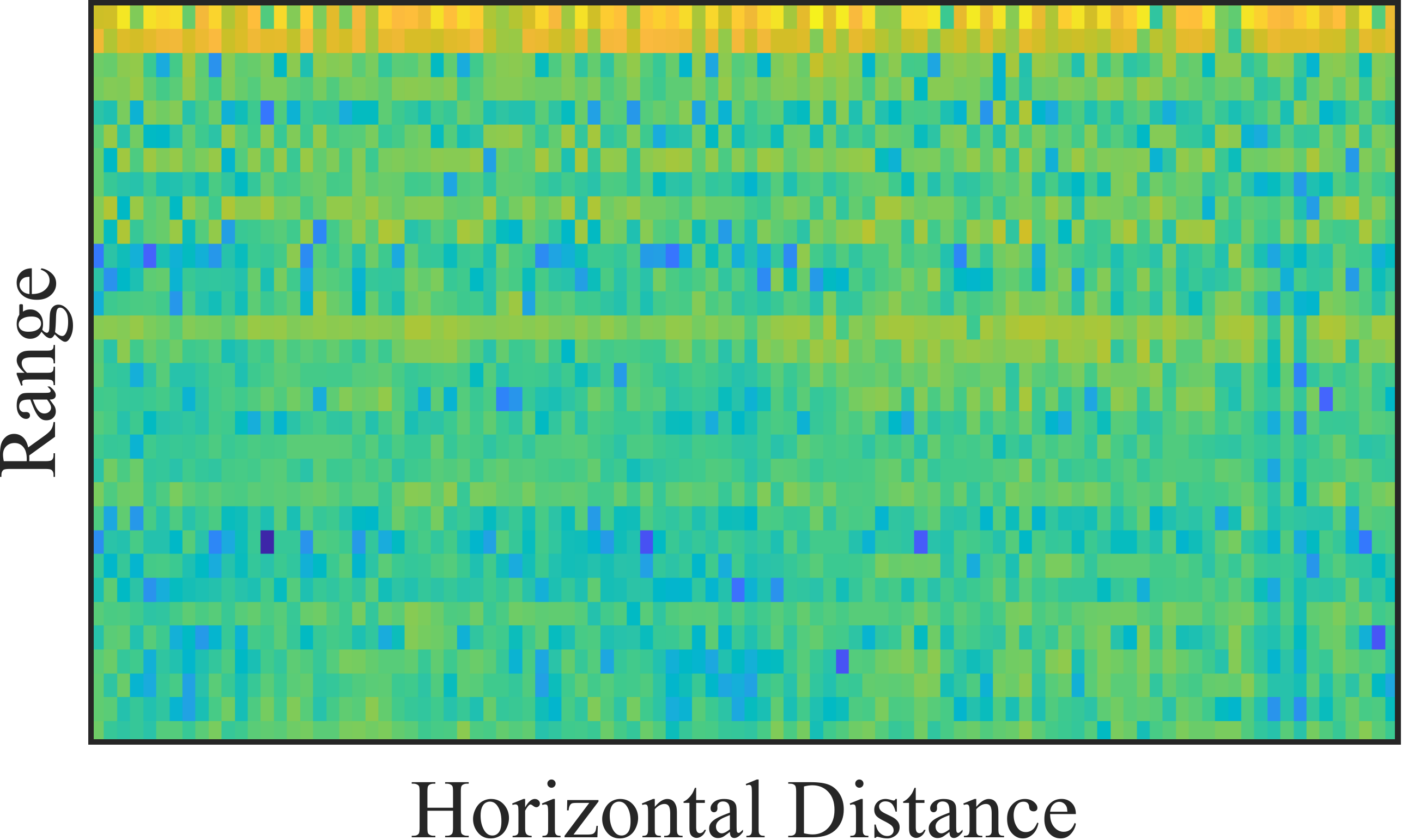}
			    \label{fig:wificross}
			\end{minipage}
		}
		\subfloat[IR-UWB: 7.29~\!GHz, 1.5~\!GHz.]{
		    \begin{minipage}[b]{0.48\linewidth}
		        \centering
			    \includegraphics[width = 0.96\textwidth]{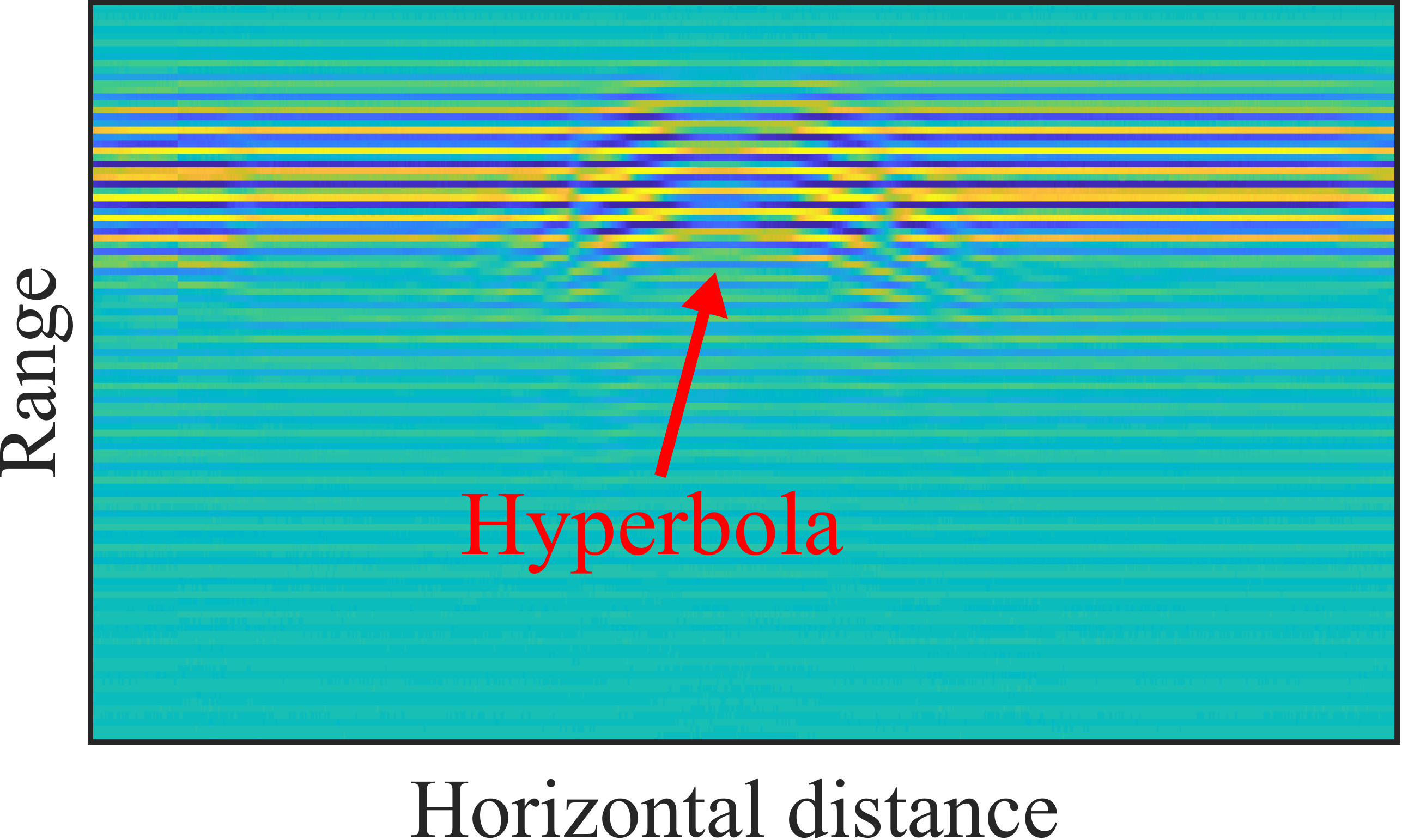}
			    \label{fig:x5hyper}
			\end{minipage}
		}
		\\
		\subfloat[FMCW Radar: 24~\!GHz, 0.2~\!GHz.]{
		    \begin{minipage}[b]{0.48\linewidth}
		        \centering
			    \includegraphics[width = 0.96\textwidth]{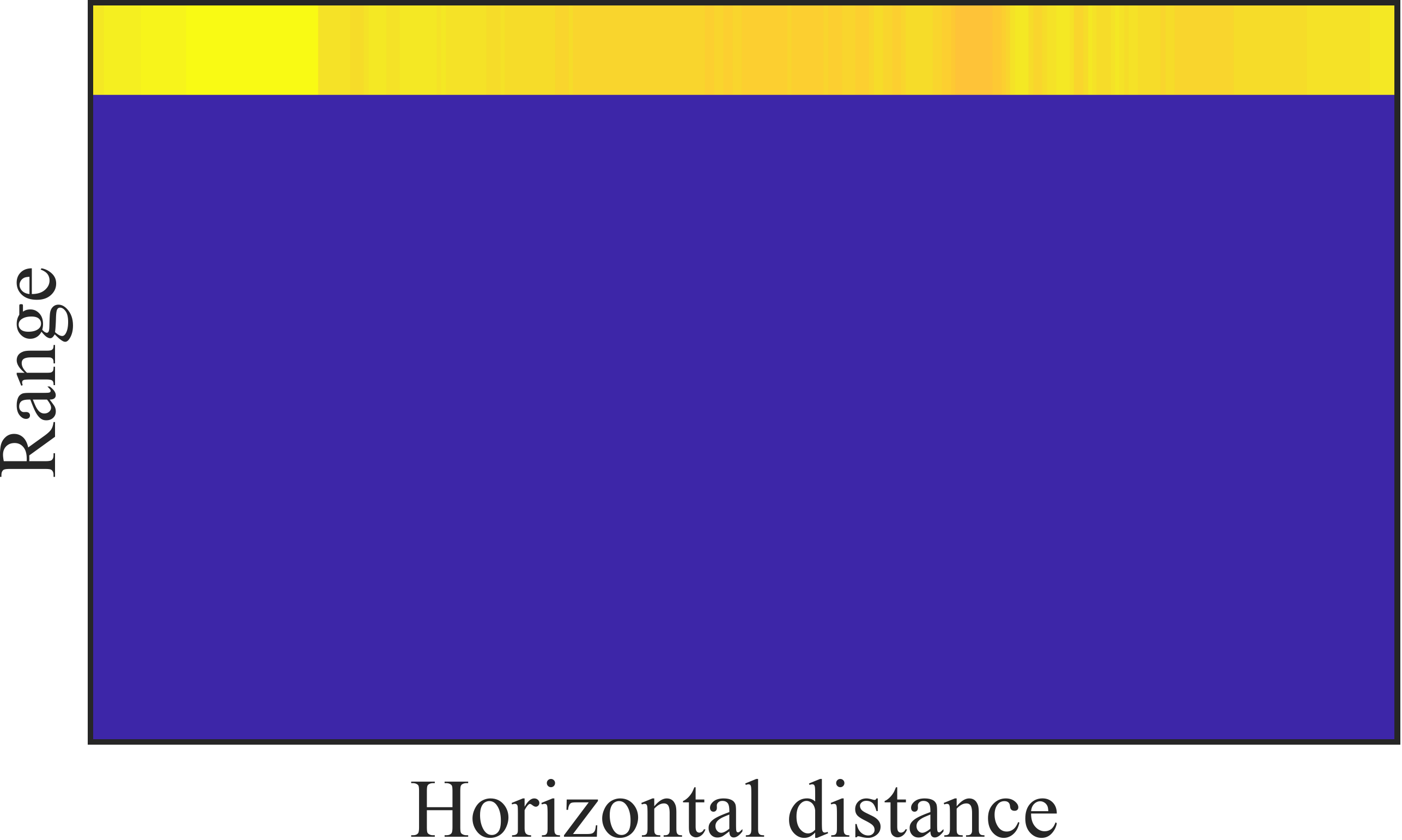}
			\end{minipage}
		}	
		\subfloat[FMCW Radar: 77~\!GHz, 4~\!GHz.]{
		    \begin{minipage}[b]{0.48\linewidth}
		        \centering
			    \includegraphics[width = 0.96\textwidth]{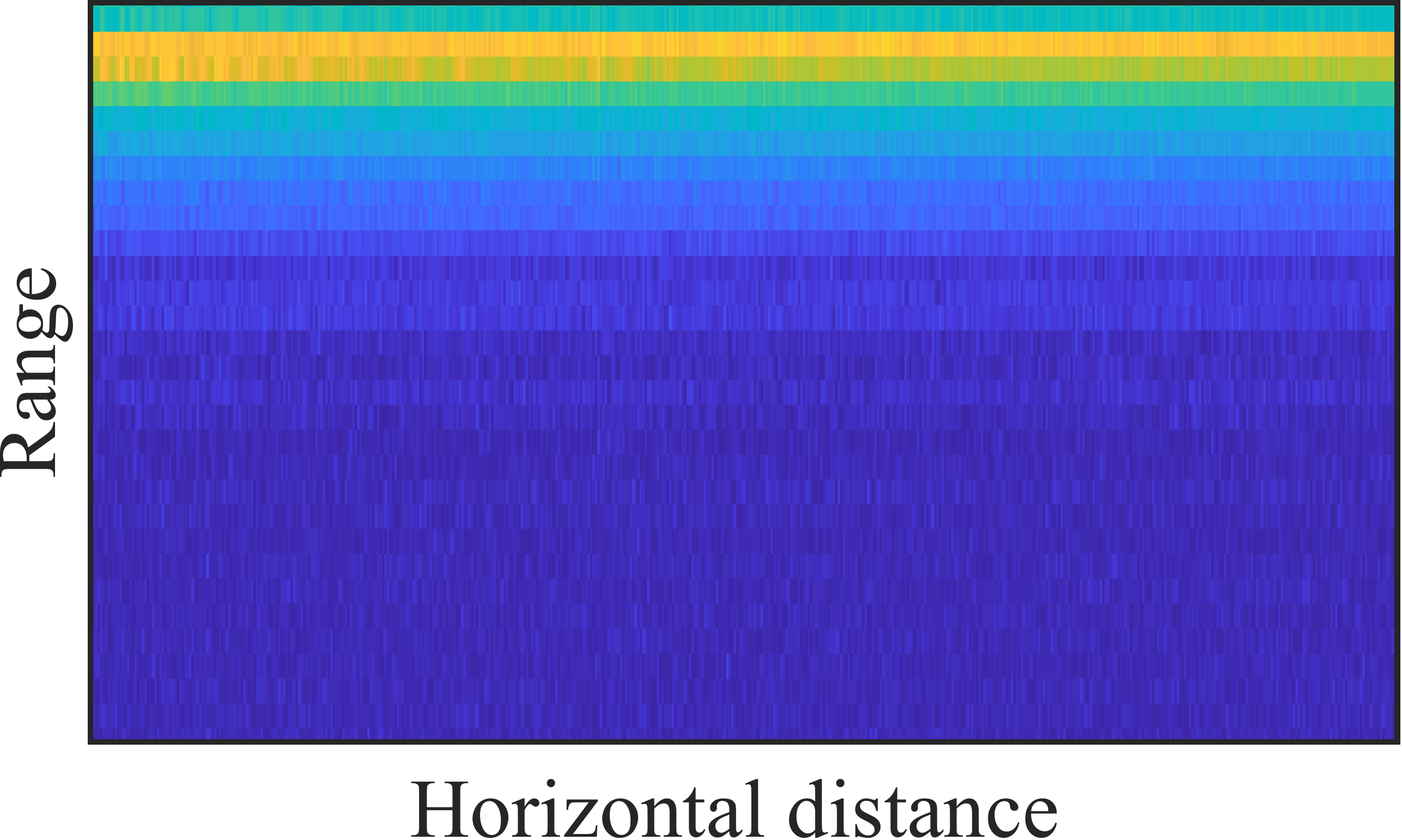}
			\end{minipage}
		}
		\caption{Preliminary study of four RF devices, with their center frequencies and bandwidths specified.}
		\label{fig:rfcomp}
\end{figure}
\subsubsection{Commodity-grade RF Sensing} \label{sssec:rf_comp}
\begin{figure}[t]
    \setlength\abovecaptionskip{8pt}    
    \centering
	   \captionsetup[subfigure]{justification=centering}
		\centering
		\subfloat[5~\!cm.]{
		    \begin{minipage}[b]{0.48\linewidth}
		        \centering
			    \includegraphics[width = 0.96\textwidth]{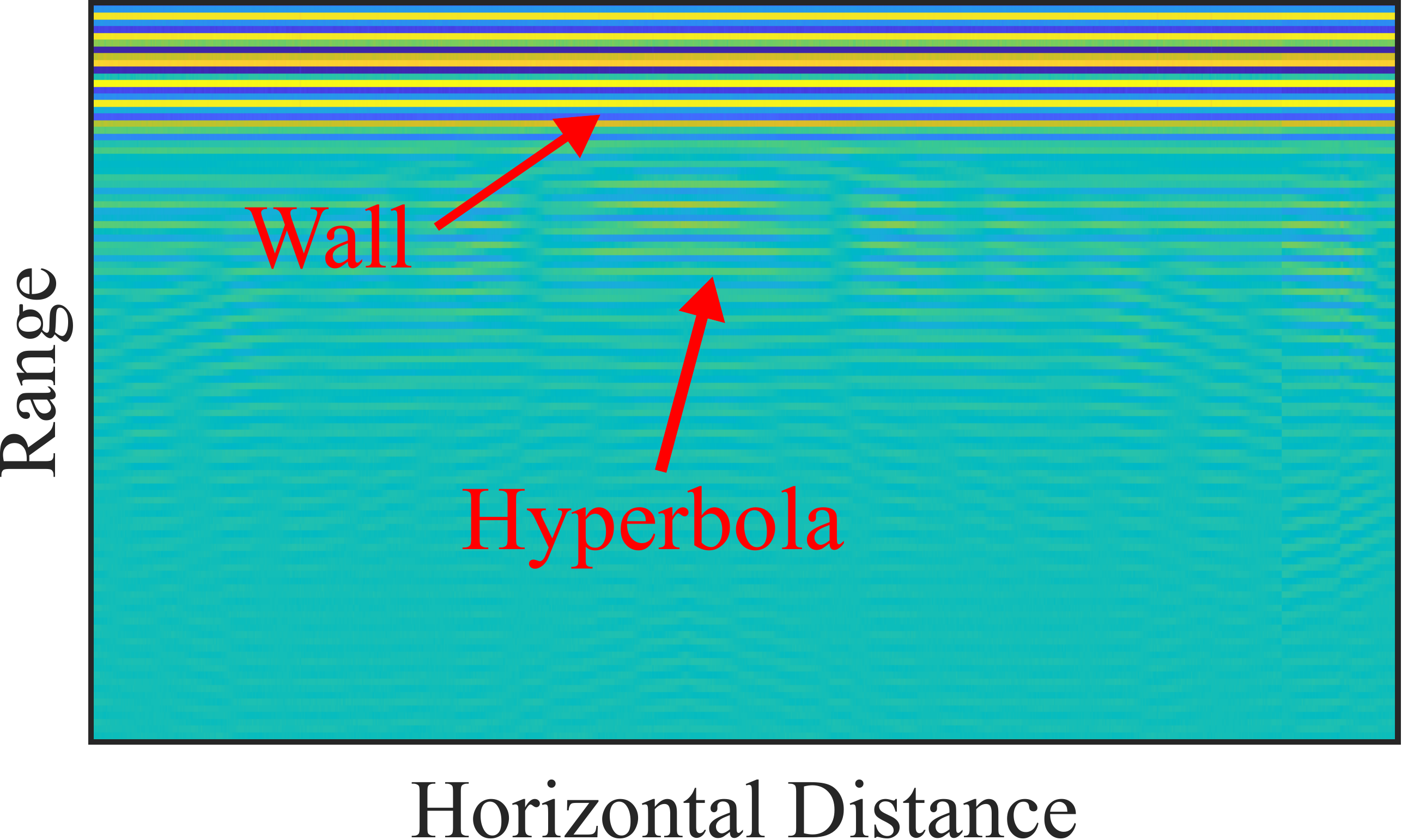}
			    \label{subfig:noncontact5}
			\end{minipage}
		}	
		\subfloat[50~\!cm.]{
		    \begin{minipage}[b]{0.48\linewidth}
		        \centering
			    \includegraphics[width = 0.96\textwidth]{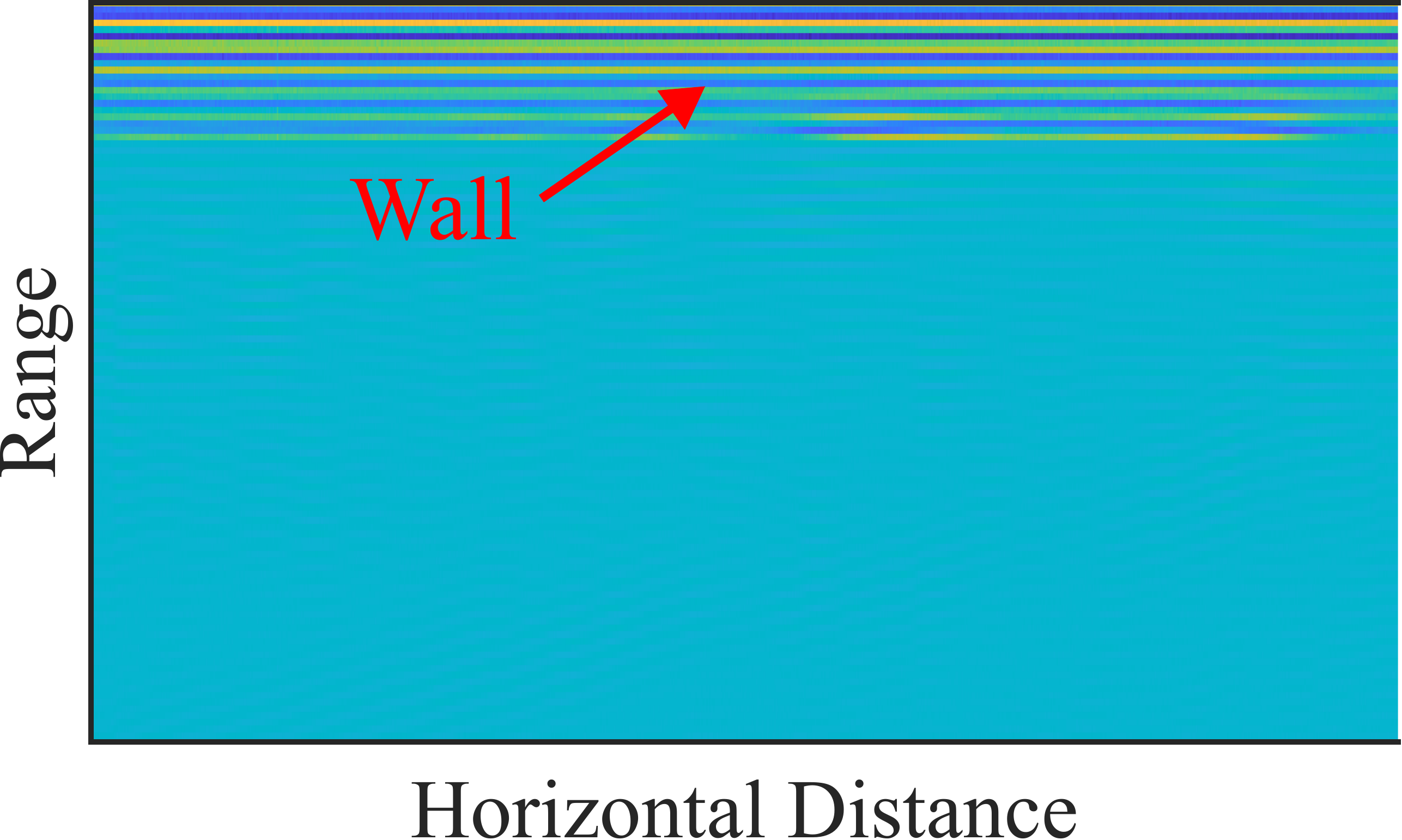}
			    \label{subfig:noncontact50}
			\end{minipage}
		}
		\caption{\rev{Non-contact scanning is vulnerable to excessive loss at the air-wall boundary and ambient interference. }} %
		\label{fig:noncontact}
	    \vspace{-2ex}
\end{figure}

Commodity-grade RF devices strike a good balance between cost and performance. Some examples include Wi-Fi, IR-UWB radar, and mmWave radar operating at different frequencies. To select an appropriate device in serving our purpose of seeing into walls, we perform a preliminary study on the capabilities of these devices. We embed a rebar into a concrete wall at a depth of 5~\!cm and swipe these devices over the wall surface. The received signals are visualized in Figure~\ref{fig:rfcomp}. We observe that, though Wi-Fi successfully penetrates the wall, excessive diffraction makes it unable to focus on and locate the rebar. Moreover, the narrow bandwidth of 20~\!MHz confines the spatial resolution of  Wi-Fi. IR-UWB radar successfully penetrates the wall and forms a typical ``hyperbola'' SAR pattern of a reflector. Two FMCW radars at 24~\!GHz and 77~\!GHz cannot penetrate into the wall; the transmitted signals are mostly reflected back by the air-wall boundary. \rev{We further test non-contact scanning performance of the IR-UWB radar. Figure~\ref{subfig:noncontact5} and~\ref{subfig:noncontact50} show the received signals when the probe is swiped respectively at 5~\!cm and 50~\!cm from the wall. Compared with the result in Figure~\ref{fig:x5hyper}, reflection of the wall becomes much stronger and the hyperbola becomes less evident at a shorter distance, but totally disappears at a further distance.}

Here are a few takeaways from this preliminary study. First, bistatic devices (e.g., Wi-Fi) whose transmitter (tx) gets separated from receiver (rx) are not suitable for our purpose anyway. On one hand, the results obtained by putting tx and rx Wi-Fi antennas on the same side of the wall fail miserably compared even with Figure~\ref{fig:wificross} (which is obtained by leaving the two antennas on different sides). On the other hand, walls with single-side access (e.g., in a tunnel) demand monostatic devices with co-located tx and rx antennas. Second, there is a trade-off in selecting a suitable center frequency: a low frequency helps penetration but may also cause too excessive diffraction to miss targets, while a high frequency can substantially limit the penetration ability. Third, a wider bandwidth is preferred to achieve a higher spatial resolution and imaging quality. \rev{Finally, a contact design swiping across wall surface is preferred over a non-contact approach scanning from a distance, because the latter faces excessive loss at the air-wall boundary and ambient interference.}
Considering the above factors, it is evident that the IR-UWB radar of 7.29~\!GHz center frequency and 1.5~\!GHz bandwidth has the best overall performance, \rev{when used in a contact manner}.

\subsubsection*{Remarks} 
Because IR-UWB and FMCW are equivalent as a time-frequency dual pair~\cite{ding2020rf}, we choose IR-UWB to be the default technology in the following discussions. \rev{We do not consider static radar imaging~\cite{amin, walabot} because SAR offers flexible extension in detection area, higher SNR after coherent integration within a synthetic aperture~\cite{chan2008introduction}, and lower cost brought by its compact design~\cite{stimson1998airborne}.}

\section{\sysname\ Design} \label{sec:design}
\begin{figure}[h]
    \vspace{-.5ex}
	\centering
	\includegraphics[width=0.9\linewidth]{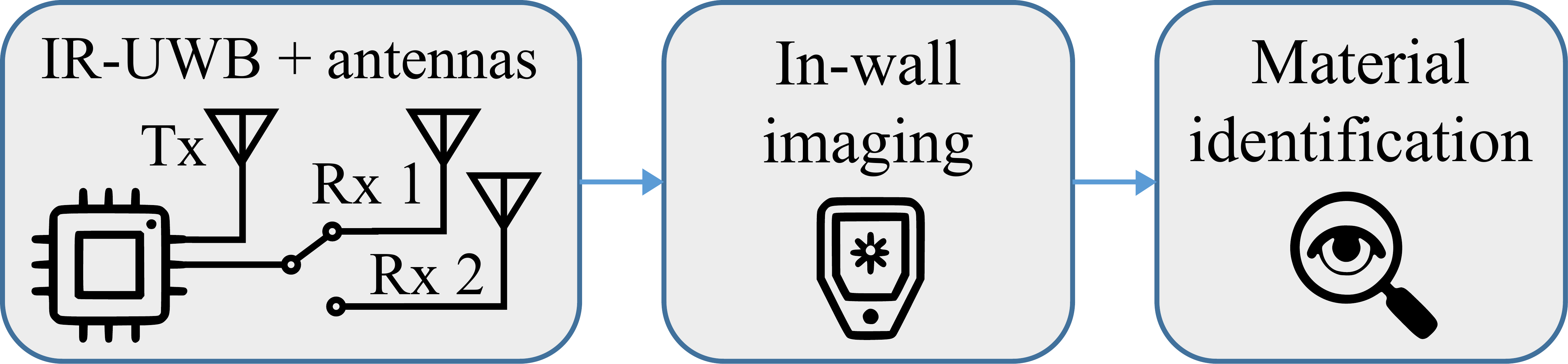}
	\caption{Overview of \sysname's design.}
	\label{fig:diagram}
	\vspace{-2ex}
\end{figure}

This section introduces the design of \sysname, whose rough diagram is shown in Figure~\ref{fig:diagram}. The hardware front-end (including an IR-UWB chip and antennas) collects SAR data matrix containing two-channel polarized and wideband signals. The SAR data matrix is then leveraged by the back-end software for in-wall imaging, which subsequently helps to locate targets and in turn identifies their respective material and status.

\subsection{Modeling Radar Signal}
\begin{figure}[t]
    \setlength\abovecaptionskip{8pt}
	\centering
	\includegraphics[width=0.83\linewidth]{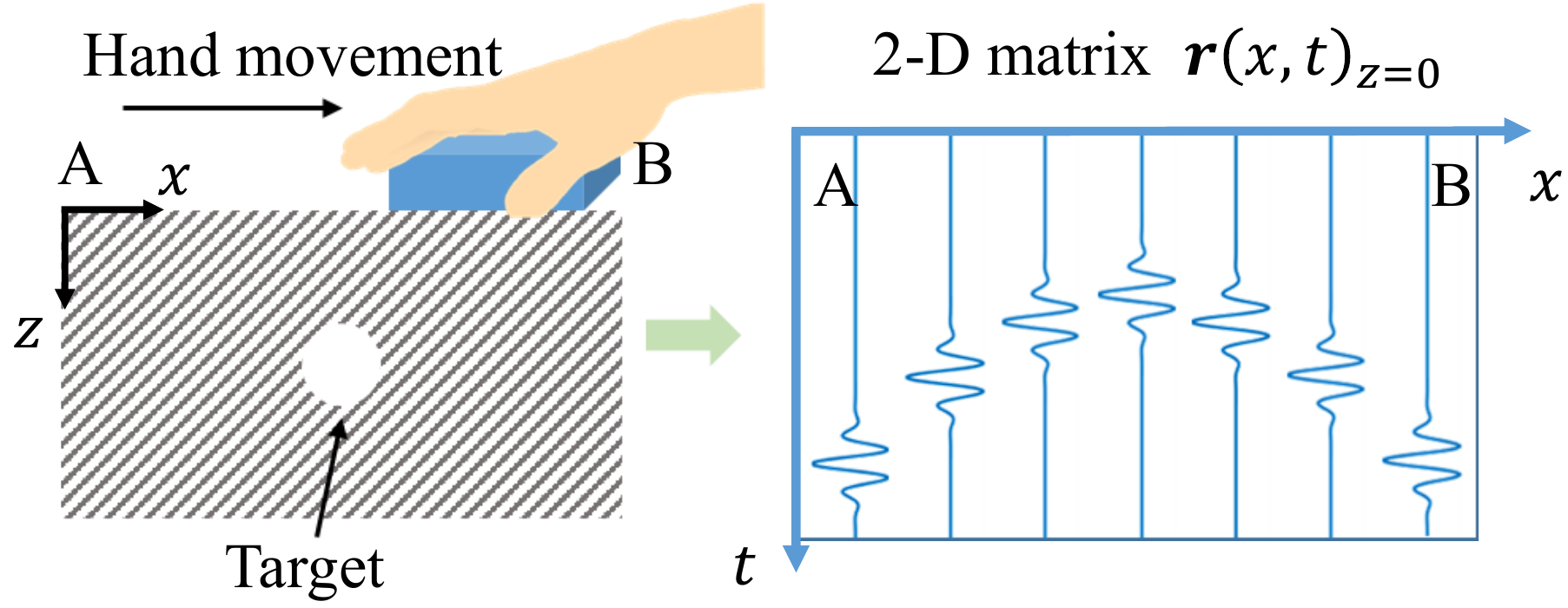}
	\caption{Formation of the 2-D signal matrix $\boldsymbol{r}$.}
	\label{fig:2dmatrix}
	\vspace{-2ex}
\end{figure}

By scanning across a wall surface,
\sysname\ works as a SAR that reconstructs a 2D image of the in-wall structure by synthesizing the reflected signals gathered at multiple positions. To model the process, we first have a look at the UWB baseband signal taking the form of a Gaussian pulse: 
\begin{equation} \label{eq:baseband}
    s(t)=\alpha_{\mathrm{tx}} \exp \left(-t^{2}/{2 \sigma^{2}}\right),
\end{equation}
and this baseband signal is then modulated onto a carrier to get:
\begin{equation} \label{eq:range}
x(t) = s(t) \cdot \cos \left( 2 \pi  f_\mathrm{c} t \right),
\end{equation}
where $\alpha_{\mathrm{tx}}$ is the pulse amplitude, $\sigma^2$ is the variance defined by the -10~\!dB bandwidth, and $f_\mathrm{c}$ is the carrier frequency,  

Assuming $K$ \rev{(point scatterer)} targets present in a wall, the tx signal travels to the $k$-th target and its echo scatters back after a time delay $\tau_k = \frac{2r_k}{c}$, where $r_k$ is the range to the $k$-th target, and $c$ is light speed in the wall. So the rx signal becomes:
\begin{equation} \label{eq:received}
r(t) = \sum_{k=1}^K \beta_k s(t-\tau_k) \cdot \cos \left(2 \pi  f_\mathrm{c} (t-\tau_k)\right) +n(t),
\end{equation}
where $\beta_k$ is a scaling term attributed to propagation and reflection, and $n(t)$ represents the Gaussian noise. 
A radar repeats this tx-rx process at a regular interval to generate 1-D sequences shown in Eq.~\eqref{eq:received}. By stacking these 1-D signals one by one, we get a 2-D signal matrix shown in Figure~\ref{fig:2dmatrix}. Generalizing Eq.~\eqref{eq:received}, we use $\boldsymbol{r}(x, t)_{z=0}$ to represent the 2-D signal matrix, where $x$ denotes the moving distance of the radar, $t$ represents the \textit{range dimension}, and $z=0$ indicates that the radar stays at zero depth. As we slide the radar on the wall surface, this signal matrix captures information for reconstructing images of in-wall structures.
\rev{This model is also valid for extended scatterers, because the received signal can be represented by replacing the summation in Eq.~\eqref{eq:received} with an integration for a scatterer extending within the $x\!-\!z$ plane; 
otherwise, as the radar antenna has a fixed beamwidth, a scatterer extending out of the $x\!-\!z$ plane is illuminated for a finite ``thickness'' and can be approximated as a point scatterer in the imaging plane~\cite{ozdemir2014review, shihab2005radius}.}

\subsection{A Conventional Design}\label{ssec:conventional}
We first study how conventional signal processing methods leverage $\boldsymbol{r}(x, t)_{z=0}$ to realize functions required by \sysname. It serves as an inspiration for our deep learning-based design elaborated later. %

\subsubsection{Imaging by SAR} \label{ssec:sar_imag}
As we move the radar, the distance from a target located at $(x_0, z_0)$ first decreases and then increases. On the received 2-D signal matrix, the trace of the Gaussian pulse forms a hyperbola as shown in Figure~\ref{fig:2dmatrix}, which can be located along range dimension by
$t\left(x, x_{0}, z_{0}\right)=\frac{2}{c} \sqrt{\left(x-x_{0}\right)^{2}+z_{0}^{2}}$~\cite{cafforio1991sar}; an example of the trace is shown in Figure~\ref{fig:x5hyper}. However, raw data in this form are often hard to interpret since a point target emerges as a hyperbola, and multiple targets may get mixed up. Existing proposals translating (temporal) range $t$ to (geometric) depth $z$ include 
the frequency-domain Range Migration Algorithm (RMA)~\cite{rocca1989synthetic, cafforio1991sar} and time-domain back-projection algorithm~\cite{munson1983tomographic, desai1992convolution}; they all aim to ``refocus'' the energy spread on the hyperbola. %
Among them, \rev{the time-wavenumber domain implementation of RMA becomes the de-facto standard, thanks to its higher computing efficiency achieved by Fast Fourier Transform (FFT)}; we briefly explain the workflow and weaknesses of RMA in the following.

A matched filter is firstly applied along the range dimension of the 2-D signal matrix to achieve range compression, 
\rev{i.e., reducing the width of the transmitted Gaussian pulse, so as to enhance range resolution.} Once filtering is completed, a 2-D FFT is performed to transform the matrix into the frequency-wavenumber domain:
\begin{equation} \label{eq:2dfft}
\boldsymbol{R}(\kappa_{x}, \omega)_{z=0} = \iint \boldsymbol{r}(x, t)_{z=0} \cdot e^{-j\left(\omega t+\kappa_{x} x\right)} d x d t,
\end{equation}
\rev{where $\omega$ is the frequency conjugate to the arrival time $t$, and analogously, $\kappa_x$ is \textit{wavenumber} (i.e., the spatial frequency) conjugate to the \textit{moving dimension} $x$.}
$\kappa_x$ is related to $\omega$ and $\kappa_z$ (wavenumber in depth dimension $z$) by
$\omega=\frac{c}{2} \cdot \sqrt{\kappa_{x}^{2}+\kappa_{z}^{2}}$.
In order to focus the energy in the current domain, we trace back the wave propagation in depth $z$ to obtain $e^{j \kappa_{z} z} \cdot \boldsymbol{R}(\kappa_{x}, \omega)_{z=0}$. As a result, the focused signal at an arbitrary depth $z$ in moving-depth domain can be recovered by a time reversal of $t_0 = 2 z_0/c$ and an inverse FFT: 
\begin{align} 
\tilde{\boldsymbol{r}} (x, z)_{t=-t_{0}} =~~& \frac{1}{(2 \pi)^{2}} \iint \boldsymbol{R}(\kappa_{x}, \omega)_{z=0}\nonumber \cdot e^{j\left(\kappa_{z} z - \omega t_{0}\right)} e^{j \kappa_{x} x} d \omega d \kappa_{x},
\end{align}
\rev{and the resulting $\tilde{\boldsymbol{r}} (x, z)_{t=-t_{0}}$ is indeed the image on $x\!-\!z$ plane.}

This seemingly straightforward process of RMA actually requires several parameters to be specified in advance. For example, the speed $v$ of the moving probe determines the eccentricity of the hyperbola, and the permittivity $\epsilon$ of the wall (related to the light speed in the wall) is used to perform matched filtering and compute wavenumbers $\kappa_x$ and $\kappa_z$.
\begin{figure}[b]
    \setlength\abovecaptionskip{6pt}
    \vspace{-2.5ex}
	   \captionsetup[subfigure]{justification=centering}
		\centering
		
		\subfloat[$\hat{v}$=2~\!cm/s (actual $v$).]{
		    \begin{minipage}[b]{0.48\linewidth}
		        \centering
			    \includegraphics[width = 0.98\textwidth]{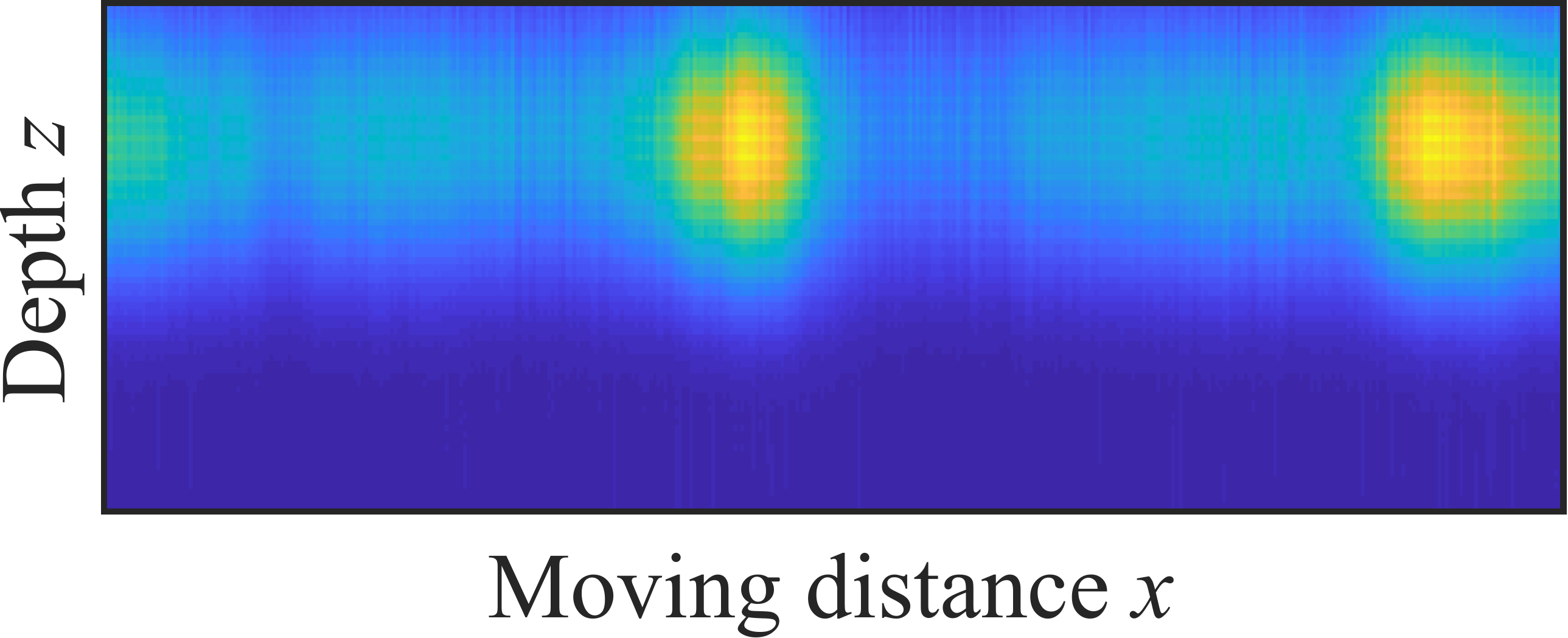}
			\end{minipage}
		}
		\subfloat[$\hat{v}$=1~\!cm/s.]{
		    \begin{minipage}[b]{0.48\linewidth}
		        \centering
			    \includegraphics[width = 0.98\textwidth]{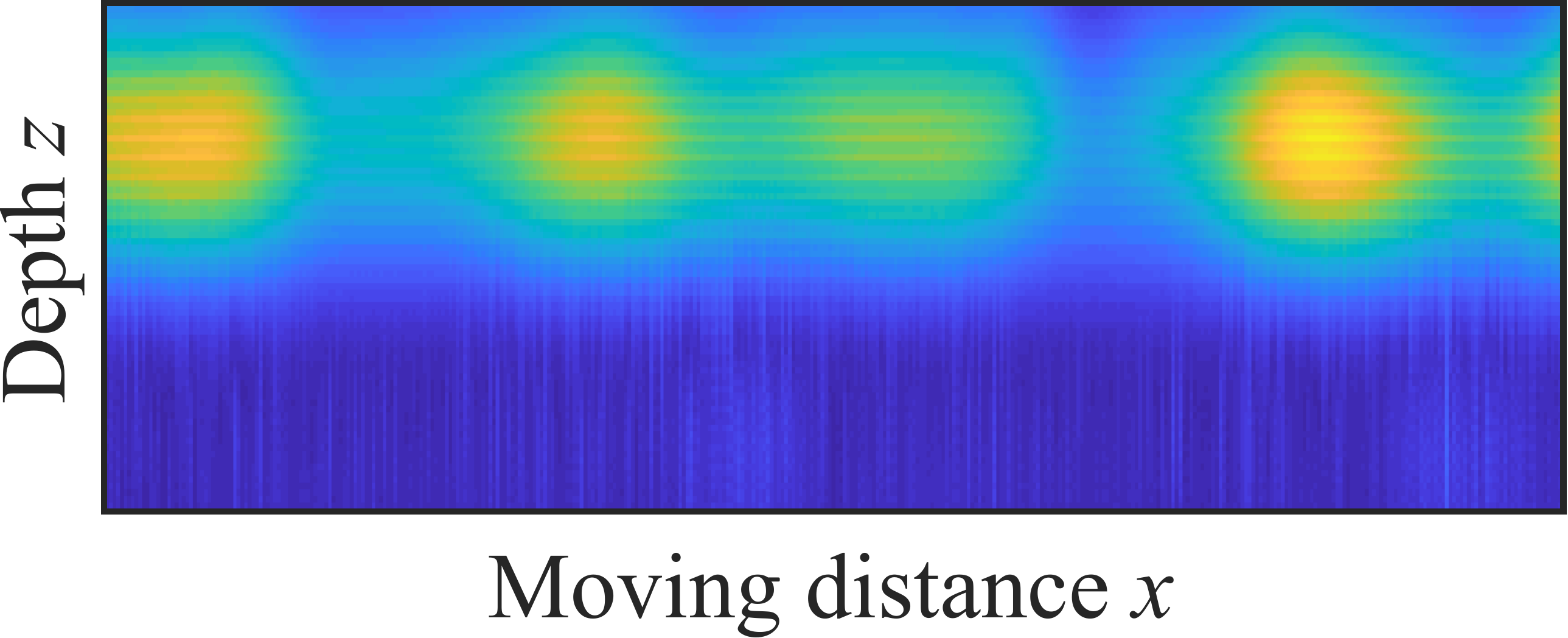}
			\end{minipage}
		}
		\\
	    \subfloat[$\hat{\epsilon}=9$ (actual $\epsilon$).]{
		    \begin{minipage}[b]{0.48\linewidth}
		        \centering
			    \includegraphics[width = 0.98\textwidth]{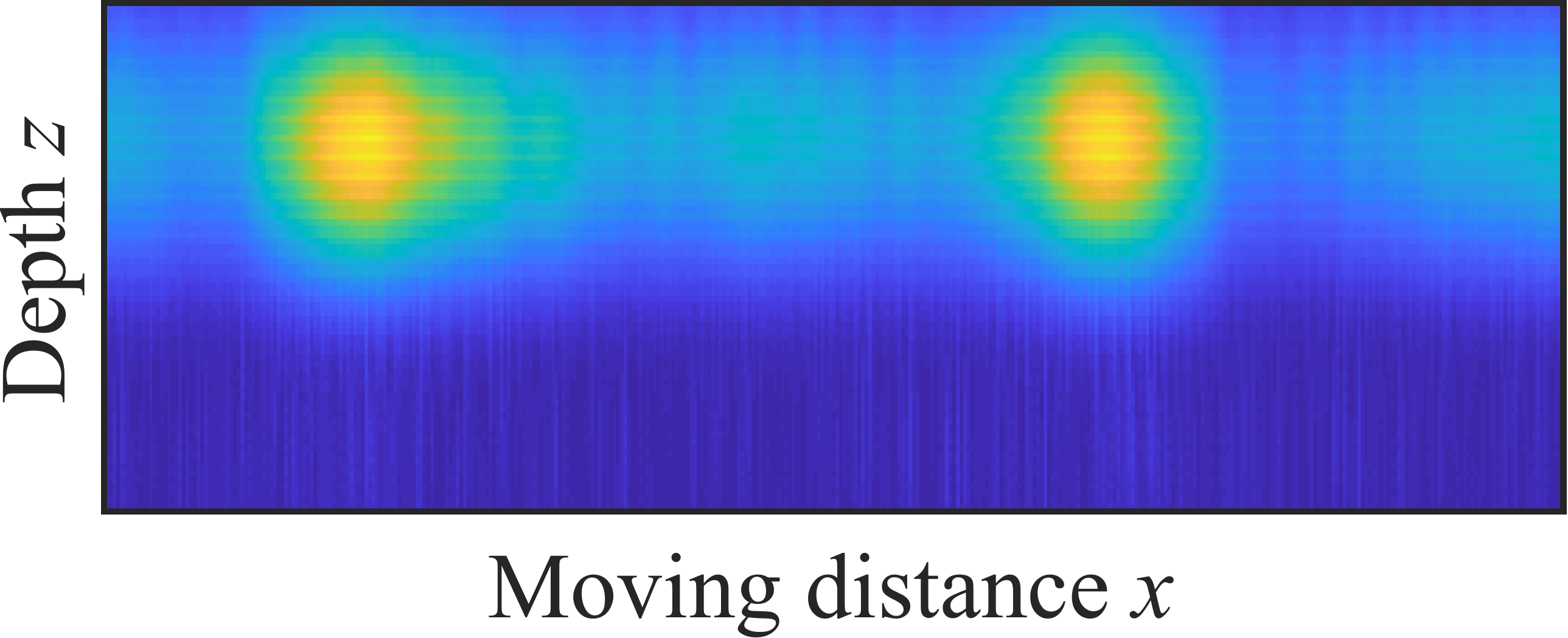}
			\end{minipage}
		}
		\subfloat[$\hat{\epsilon}=7$.]{
		    \begin{minipage}[b]{0.48\linewidth}
		        \centering
			    \includegraphics[width = 0.98\textwidth]{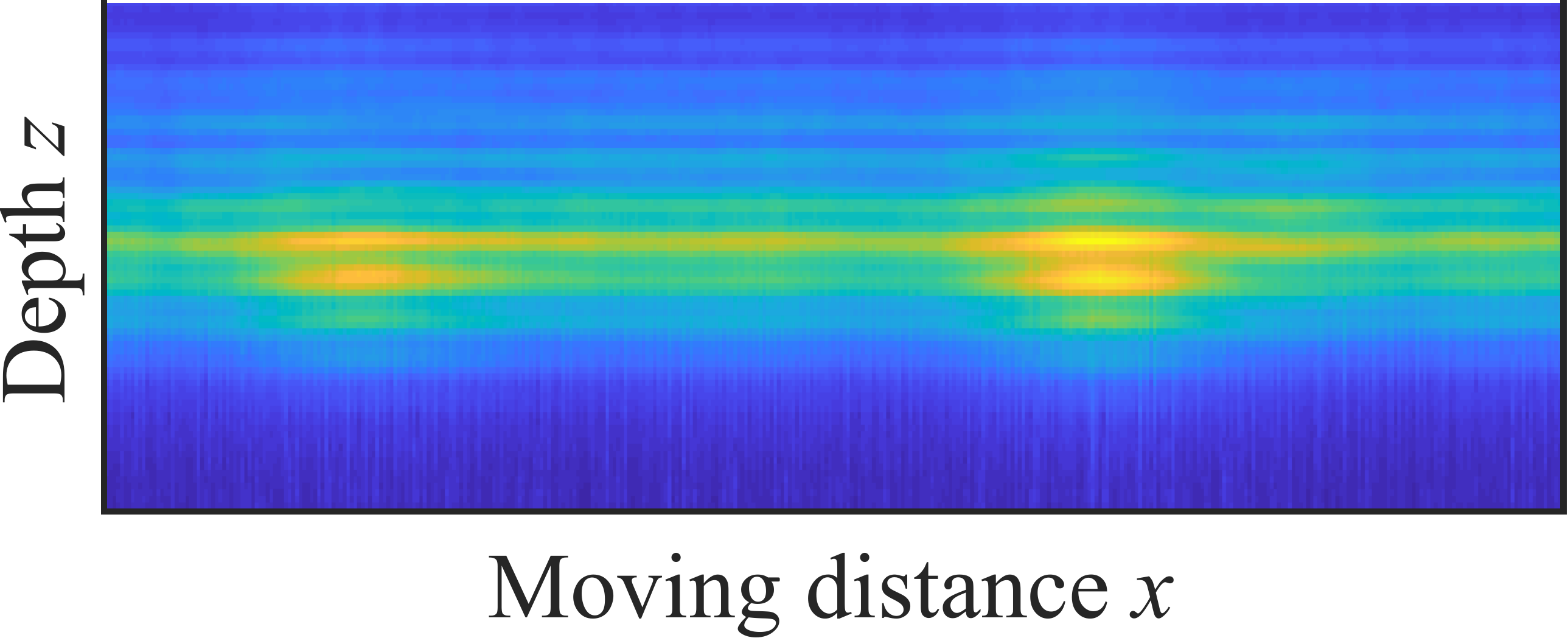}
			\end{minipage}
		}
		\caption{RMA imaging results affected by different parameter settings. (a) and (b) show the effect of the (radar) probe swiping speed $v$. (c) and (d) show the effect of relative permittivity $\epsilon$ of the wall. }
		\label{fig:velocity}
\end{figure}
However, we cannot count on a \sysname\ user to move his/her hand at a pre-determined speed. In addition, 
the diversity of wall materials can lead to very different permittivities.
To make things worse, environment conditions such as humidity may also alter the properties of a wall drastically (e.g., the permittivity may vary from 5 to 14 at different moisture levels~\cite{ogunsola2005shielding}). If these parameters are not specified in advance, existing algorithms cannot perform imaging successfully. 

To illustrate this difficulty, we use RMA for imaging two rebars in a concrete wall, whose actual permittivity $\epsilon$ is 9, and the speed of swiping the probe is about 2~\!cm/s. 
The results shown in Figure~\ref{fig:velocity} clearly demonstrate that, if correct parameters are provided, the energy distribution is concentrated and the rebar boundaries are sharp. However, if the estimated velocities $\hat{v}$ and permittivity $\hat{\epsilon}$ deviate from the actual values, the imaging results become blurred and may even deliver the wrong location information. As these weaknesses are inherent to existing model-based methods, \textit{it is imperative to equip \sysname\ with a calibration-free imaging module}.

\subsubsection{Material Identification}\label{sssec:material_id}
Suppose a map of the in-wall structure has been obtained by imaging, $\boldsymbol{r}(x,t)$ may further enable material identification and hence structural diagnosis. 
Two conventional methods for material identification are spectroscopy and ellipsometry, which respectively exploit dispersion and polarization.

\begin{figure}[b]
    \setlength\abovecaptionskip{6pt}
    \vspace{-2ex}
	   \captionsetup[subfigure]{justification=centering}
		\centering
		\subfloat[Time domain.]{
		  \begin{minipage}[b]{0.47\linewidth}
		        \centering
			    \includegraphics[width = 0.96\textwidth]{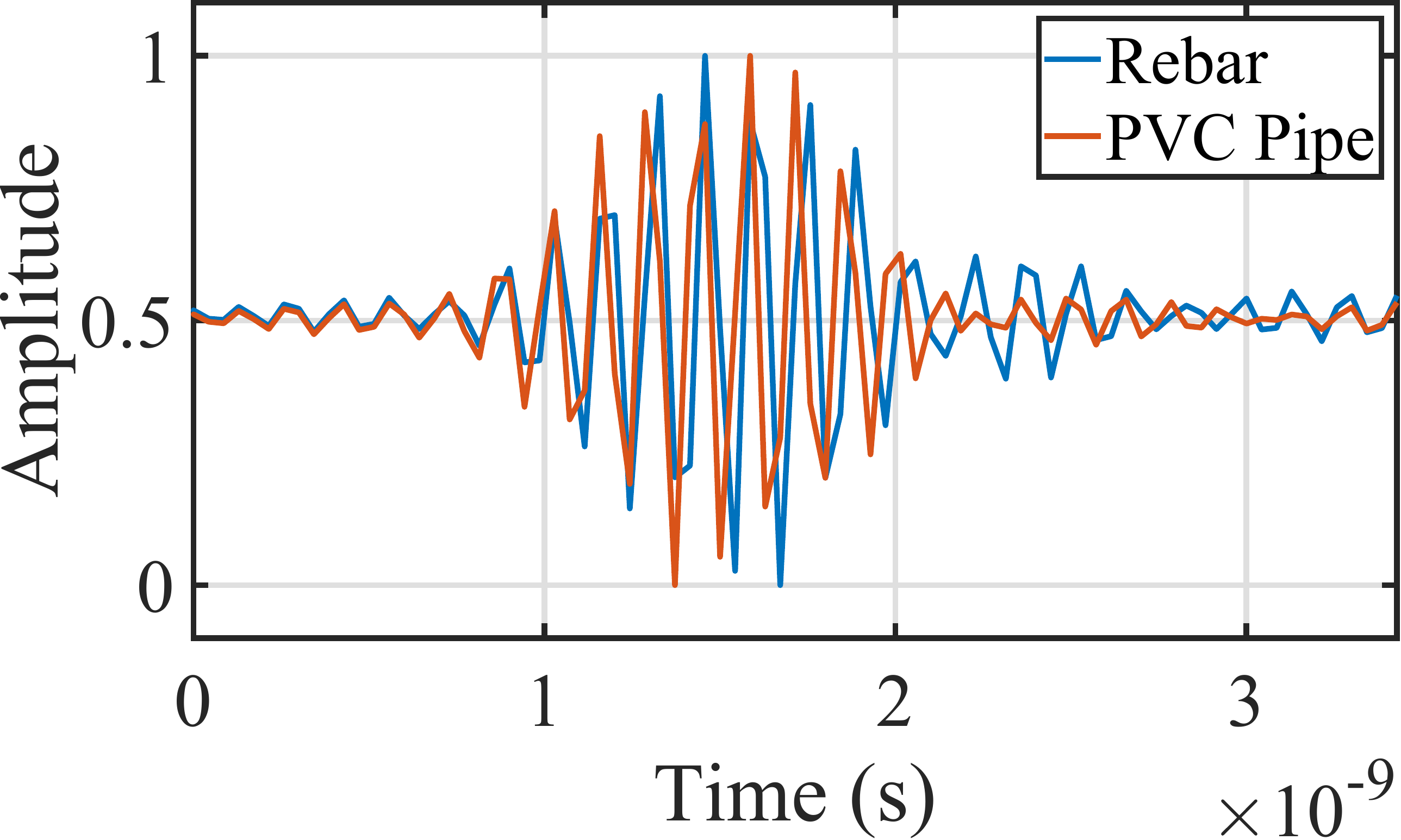}
			\end{minipage}
		}
		\subfloat[Frequency domain.]{
		    \begin{minipage}[b]{0.47\linewidth}
		        \centering
			    \includegraphics[width = 0.96\textwidth]{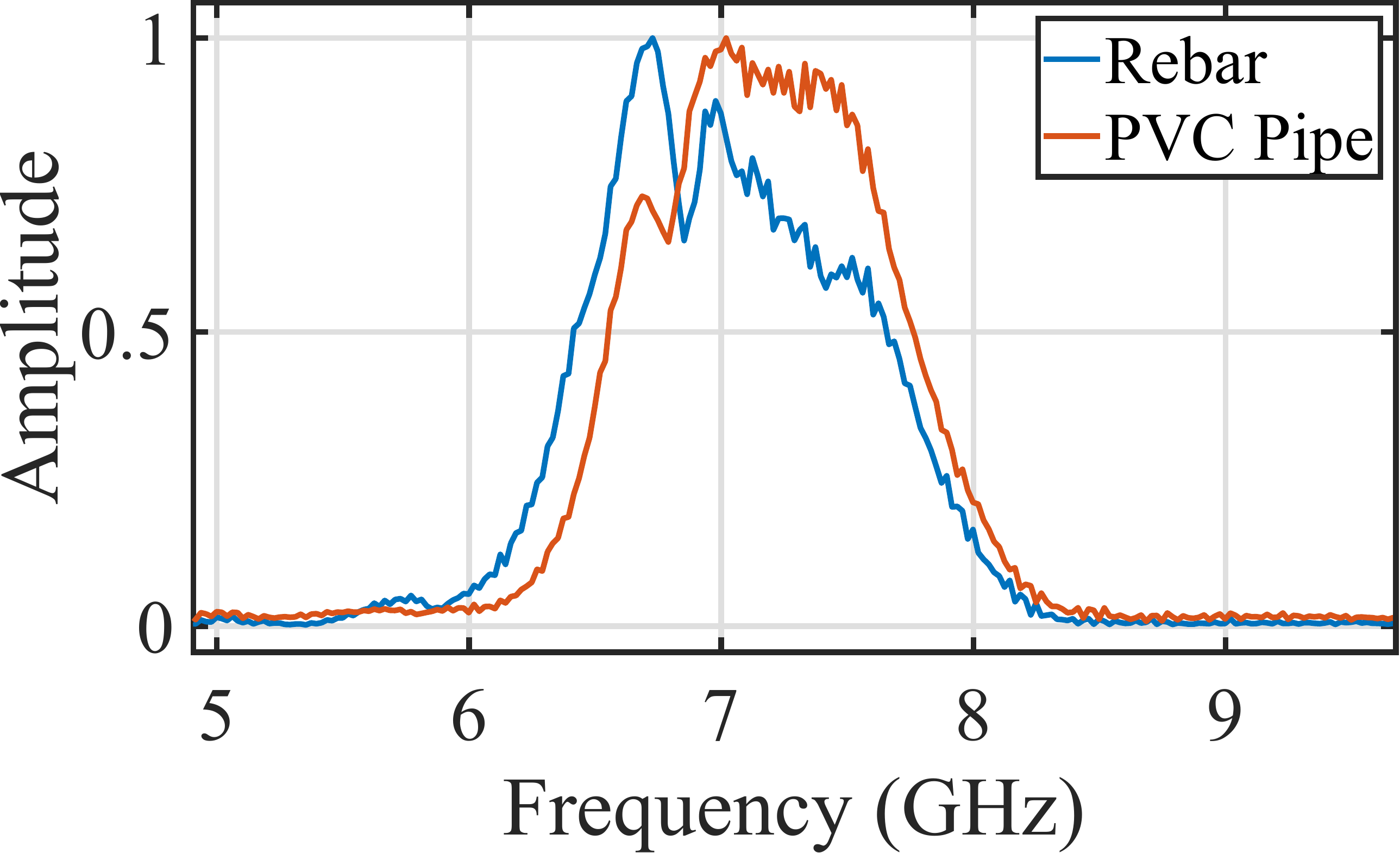}
			\end{minipage}
		}
		\caption{\rev{Dispersions caused by different materials.}}
		\label{fig:dispersion}
\end{figure}

Dispersion is the phenomenon of frequency differentiation of propagation speed. Because different materials exhibit distinct dispersion properties, this phenomenon can be leveraged to characterize respective materials via \textit{spectroscopy}. The effect of dispersion manifests both in the time domain (broadening or narrowing of pulses) and frequency domain (skewed spectra). As a result, time-domain~\cite{dorney2001material} and frequency-domain spectroscopy~\cite{hollas2004modern} have been developed to identify materials. In our case, the wall materials (e.g., concrete), subsurface structure (e.g., rebar), 
and seeping substances (e.g., water) together yield rich dispersion information. As an illustration, we probe a rebar and PVC pipe buried under concrete; the results in Figure~\ref{fig:dispersion} clearly showcase the distinct dispersion characteristics of the two materials.

We further study \textit{polarimetry} (a.k.a. \textit{ellipsometry}) founded on polarization~\cite{azzam1978ellipsometry,zhang2019feasibility,li2020noninvasive,polardraw}. Essentially, when electromagnetic waves are reflected from the boundary of a target (of a certain material), their polarization is altered by the target's properties (e.g., refractive index, thickness, and dielectric function tensor)~\cite{azzam1978ellipsometry}. Therefore, if we deploy two antennas with linear orthogonal polarization, the readings from them should indicate the polarization rotation of the reflected waves, 
which in turn indicates the reflecting material. Normally, \textit{reflection coefficient}s $\gamma_{\mathrm{p}}$ and $\gamma_{\mathrm{s}}$ in the co-polarization and cross-polarization directions are adopted by polarimetry as features, they are represented as~\cite{lindon2016encyclopedia}:
\begin{equation}\label{eq:ellip}
	\!\!\!\!\gamma_{\mathrm{p}} = \frac{N_{\mathrm{s}} \cos \left(\phi_{0}\right)-N_{0} \cos \left(\phi_{\mathrm{s}}\right)}{N_{\mathrm{s}} \cos \left(\phi_{0}\right)+N_{0} \cos \left(\phi_{\mathrm{s}}\right)},
	\gamma_{\mathrm{s}} = \frac{N_{0} \cos \left(\phi_{0}\right)-N_{\mathrm{s}} \cos \left(\phi_{\mathrm{s}}\right)}{N_{0} \cos \left(\phi_{0}\right)+N_{\mathrm{s}} \cos \left(\phi_{\mathrm{s}}\right)},
\end{equation}
where $N_{0}$ and $N_{\mathrm{s}}$ are the refractive indices of the wall and the target material, and $\phi_{0}$ and $\phi_{\mathrm{s}}$ are the complex incidence and refraction angles determined by $\frac{\sin \phi_{0}}{\sin \phi_{\mathrm{s}}}=\frac{N_{\mathrm{s}}}{N_{0}}$~\cite{iizuka2008engineering}. By measuring the reflection coefficients $\gamma_{\mathrm{p}}$ and $\gamma_{\mathrm{s}}$, the refractive indices $N_{\mathrm{s}}$ of the target material can be inferred and in turn used to characterize the material. 

\begin{figure}[t]
    \setlength\abovecaptionskip{6pt}
	   \captionsetup[subfigure]{justification=centering}
		\centering
		\subfloat[Rebar.]{
		  \begin{minipage}[b]{0.47\linewidth}
		        \centering
			    \includegraphics[width = 0.96\textwidth]{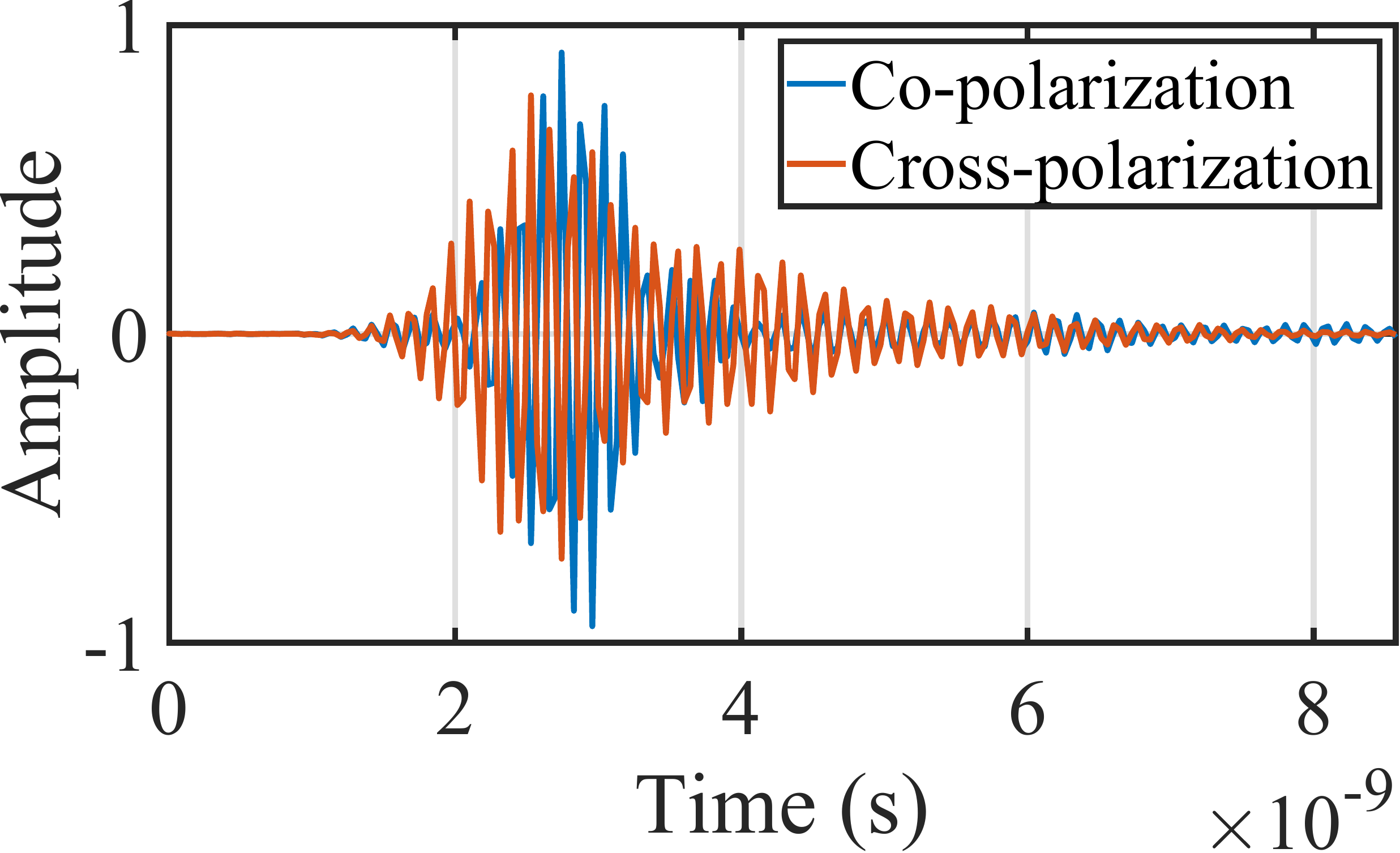} \\
			    \vspace{6pt}
			    \includegraphics[width = 0.96\textwidth]{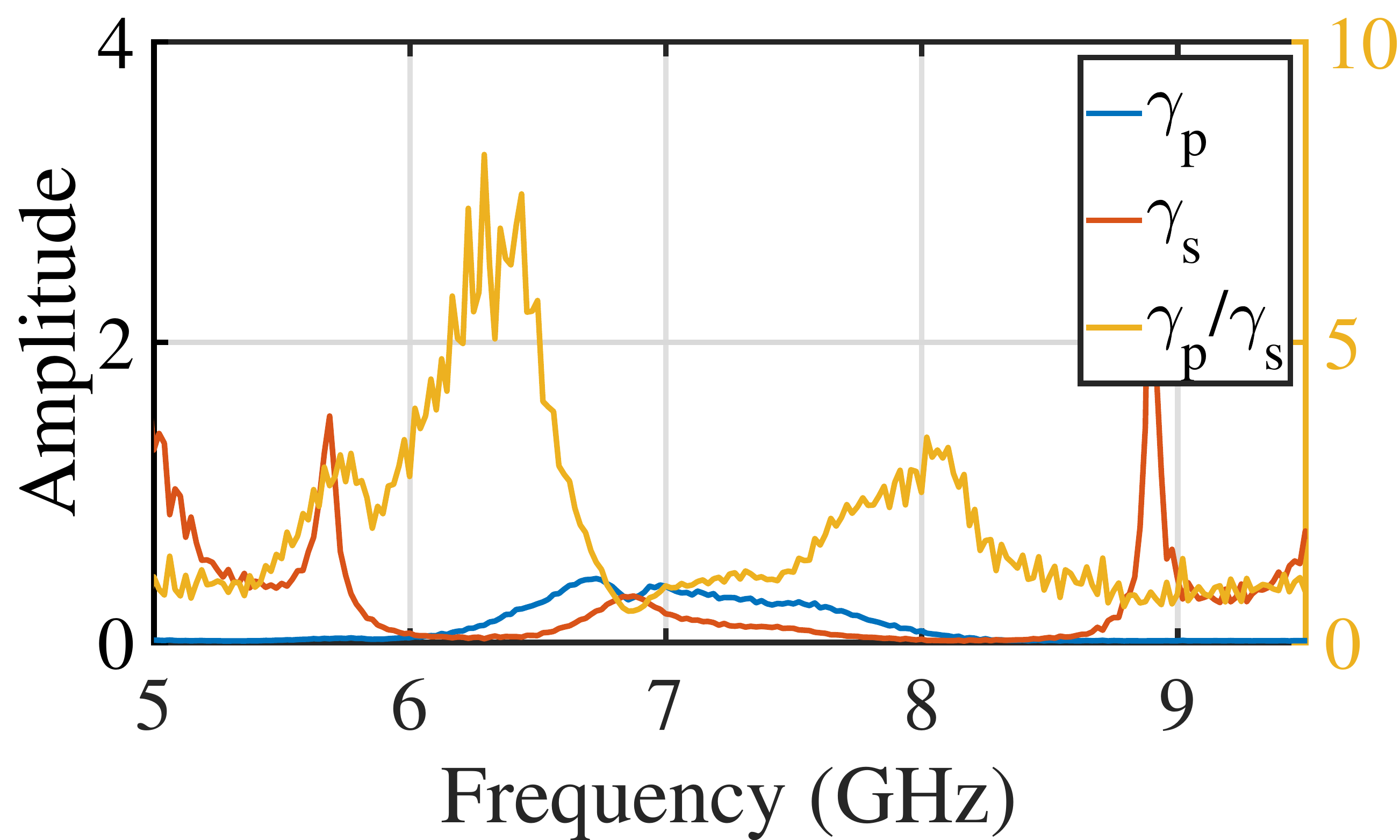}
			    \label{subfig:rebarsignal}
			\end{minipage}
		}
		\subfloat[PVC pipe.]{
		    \begin{minipage}[b]{0.47\linewidth}
		        \centering
			    \includegraphics[width = 0.96\textwidth]{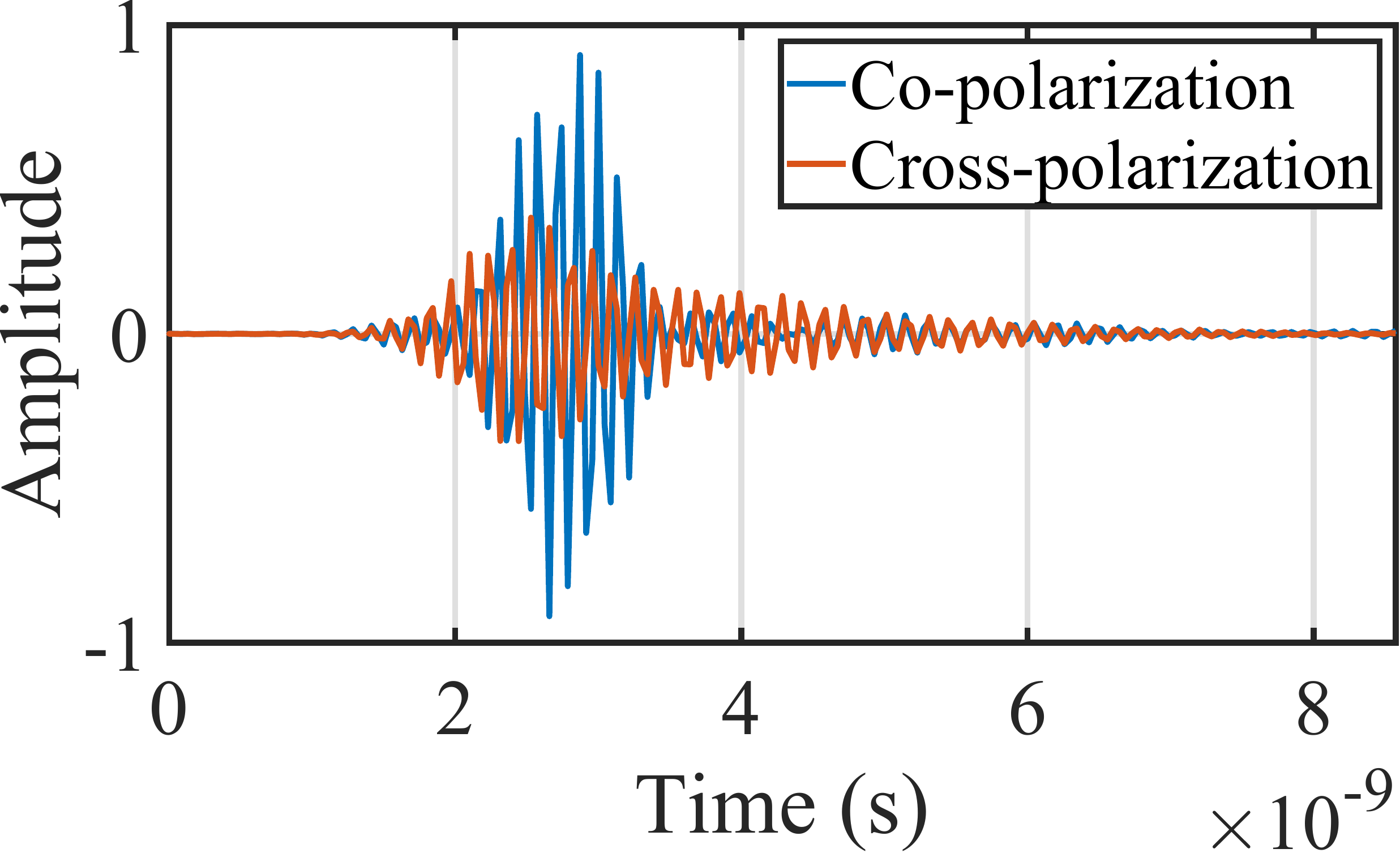} \\
			    \vspace{6pt}
			    \includegraphics[width = 0.96\textwidth]{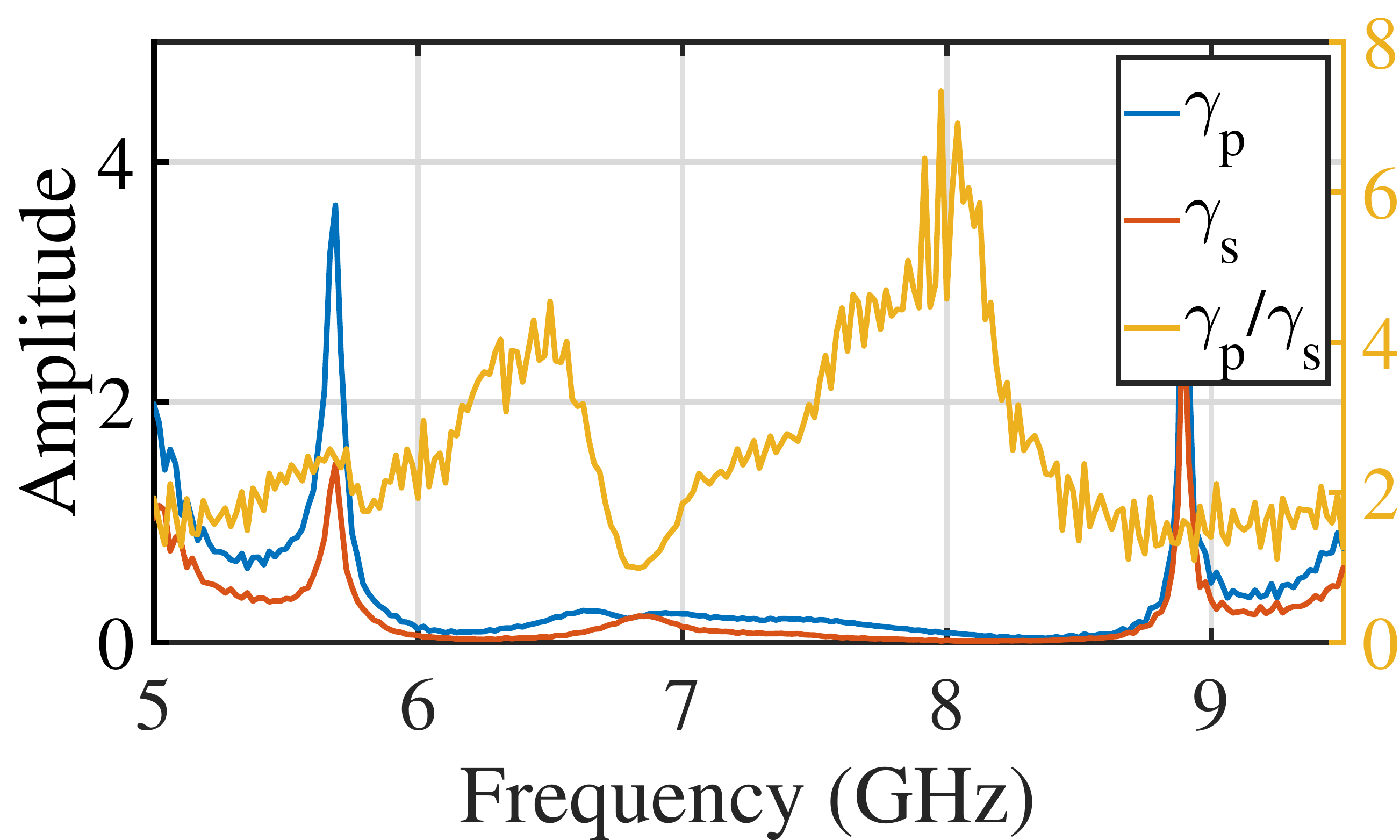}
			    \label{subfig:pvcsignal}
			\end{minipage}
		}
		\caption{\rev{Reflected signals and reflection coefficients from orthogonally polarized antennas
	    for different materials.}}
		\label{fig:polarization}
	    \vspace{-2ex}
\end{figure}

We briefly validate the sensitivity of reflection polarization to materials.
We record reflected signals from two orthogonally polarized antennas (see Figure~\ref{fig:hardware}) for rebar and PVC pipe buried in a concrete wall, and we also compute the reflection coefficients at all frequencies by dividing the spectrum of the received signal by that of the transmitted Gaussian pulse. As shown in Figure~\ref{fig:polarization}, 
distinct materials indeed cause different patterns in the waveforms. In accordance with Eq.~\eqref{eq:ellip}, we also observe different ratios of $\gamma_p$ to $\gamma_s$, evidently confirming the distinct rotations of reflected waves by two materials.
\rev{We refer to~\cite{sagnard2005situ2} for a more detailed benchmark study on RF polarimetry for material identification.}

Unfortunately, these promising techniques face two major issues for our application. On one hand, the signal difference may be conspicuous for two distinct materials, but 
what features to be used for differentiating a variety of materials is a tricky question. On the other hand, the above testing subjects are put at the same depth, but such ideal conditions do not exist in practice, hence environment influence such as wall materials and depth may greatly affect the reliability of these model-based techniques.
Therefore, \textit{one has to identify novel methods that single out the relevant features automatically while masking the environment influence}.

\setcounter{figure}{11}
\begin{figure*}[b]
    \setlength\abovecaptionskip{6pt}
	\centering
	\includegraphics[width=0.92\linewidth]{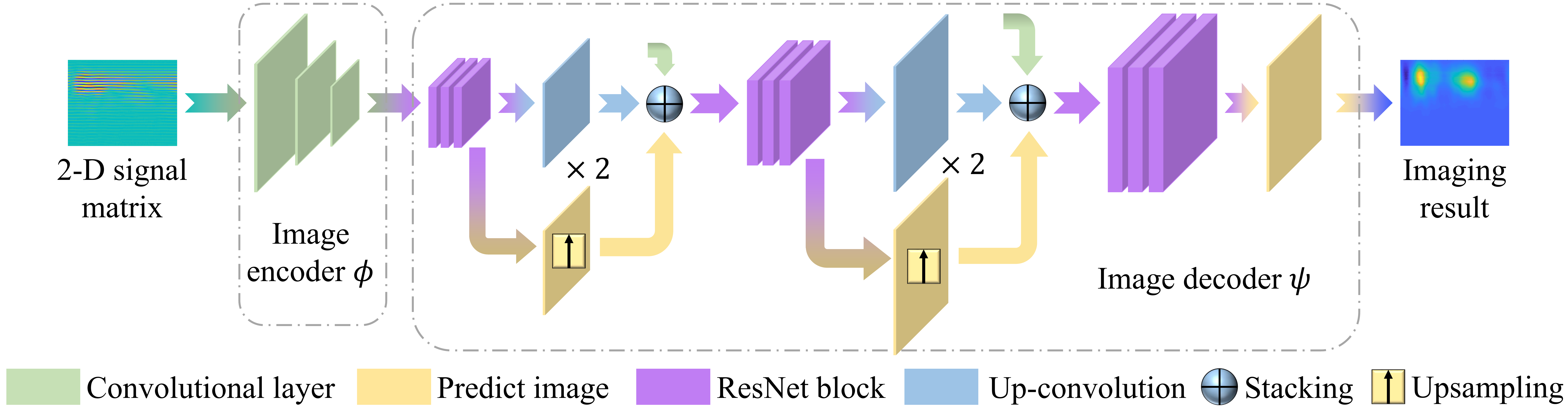}
	\caption{The encoder-decoder architecture of I-Net; the details are shown only for the decoder part.}
	\label{fig:imaging}
\end{figure*}
\subsection{SiWa: A Deep Learning Approach} \label{ssec:siwadl}
According to Section~\ref{ssec:conventional}, model-based methods do not offer usable adaptability for our applications, although the signal matrix $\boldsymbol{r}(x,t)$ does contain sufficient information. To this end, we propose to apply a deep learning framework to integrate structural imaging and material identification functions while combating the challenges faced by model-based approaches. We refine the \sysname\ workflow in Figure~\ref{fig:diagram2}, in which the hardware front-end remains (as in Figure~\ref{fig:diagram}), but $\boldsymbol{r}(x,t)$ is fed into the Imaging Network (I-Net) for producing structural imaging, and the output of I-Net guides the further processing of 1-D signal samples via 
Material identification Network (M-Net) for material identification and diagnosis.

\setcounter{figure}{10}
\begin{figure}[t]
    \setlength\abovecaptionskip{6pt}
    \vspace{2ex}
	\centering
	\includegraphics[width=0.96\linewidth]{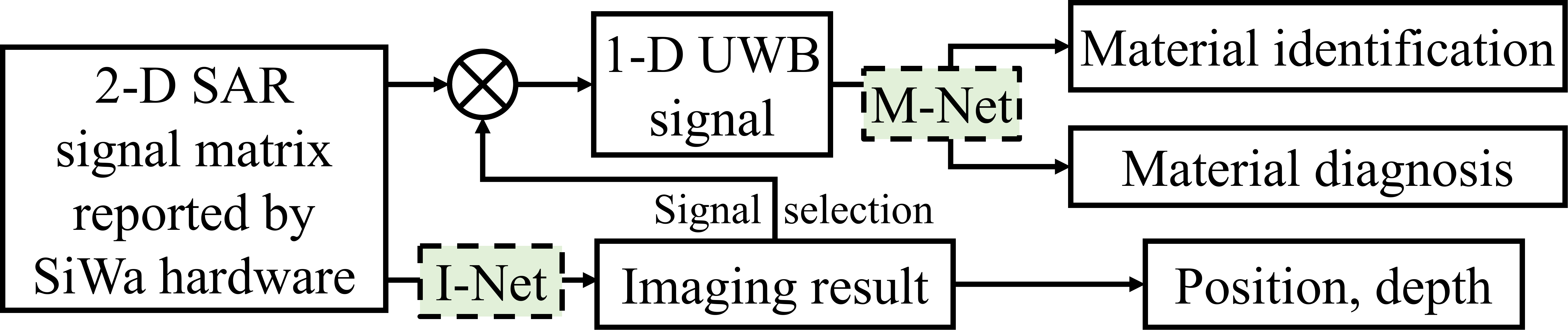}
	\caption{A deep learning view of SiWa’s design.}
	\label{fig:diagram2}
	\vspace{-2ex}
\end{figure}
\subsubsection{\sysname\ Imaging Network (I-Net).}\label{I-Net}
Let us first recapitulate the limitations of SAR imaging algorithms: the speed of hand movement and the permittivity of a wall have to be specified beforehand; otherwise, the imaging results can be blurred and excessive noise can be caused by the wall, as shown in Figure~\ref{fig:velocity}. 
Fortunately, it is well-known that a \textit{deep neural network} (DNN) can approximate arbitrary functions~\cite{csaji2001approximation} and it can be trained to be robust against parameter changes~\cite{Orabona20141116}. Therefore, it makes sense for us to replace SAR imaging with a DNN that approximates the SAR capability while being adaptive to for example, wall properties. 
In fact, existing studies have already confirmed that commonly used filters 
can be approximated by DNNs~\cite{dao2019learning, george2018deep}. Moreover, recent advances in optics~\cite{wu2019three, jin2018machine} also support our idea that deep learning can be used for imaging purposes free of parameter tuning. 

Specifically, we aim to design an encoder-decoder model~\cite{ronneberger2015u} for ``focusing'' the RF signal matrix $\boldsymbol{r}(x,t)$ (in moving-range domain) into an image of the subsurface structure $\tilde{\boldsymbol{r}}(x,z)$ (in moving-depth domain). 
The encoder-decoder module 
consists of two parts: encoder $\phi$ and decoder $\psi$, with respective parameters $\theta_\phi$ and $\theta_\psi$. The former maps the input space $\mathcal{R}$ to the (hidden) feature representation space $\mathcal{H}$, while the latter leverages $\mathcal{H}$ to generate the output space $\tilde{\mathcal{R}}$. The training of I-Net is implemented by minimizing a loss function as follows:
\begin{align}
\hat{\theta}_\phi, \hat{\theta}_\psi&=\underset{\theta_\phi, \theta_\psi}{\arg \min }\lVert\hat{\boldsymbol{r}}-\psi(\phi(\boldsymbol{r}))\rVert_{1}; \label{eq:encoder-decoder} 
\end{align}
the loss function measures the difference between a ground truth image (produced by RMA with correct parameters) $\hat{\boldsymbol{r}}\in \mathbb{R}^{N_{x} \times N_{z}}$ and an output image $\tilde{\boldsymbol{r}}=\psi( \phi(\boldsymbol{r}))$, where $\boldsymbol{r} \in \mathbb{R}^{N_{x} \times N_{t}}$, and $N_{x}$, $N_z$, and $N_{t}$ are the number of samples along the moving, depth, and range axes, respectively.

The architecture of I-Net is shown in Figure~\ref{fig:imaging}. First, $\boldsymbol{r}(x,t)$ is fed into the multi-level encoder $\phi$, a contraction network adopting regular Convolutional Neural Network (CNN)~\cite{lecun1999object}. 
The feature representations distilled at every level of $\phi$ are subsequently fed to the peer level in the decoder $\psi$, which reverses the contraction and translates the feature representations to the imaging results. The decoder adopts more powerful ResNet blocks~\cite{he2016deep}.
Replacing pooling operators with upsampling operators, the resolutions of intermediate images are gradually increased. At each level of $\psi$, the features from i) skip connection with the peer level in $\phi$ (green), ii) up-convolutional layer (blue), and iii) upsampling layer (yellow) are stacked together
and forwarded to the next level for further processing. Multiple levels of these functional blocks are connected in series to achieve a deeper network for better performance. As the last stage, a ResNet block followed by a convolutional layer gathers the intermediate image and generates the final result $\tilde{\boldsymbol{r}}(x,z)$. \rev{The sufficient capacity provided by the encoder-decoder model enables I-Net to discover environment parameters implicitly via $\phi$, rendering structural imaging robust to changes of multiple environment conditions.} With the structural imaging results provided at the output of I-Net, \sysname\ can now locate subsurface structures and further conduct material identification and diagnosis.

\setcounter{figure}{12}
\begin{figure*}[t]
	\centering
	\includegraphics[width=0.96\linewidth]{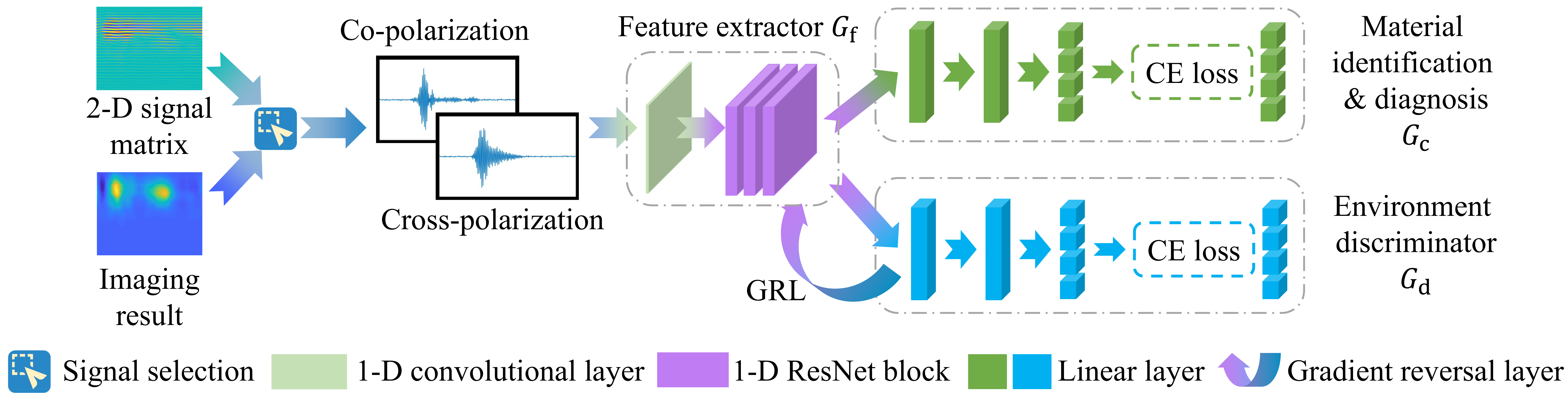}
	\caption{The architecture of M-Net along with its adversarial learning strategy.}
	\label{fig:material_arch}
\end{figure*}
\subsubsection{\sysname\ Material Identification Network (M-Net).}\label{M-Net} 
In Section~\ref{ssec:conventional}, we have explored the possibilities of employing the dispersion and polarization of the reflected signals for material identification. We also point out that these model-based methods face major difficulties in practice given the environment influence. To tackle this challenge,
we aim to design M-Net as an effective classifier robust to environment influence.
\sysname\ obtains the position of subsurface structures using the Constant False Alarm Rate (CFAR) algorithm~\cite{levanon1988radar} from the images produced by I-Net. It then extracts the 1-D signal sequence $r(t)\in \mathbb{R}^{N_{t}\times 1} $ from the 2-D radar signal matrix $\boldsymbol{r}(x,t)\in \mathbb{R}^{N_{x} \times N_{t}}$. Ideally, \sysname\ could perform beamforming to synthesize multiple $r(t)$ into one with stronger SNR, but that would require estimating the radar swiping speed $v$. Therefore, our current prototype takes the highest SNR $r(t)$ among those close to the known location. 

As a classifier, M-Net $G$ maps the input 1-D signal space $\mathcal{R}^1$ to the label space $\mathcal{Y}$ that 
is a finite set indicating materials of different conditions (e.g., leaked PVC pipe, non-corroded rebar, etc.). Ideally, the parameters $\theta_G$ of M-Net can be trained by minimizing the cross-entropy loss $L(y, G(r))$ between the estimated and actual label:
\begin{align}
\hat{\theta}_G&=\underset{\theta_G}{\arg \min } \, L(y, G(r)).\label{eq:naive}
\end{align}
However, the above formulation neglects the environment influence: it works only when the training and testing environments remain the same, because the training may overfit the features offered by the environment (e.g., wall thickness). 
To remove the environment influence on the classification results, we leverage the power of adversarial learning~\cite{ganin2015unsupervised}, and instead train M-Net as follows:
\begin{align}
(\hat{\theta}_\mathrm{f}, \hat{\theta}_\mathrm{c}) = \underset{\theta_\mathrm{f}, \theta_\mathrm{c}}{\arg \min } \,  L(y,e,r), \quad
\hat{\theta}_\mathrm{d} = \underset{\theta_\mathrm{d}}{\arg \max } \, L(y,e,r),
\label{eq:adv}
\end{align}
where $L(y,e,r) = L_\mathrm{c}\left(y, G_\mathrm{c}(G_f(r))\right)-\lambda L_\mathrm{d} \left(e, G_\mathrm{d}(G_\mathrm{f}(r)\right),G_\mathrm{f}$ is the feature extraction network, $G_\mathrm{c}$ is the material classifier, and $G_\mathrm{d}$ is the environment discriminator; their corresponding parameters are $\theta_\mathrm{f}$, $\theta_\mathrm{c}$, and $\theta_\mathrm{d}$. $L_\mathrm{c}$ and $L_\mathrm{d}$ are cross-entropy losses for the classifier and discriminator, with $\lambda$ controlling the trade-off between them.

The rationale behind this formulation is that, as M-Net is trained to predict material class labels using samples collected from different environments,
the features used for classification should come only from materials but not from environments. 
To this end, an environment discriminator $G_\mathrm{d}$ is introduced to predict environment class labels $e$. While the training process aims to improve the environment classification accuracy by optimizing $\hat{\theta}_\mathrm{d}$, an adversarial learning method is also taken to ``cheat'' $G_\mathrm{d}$ by negating the loss function via $-\lambda$. This procedure suppresses material-unrelated features and thus prevents M-Net from overfitting to specific environments. 
The network and its training strategy are illustrated in Figure~\ref{fig:material_arch}. The feature extractor $G_\mathrm{f}$ takes in two data sequences with different polarization directions, and the extracted features are fed into the material classifier $G_\mathrm{c}$ and environment discriminator $G_\mathrm{d}$. The loss negation via $-\lambda$ is achieved by reversing the gradient during the backpropagation process with the Gradient Reversal Layer (GRL), which only works for training purposes.
In order to prevent conflicting gradients~\cite{yu2020gradient}, we augment the backpropagation by projecting the gradients of $G_\mathrm{f}$ and $G_\mathrm{c}$ onto each other.

\section{Implementations} \label{sec:implementation}

\subsubsection*{Hardware Implementations}
\begin{figure}[t]
    \setlength\abovecaptionskip{8pt}
	\centering
	\includegraphics[width=0.9\linewidth]{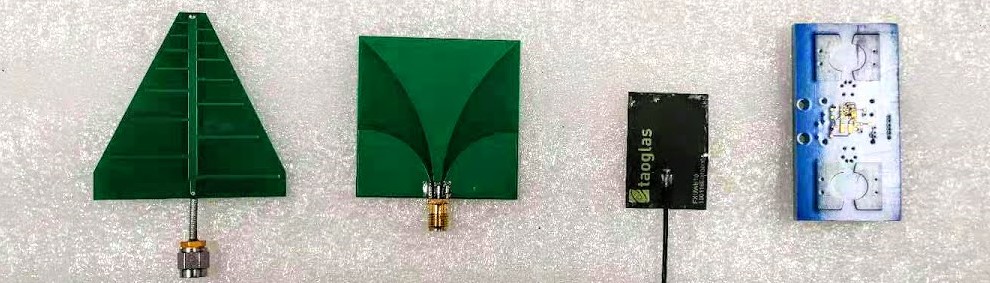}
	\caption{Four antennas from left to right: log-periodic, Vivaldi, patch, and planar dipole antenna.}
	\label{fig:ant}
	\vspace{-2ex}
\end{figure}

\begin{figure}[b]
    \setlength\abovecaptionskip{8pt}
    \vspace{-2ex}
	\centering
	\includegraphics[width=0.88\linewidth]{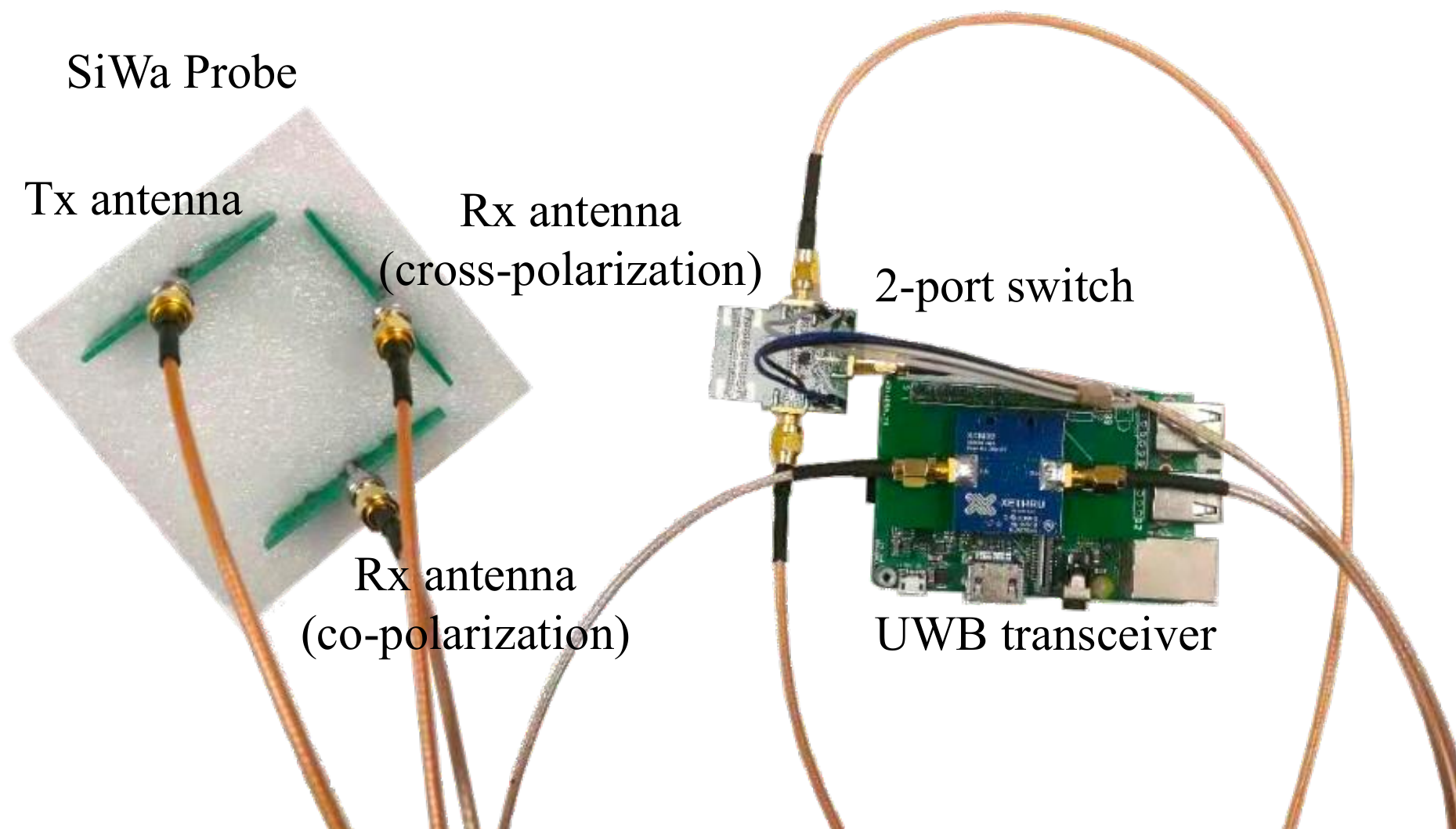}
	\caption{Hardware prototype of \sysname.}
	\label{fig:hardware}
\end{figure}

\sysname\ performs imaging and material identification using UWB signals. The core component is a compact and low-cost Novelda X4M05~\cite{xethru} IR-UWB transceiver. \rev{It transmits a baseband signal with a bandwidth of 1.5~\!GHz, and the signal is further modulated onto a 7.29~\!GHz carrier.} A Raspberry Pi single-board computer~\cite{rpi} is used to control the transceiver and interface with a laptop computer. The antennas are critical hardware components and they should have a good coupling ability with walls so as to deliver the maximum amount of power $P_{\text{in}}$ into walls and minimize $P_{\text{ref}}$ reflected by the air-wall boundary.
Given this requirement, we define the SNR of an antenna as $\frac{P_{\text {in}}}{P_{\text {ref }}}$ to measure the performance of the antenna. We choose four common antenna types: log-periodic, Vivaldi, patch, and planar dipole antennas, as shown in Figure~\ref{fig:ant}, and their respective SNRs are -4.17~\!dB, -17.23~\!dB, -22.21~\!dB, and -7.09~\!dB. Therefore, we choose log-periodic antennas for \sysname\ as they deliver the highest SNR.
To utilize the information carried by reflected signals of different polarization states, as described in Section~\ref{sssec:material_id}, we design a probe with 1 tx antenna, and 2 perpendicularly placed rx antennas (connected via a 2-port switch). The antennas are fixed into a foam block and spaced 4~\!cm (about 1 wavelength) apart to avoid mutual coupling. The whole hardware prototype is shown in Figure~\ref{fig:hardware}. %

\begin{figure*}[t]
	\centering
	\includegraphics[width=0.95\linewidth]{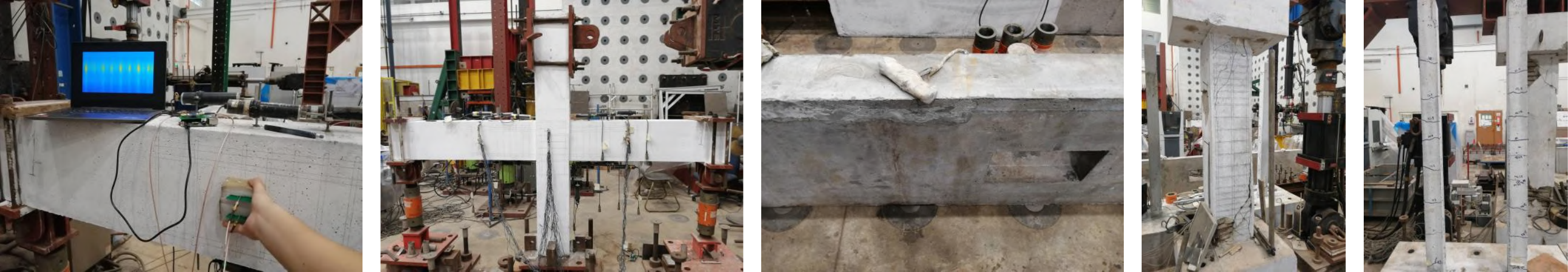}
	\caption{We perform extensive field tests in a precast concrete plant. Concrete blocks with different shapes, mix ratios, water contents, and aging degrees are investigated under our tests.}
	\label{fig:plant}
	\vspace{-1.5ex}
\end{figure*}
\subsubsection*{Software Implementations}
\sysname's imaging and material identification modules are implemented using Python 3.7 and PyTorch 1.7.1, and trained using GeForce RTX 2080 Ti graphics card. %
CPUs from PC workstations and single-board computers 
are used during the inference stage of structural imaging and material identification. The software module for controlling the X4 transceiver and antennas is implemented in C\texttt{++}. The frame rate of the X4 transceiver is set to 40 frames per second. To get the most out of the X4 chip, we modify its driver program to directly obtain the raw RF signal, instead of the downsampled baseband signal accessible by default. 

The parameters of I-Net are set as follows: for the 2-D convolutional layer, we set the kernel size to 3, stride to 1, padding size to 1, and the number of output channels is 2 times the number of input channels; for the 2-D up-convolutional layer, we set the kernel size to 3, stride to 1, scale factor to 2, and the number of input channels is 2 times of the number of output channels. All weights are initialized by the Xavier uniform initializer~\cite{kumar2017weight}. We divide the collected dataset into training and test sets, \rev{based on the principle that they come from different environments.} The training set contains 8,000 pairs of 2-D signal matrix and imaging ground truth, and the test set contains 6,000 pairs. The size of a training image is $200 \times 200$ (though I-Net can handle variable-length inputs during inference), and the ground truth images are obtained by RMA. 
As a preprocessing, the 2-D radar signal matrices are normalized to have zero mean and a standard deviation of 0.017, the ground truth images are normalized to a mean of 0.092 and standard deviation of 0.2423. For the training process, we set the batch size to 64, adopt the $L^1$ loss, and use the Stochastic Gradient Descent optimizer~\cite{bottou2012stochastic}, whose learning rate and momentum are set to 0.001 and 0.9, respectively.

The parameters of M-Net are set as follows: the kernel size of the convolutional layer is 3, and the stride is 2, $\lambda$ that controls the trade-off between $G_\mathrm{c}$ and $G_\mathrm{d}$ is 1. $G_\mathrm{d}$ consists of 3 consecutive fully-connected layers, transforming the 15,360 features output by $G_\mathrm{f}$ to the final layer representing different environments. We divide the collected dataset to a training set containing 8,000 1-D signals collected directly and a test set contains 6,000 1-D signals extracted from 2-D signal matrices, \rev{and the training and test sets are so partitioned that they come from different environments.} The length of each signal sequence is 1120. %
In the preprocessing stage, the two-channel signals are normalized to a mean of $(0, 0)$ and a standard deviation of $(0.0052, 0.0021)$. During the training process, we set the batch size to 128, adopt cross-entropy loss for both the material classifier and environment discriminator, and use the Adam optimizer~\cite{kingma2014adam} with a learning rate of 0.001. 

\section{Evaluation} \label{sec:evaluation}
In this section, we evaluate the performance of \sysname. We embed different objects (corroded rebar, non-corroded rebar, PVC pipe, and leaked PVC pipe) in various walls. \rev{The rebar is HRB400-grade and has a nominal diameter of 10~\!mm, and the PVC pipe has a diameter of 20~\!mm and a wall thickness of 4~\!mm.} Aside from common drywall (made of gypsum), brick, and wooden walls that can be readily handled in our lab, we pay special attention to concrete walls since they are the most important building material, yet they cause the highest loss to RF signals. We have conducted extensive field tests on concrete blocks with different shapes, mix ratios, water contents, aging degrees, and embedded materials in a precast concrete plant, as shown in Figure~\ref{fig:plant}; the ground truth positions, depths, materials of the embedded objects are all clearly labeled. 

\rev{In the experiment, the permittivity of a wall is measured using a Keysight N5234B PNA-L vector network analyzer~\cite{keysight}. We maintain a relatively stable hand moving speed by following the visual marks on the wall and the auditory beats generated by a metronome. The hand moving speed is then derived from distance changes per unit time measured by a Bosch GLM 50C laser measure~\cite{bosch2}. The outcome is that every collected signal matrix is induced by a different nominal speed along with minor (unknown) speed variations. We shall refrain from comparing I-Net with through-wall~\cite{WiFiThruW, amin} and in-air~\cite{WiFiThruW, amin, ThruFog, deepmeet} RF imaging techniques, because i) through-wall Wi-Fi sensing
fails to support in-wall imaging, as proven by Figure~\ref{fig:wificross}, and ii) in-air imaging faces challenges such as excessive air-wall boundary loss and varying environment parameters, as demonstrated in Figures~\ref{fig:noncontact} and~\ref{fig:velocity}. Proposals for material identification (e.g.,~\cite{material}) are also excluded from comparing with M-Net, as they often classify materials based on photos and are thus totally irrelevant to RF-sensing.
In the following, we specify comparison baselines for structural imaging and material identification and report their evaluation results respectively.}

\vspace{-.5ex}
\subsection{Structural Imaging}
\begin{figure}[t]
    \setlength\abovecaptionskip{6pt}
	   \captionsetup[subfigure]{justification=centering}
		\centering
		\subfloat[Concrete specimen with rebars embedded.]{
		    \begin{minipage}[b]{0.8\linewidth}
		        \centering
			    \includegraphics[width = 0.96\textwidth]{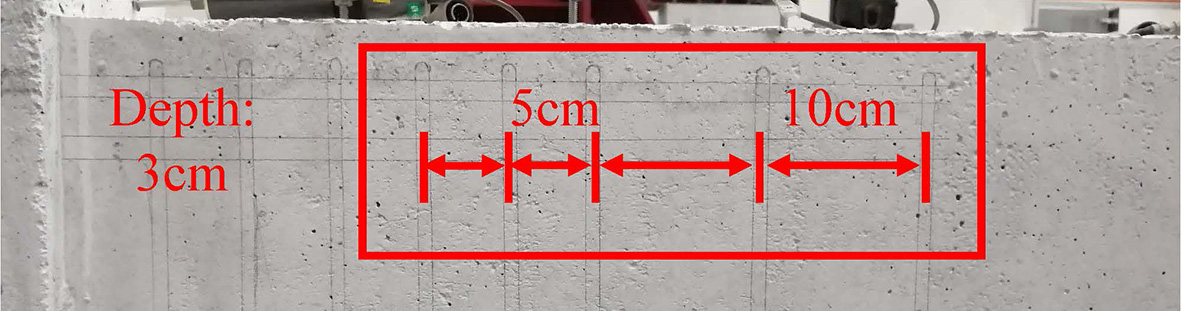}
			\end{minipage}
			\label{fig:quality_l}
		}
		\\
	    \subfloat[Imaging results.]{
		    \begin{minipage}[b]{1\linewidth}
		        \centering
			    \includegraphics[width = 0.9\textwidth]{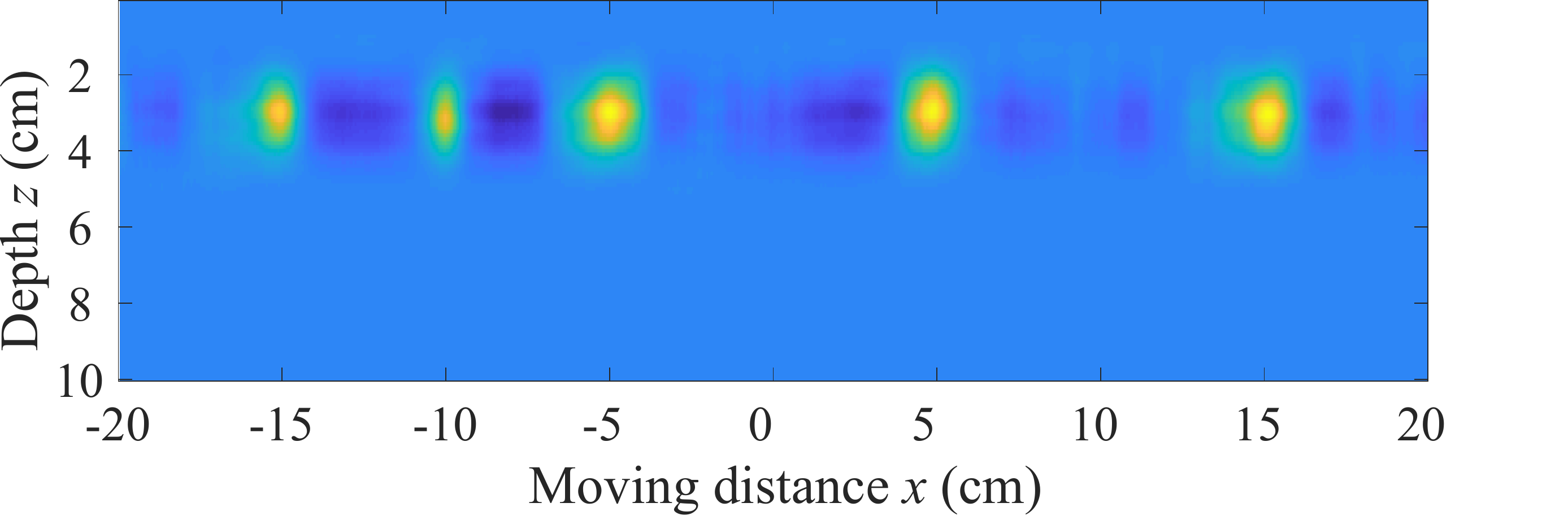}
			\end{minipage}
			\label{fig:quality_i}
		}
		\caption{A qualitative test on a concrete block with five rebars embedded. 
		}
		\label{fig:quality}
	    \vspace{-2ex}
\end{figure}

We start with evaluating the performance of the I-Net 
proposed in Section~\ref{I-Net}; a qualitative result is provided first for a concrete block with rebars embedded, where we mark the ground truth labels in Figure~\ref{fig:quality_l} and report the imaging results obtained by swiping the probe for a 40~\!cm distance along the concrete surface in Figure~\ref{fig:quality_i}.
We can see that \sysname\ accurately produces the structural image within the concrete. Note that what we see are the cross-sections of rebars; multiple parallel scans are needed to obtain a 3-D tomography map that extends along their axial direction. We hereafter report more on I-Net's quantitative evaluations. \rev{Moreover, we compare I-Net with RMA~\cite{cafforio1991sar} instead of other methods summarized in~\cite{deepmeet}, as it is the most practiced in-air SAR imaging algorithm capable of obtaining high
resolution 
if all parameters are properly specified. We also refrain from comparing with Walabot~\cite{walabot}, because it is a closed-source system that cannot be customized.}

\vspace{-.5ex}
\subsubsection{Horizontal Distance Accuracy} \label{sssec:distance_acc}
\begin{figure}[ht]
    \setlength\abovecaptionskip{6pt}
    \vspace{-2.5ex}
	   \captionsetup[subfigure]{justification=centering}
		\centering
		\subfloat[I-Net.]{
		  \begin{minipage}[b]{0.49\linewidth}
		        \centering
			    \includegraphics[width = \textwidth]{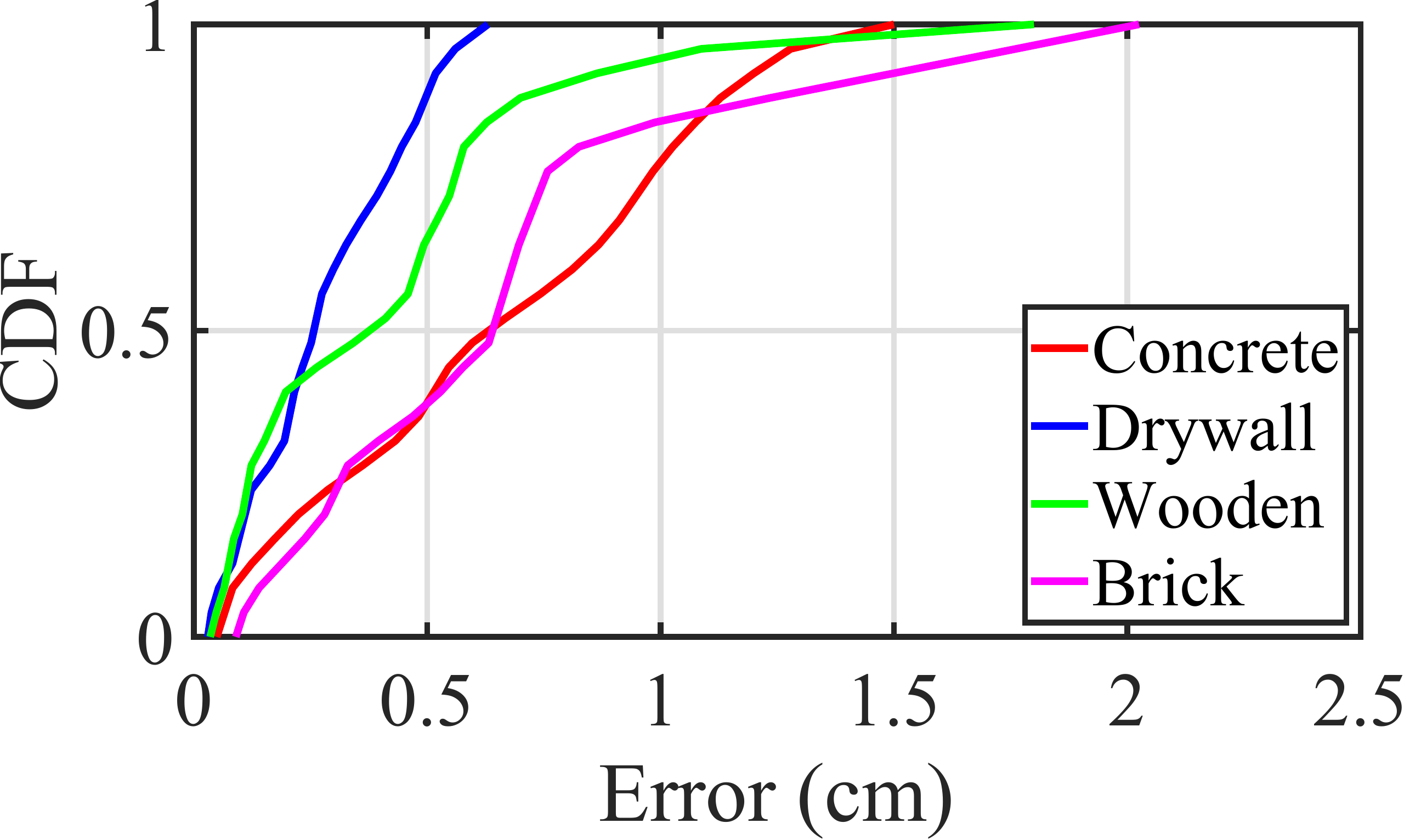}
			    \label{fig:azimuth_inet}
			\vspace{-3ex}
			\end{minipage}
		}
		\subfloat[RMA.]{
		    \begin{minipage}[b]{0.49\linewidth}
		        \centering
			    \includegraphics[width = \textwidth]{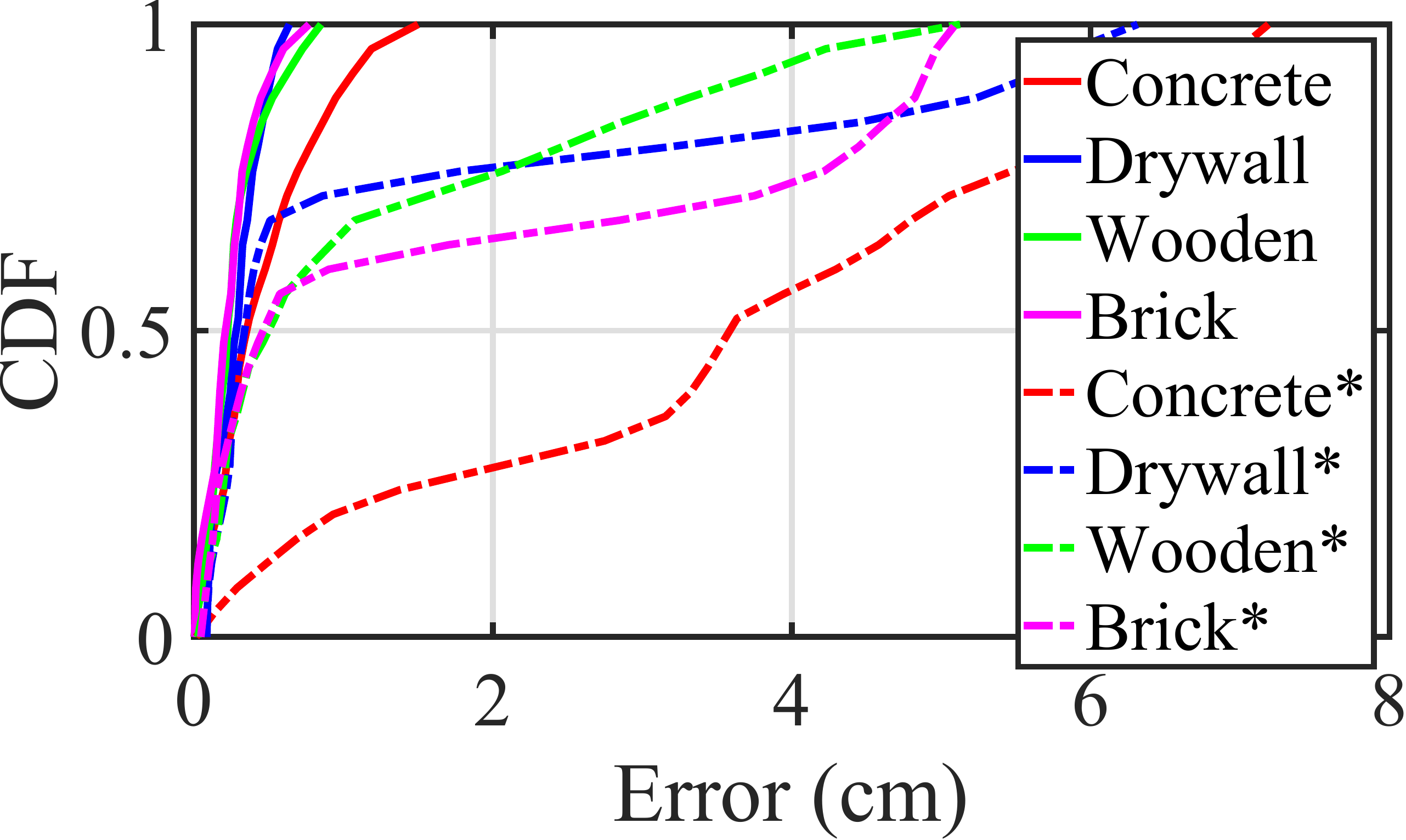}
			    \label{fig:azimuth_rma}
			\vspace{-3ex}
			\end{minipage}
		}
		\caption{CDFs of horizontal distance errors.}
		\label{fig:azimuth_cdf}
	    \vspace{-2ex}
\end{figure}

Horizontal distance accuracy along the radar scanning direction $x$
is first evaluated. The error is defined as the absolute difference between the estimated horizontal distance $x_\mathrm{e}$ and the actual horizontal distance $x_\mathrm{a}$, namely, $|x_\mathrm{e}-x_\mathrm{a}|$. For RMA, we consider two cases: i) the parameters (i.e., the permittivity of a wall and the speed of hand movement) of the algorithms are accurate, and ii) the parameters have errors within $\pm 20 \%$, representing common parameter estimation situations
in real-life usage.  
Figure~\ref{fig:azimuth_cdf} plots the Cumulative Distribution Functions (CDFs) of horizontal distance error given different walls. For I-Net, we observe from Figure~\ref{fig:azimuth_inet} that the median errors are 0.2~\!cm, 0.4~\!cm, 0.7~\!cm, and 0.7~\!cm for drywall, wooden wall, brick wall, and concrete wall, respectively. These errors are in line with common sense that concrete and brick walls cause higher attenuation to RF signals, hence larger localization errors. As a comparison, RMA, whose CDFs shown in Figure~\ref{fig:azimuth_rma}, has slightly better performance: median errors 0.2~\!cm, 0.2~\!cm, 0.3~\!cm, and 0.4~\!cm given the same four cases, if the parameters are set correctly.
However, if the estimated parameters deviate from the actual values, the median errors increase to 0.3~\!cm, 0.4~\!cm, 0.5~\!cm, and 3.6~\!cm (legend bearing asterisk in the plot), and the curves (dashed lines) have very long tails, indicating a serious performance degradation. 

\begin{figure}[!t]
    \setlength\abovecaptionskip{6pt}
	   \captionsetup[subfigure]{justification=centering}
		\centering
		\subfloat[I-Net.]{
		  \begin{minipage}[b]{0.49\linewidth}
		        \centering
			    \includegraphics[width = \textwidth]{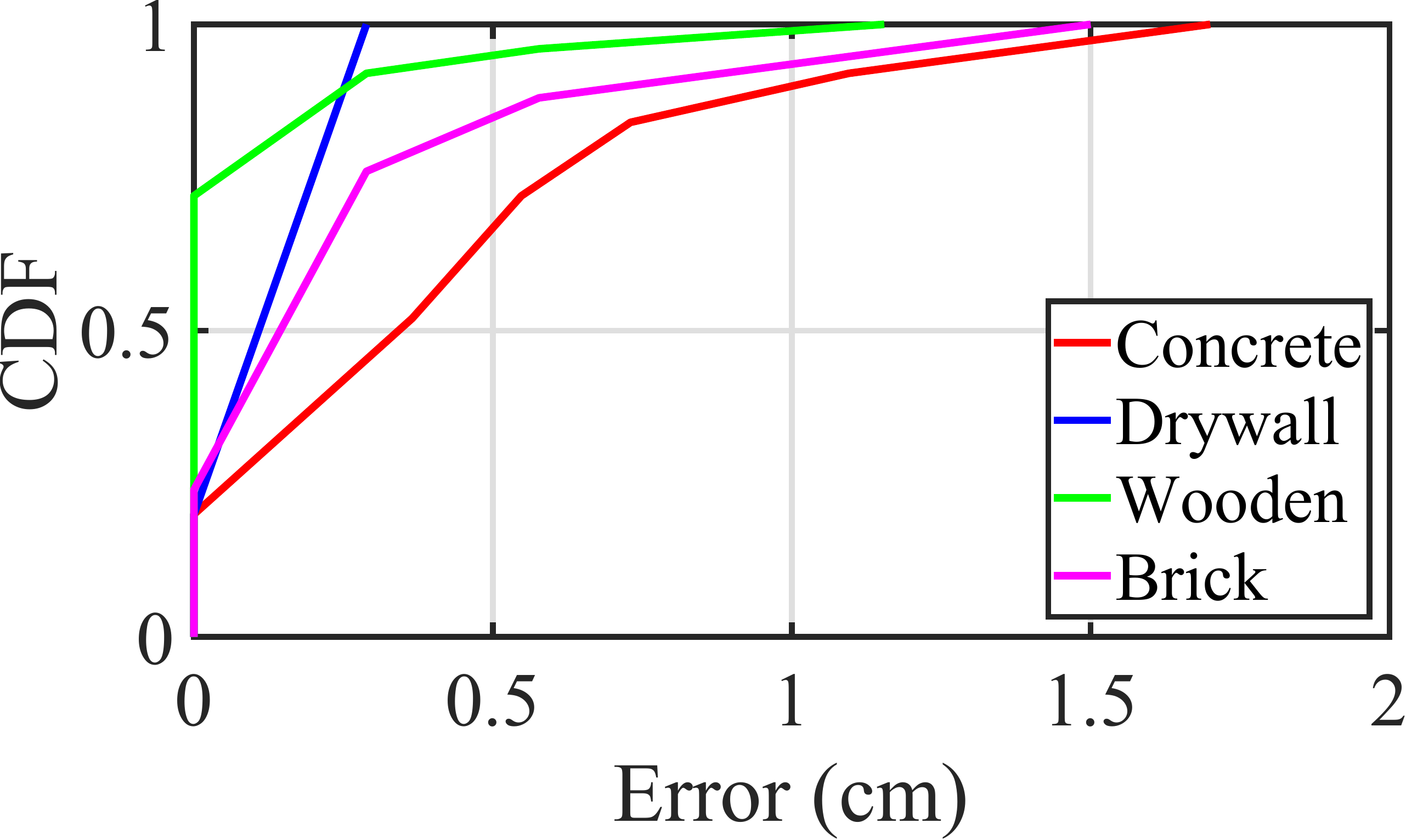}
			    \label{fig:depth_inet}
			    \vspace{-3ex}
			\end{minipage}
		}
		\subfloat[RMA.]{
		    \begin{minipage}[b]{0.49\linewidth}
		        \centering
			    \includegraphics[width = \textwidth]{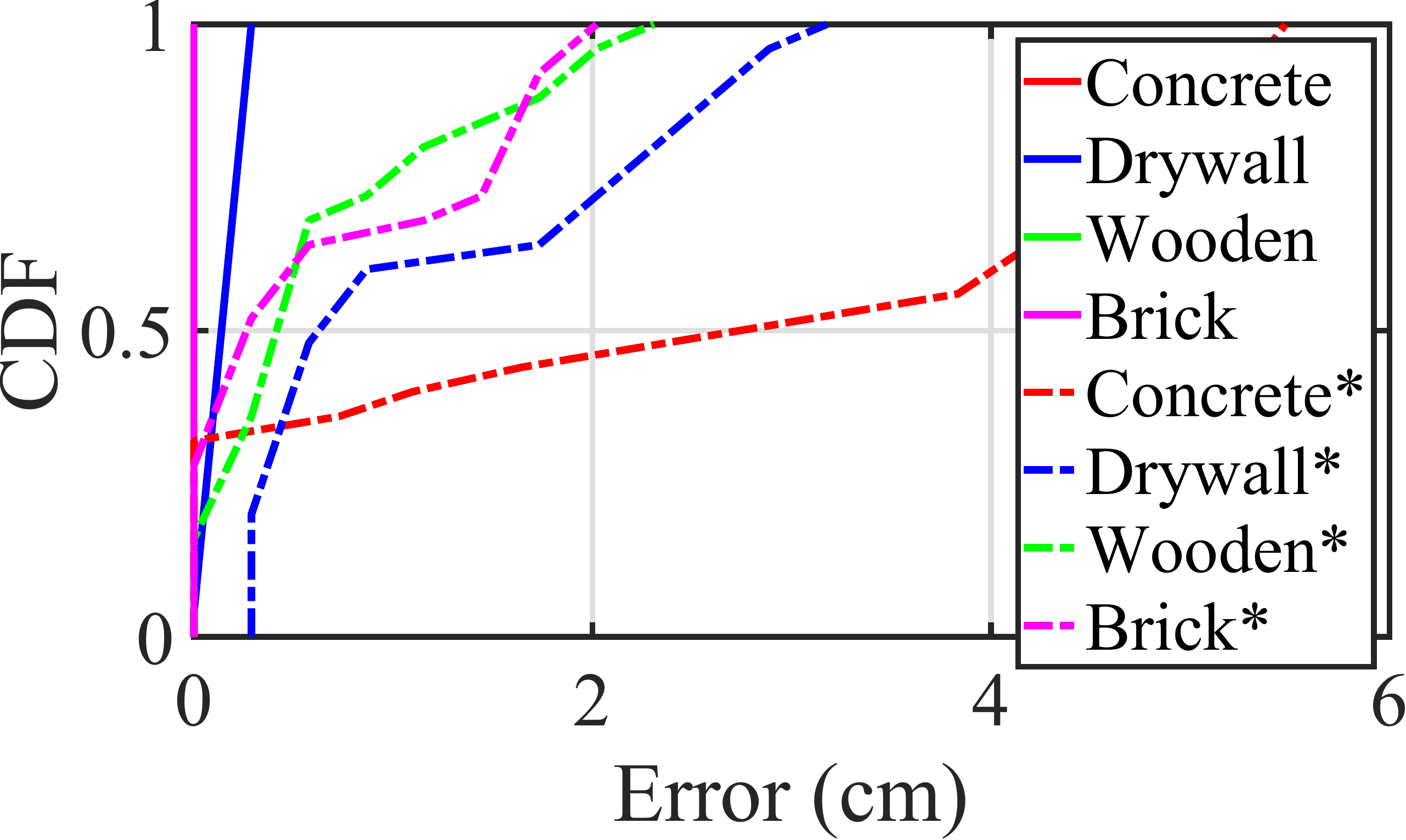}
			    \label{fig:depth_rma}
			    \vspace{-3ex}
			\end{minipage}
		}
		\caption{CDFs of depth errors.}
		\label{fig:depth_cdf}
	    \vspace{-2ex}
\end{figure}
\subsubsection{Depth Accuracy}\label{sssec:depth_acc}
Next, we evaluate the performance of \sysname\ with respect to the depth of subsurface structures. The error is defined as the absolute difference between the estimated depth $z_\mathrm{e}$ and the actual depth $z_\mathrm{a}$, namely, $|z_\mathrm{e}-z_\mathrm{a}|$. We again compare I-Net with RMA, based on settings similar to Section~\ref{sssec:distance_acc}. Figure~\ref{fig:depth_inet} shows I-Net's median errors of 0~\!cm, 0.1~\!cm, 0.2~\!cm, and 0.4~\!cm for wooden wall, drywall, brick wall, and concrete wall, respectively; the same intuition on horizontal distance applies here too. Note that the non-smoothness of the CDF curve is caused by the limited range resolution of a microwave radar, yet mmWave radars delivering a higher range resolution fail to work (see Section~\ref{sssec:rf_comp}). RMA (see Figure~\ref{fig:depth_rma}) again leads to negligible errors given correct parameters, but non-accurate parameters cause it to deliver totally wrong results with median error up to 3~\!cm, definitely not suitable for practical adoption. \rev{The results in Sections~\ref{sssec:distance_acc} and~\ref{sssec:depth_acc} clearly indicate that I-Net is robust to multi-parameter changes.
}

\subsubsection{Imaging Quality}
\begin{figure}[h]
    \setlength\abovecaptionskip{6pt}
    \vspace{-3ex}
	   \captionsetup[subfigure]{justification=centering}
		\centering
		\subfloat[I-Net.]{
		  \begin{minipage}[b]{0.47\linewidth}
		        \centering
			    \includegraphics[width = 0.96\textwidth]{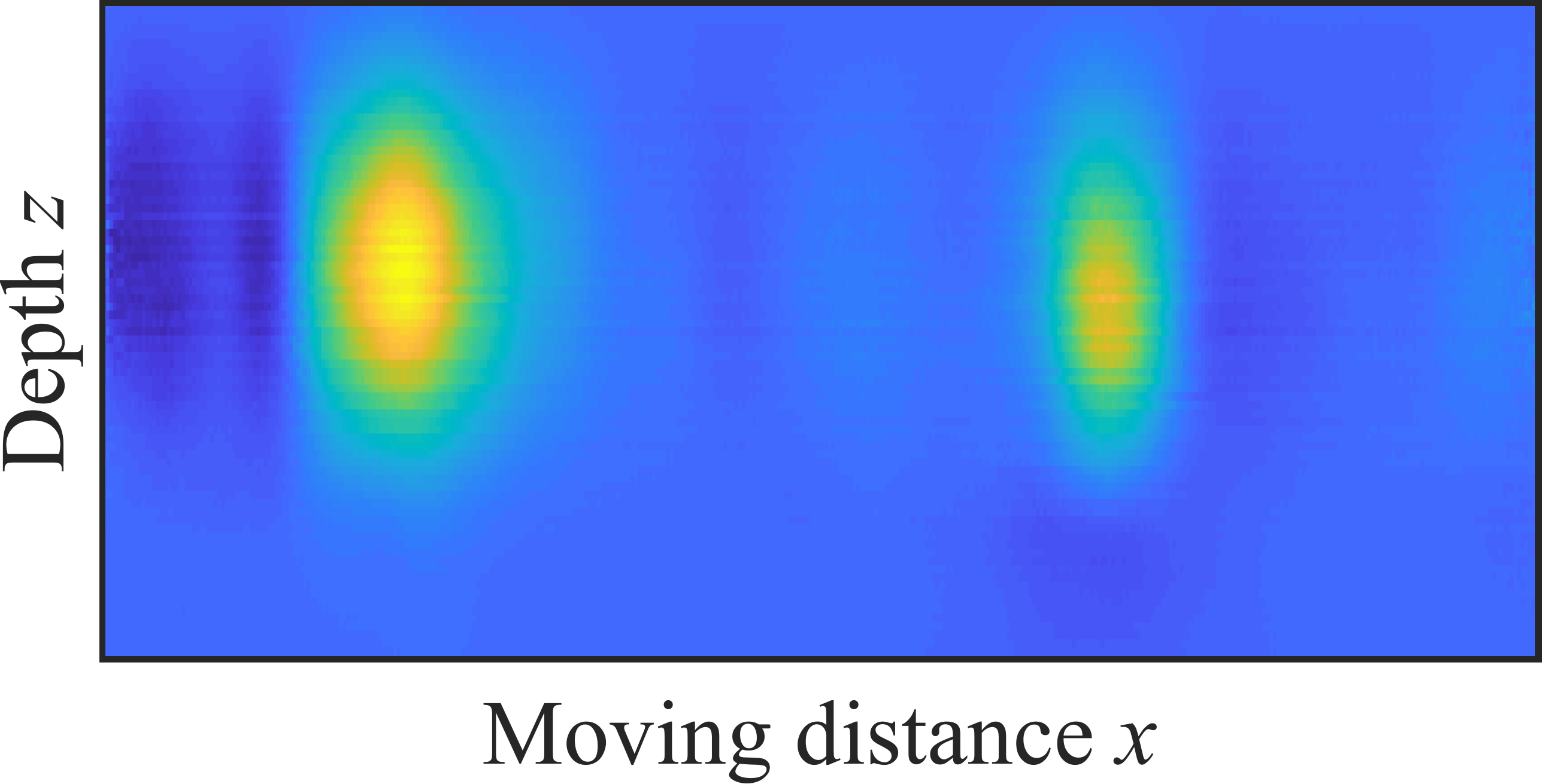}
			\end{minipage}
		}
		\subfloat[RMA.]{
		    \begin{minipage}[b]{0.47\linewidth}
		        \centering
			    \includegraphics[width = 0.96\textwidth]{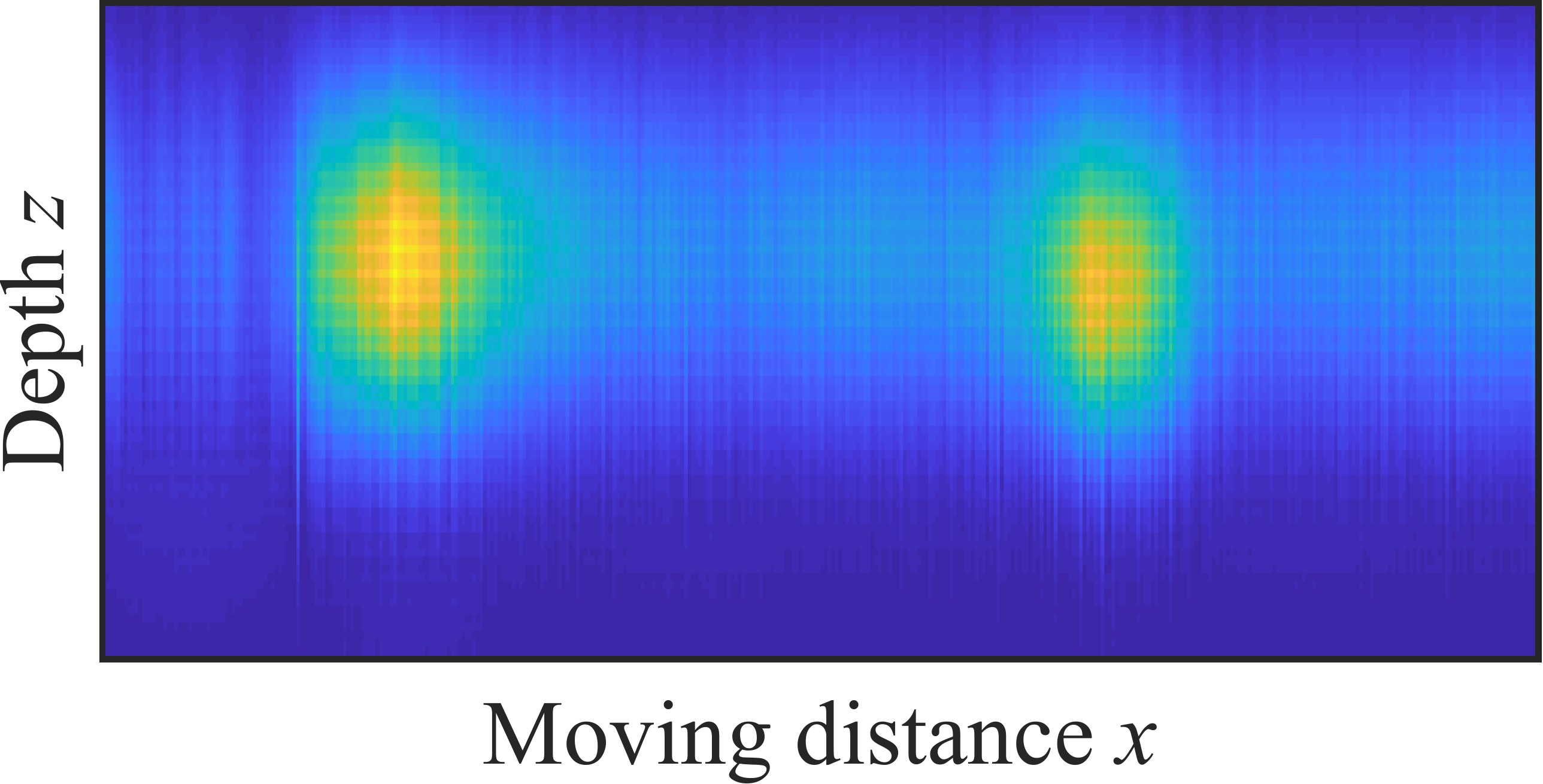}
			\end{minipage}
		}
		\caption{Imaging results of two algorithms. The result of I-Net is smoother than that of RMA.}
		\label{fig:comp_imag}
	    \vspace{-3ex}
\end{figure}

Since the imaging results will be used to assist inspectors to locate subsurface structures, it should be clear and visually appealing. Figure~\ref{fig:comp_imag} illustrates the imaging quality difference between I-Net and RMA with correct parameters. The I-Net imaging result is smoother and mostly free of noise, while the result of RMA is noisy due to the existence of high-frequency components. 
To measure the imaging quality quantitatively, we cannot use conventional SNR, since we do not have a reference image. Therefore, we adopt the Blind/Referenceless Image Spatial Quality Evaluator (BRISQUE)~\cite{mittal2012no} score. The BRISQUE method compares an imaging result to a default model computed from images of natural scenes with similar distortions; \rev{it yields a unitless score between 0 and 100, with a smaller score value indicating a} better perceptual quality. The BRISQUE scores of I-Net and RMA are shown in Table~\ref{tab:brisque}, confirming a slight advantage of I-Net over RMA in terms of overall image quality given all wall materials. 
\begin{table}[ht]
	\centering	
	\caption{BRISQUE scores of I-Net and RMA.} \label{tab:brisque}
    \begin{tabular}{|l|c|c|c|c|}
    \hline
            & Concrete & Brick & Wooden & Drywall \\ \hline
    I-Net   &  40.24 & 45.28 & 34.80& 38.57\\ \hline
    RMA &  45.44 & 49.34 & 40.15& 40.84 \\ \hline
    \end{tabular}
    \vspace{-1.5ex}
\end{table}

\begin{figure*}[b]
    \setlength\abovecaptionskip{6pt}
    \vspace{-2.5ex}
	   \captionsetup[subfigure]{justification=centering}
		\centering
		\subfloat[Water content.]{
		  \begin{minipage}[b]{0.24\linewidth}
		        \centering
			    \includegraphics[width = 0.96\textwidth]{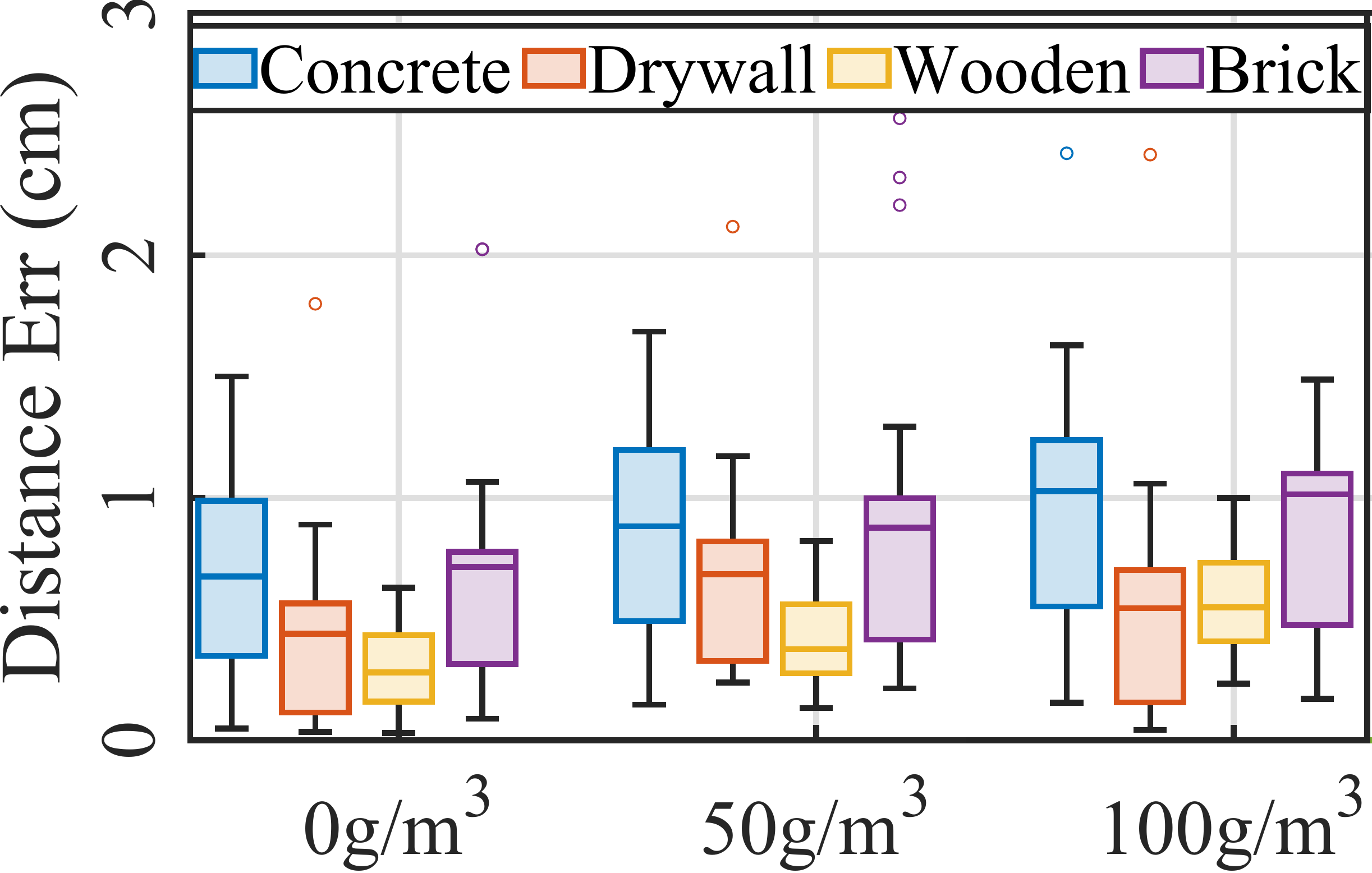}
			    \includegraphics[width = 0.96\textwidth]{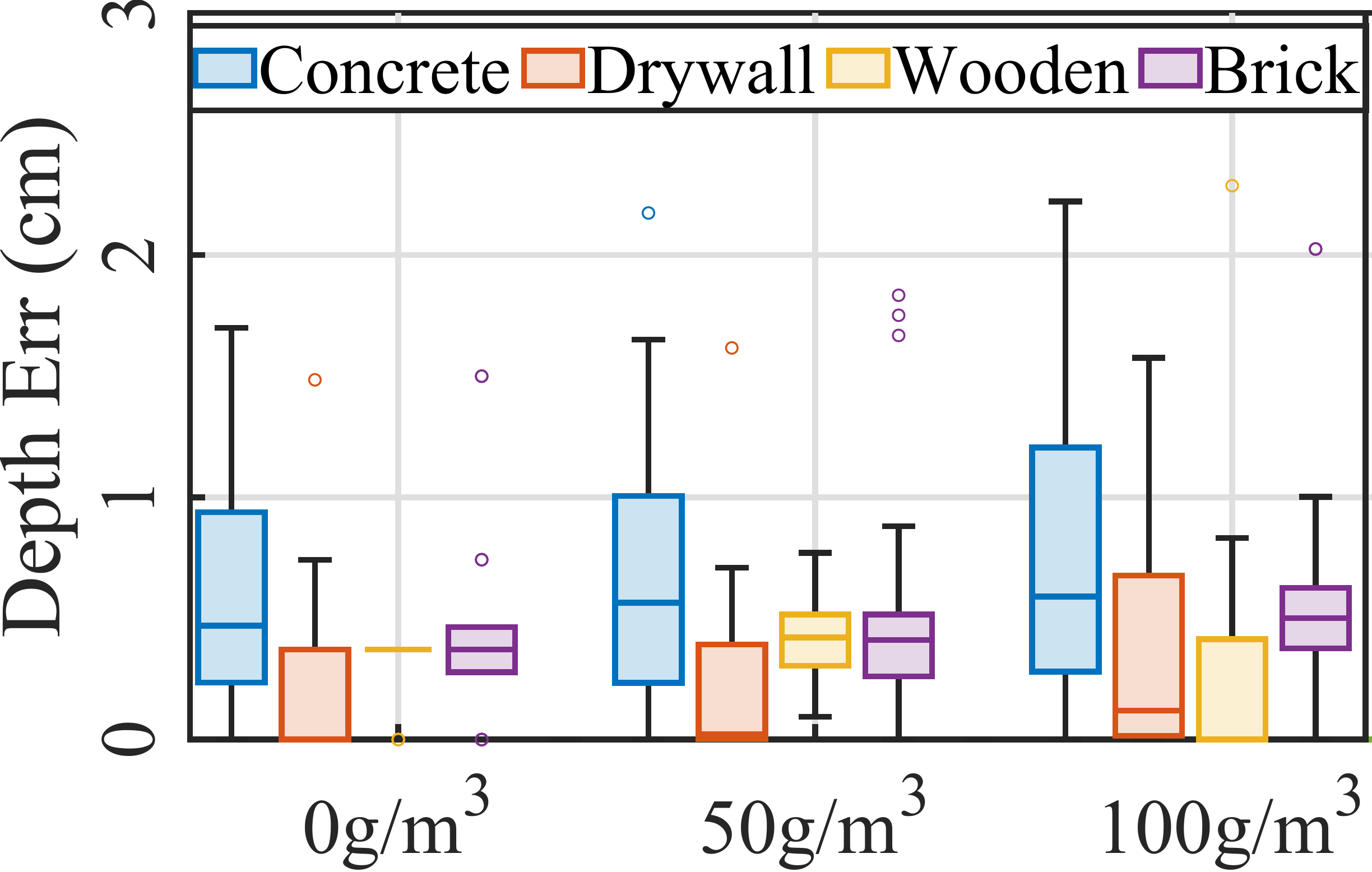}
			    \label{subfig:moisture_err}
			\end{minipage}
		}
		\subfloat[Depth.]{
		    \begin{minipage}[b]{0.24\linewidth}
		        \centering
			    \includegraphics[width = 0.96\textwidth]{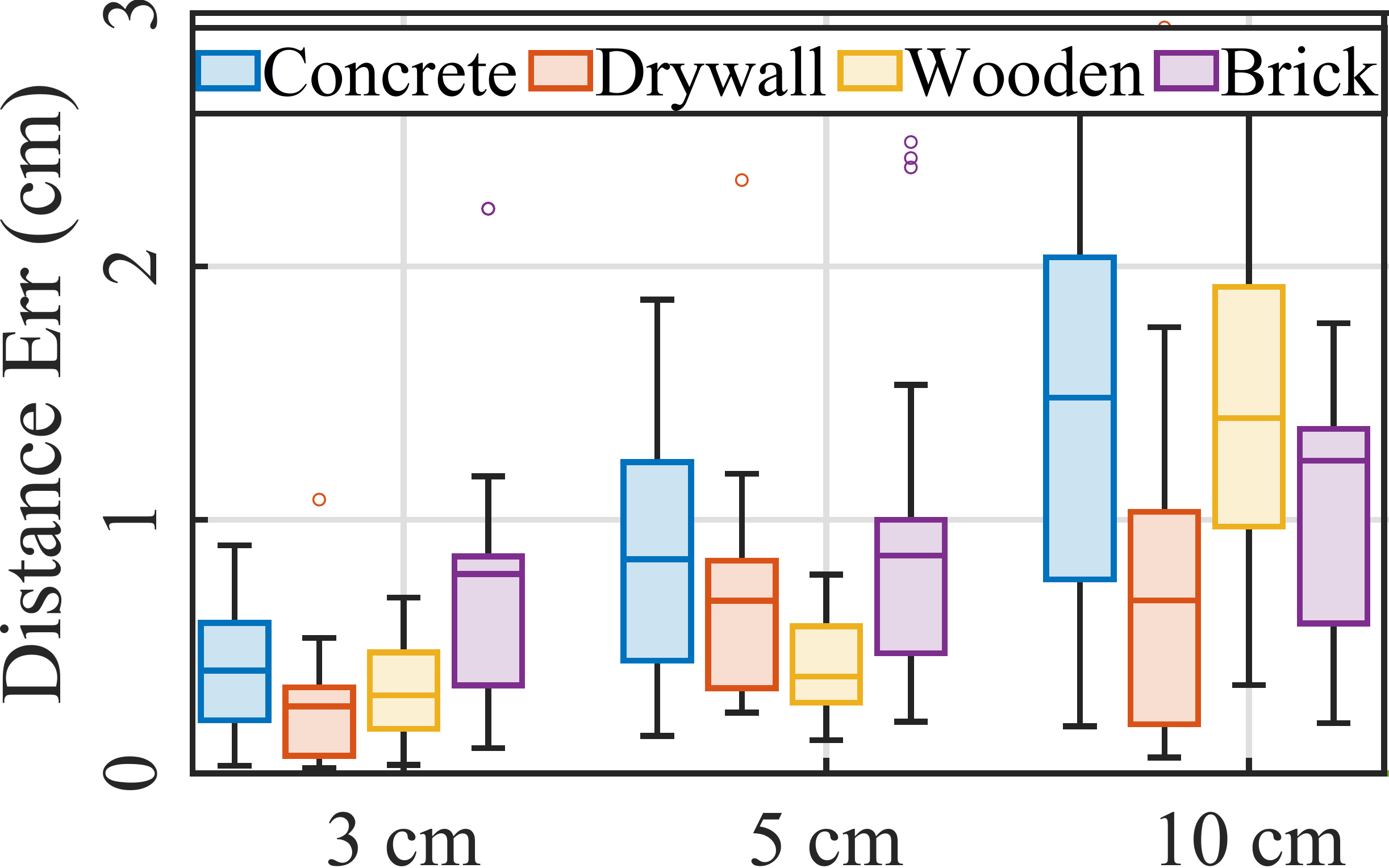}
			    \includegraphics[width = 0.96\textwidth]{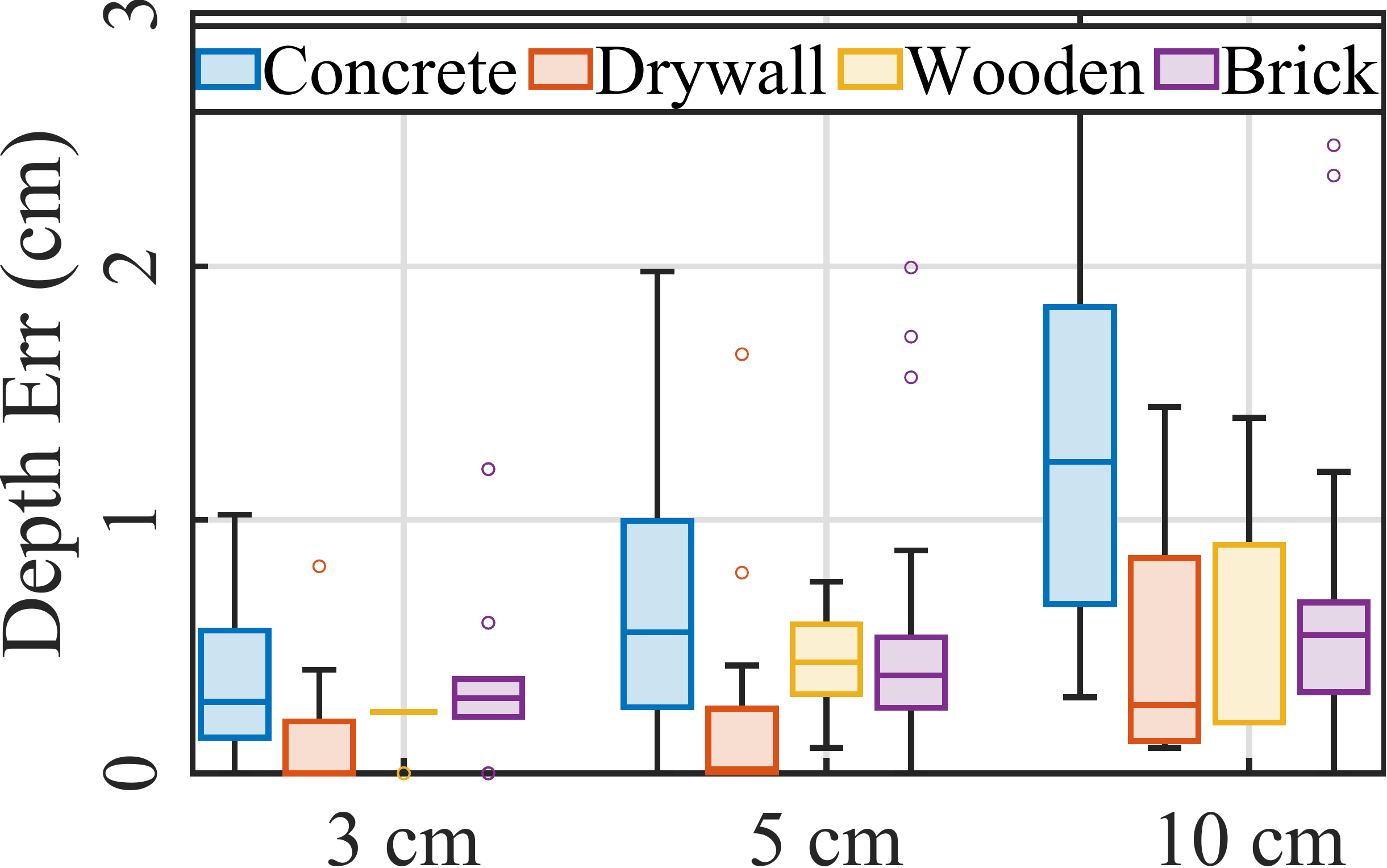}
			    \label{subfig:depth_err}
			\end{minipage}
		}
		\subfloat[Hand movement speed.]{
		  \begin{minipage}[b]{0.24\linewidth}
		        \centering
			    \includegraphics[width = 0.96\textwidth]{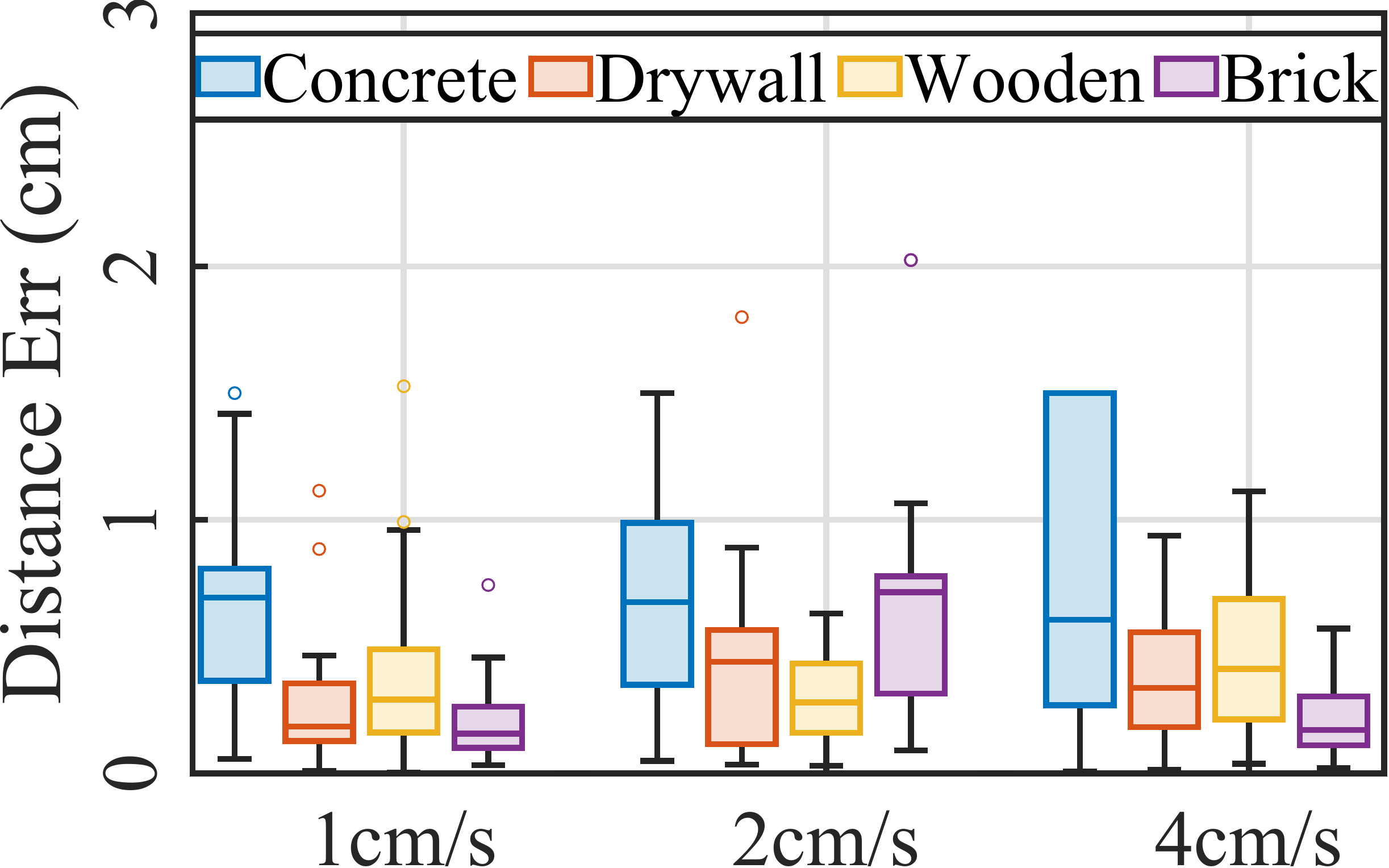}
			    \includegraphics[width = 0.96\textwidth]{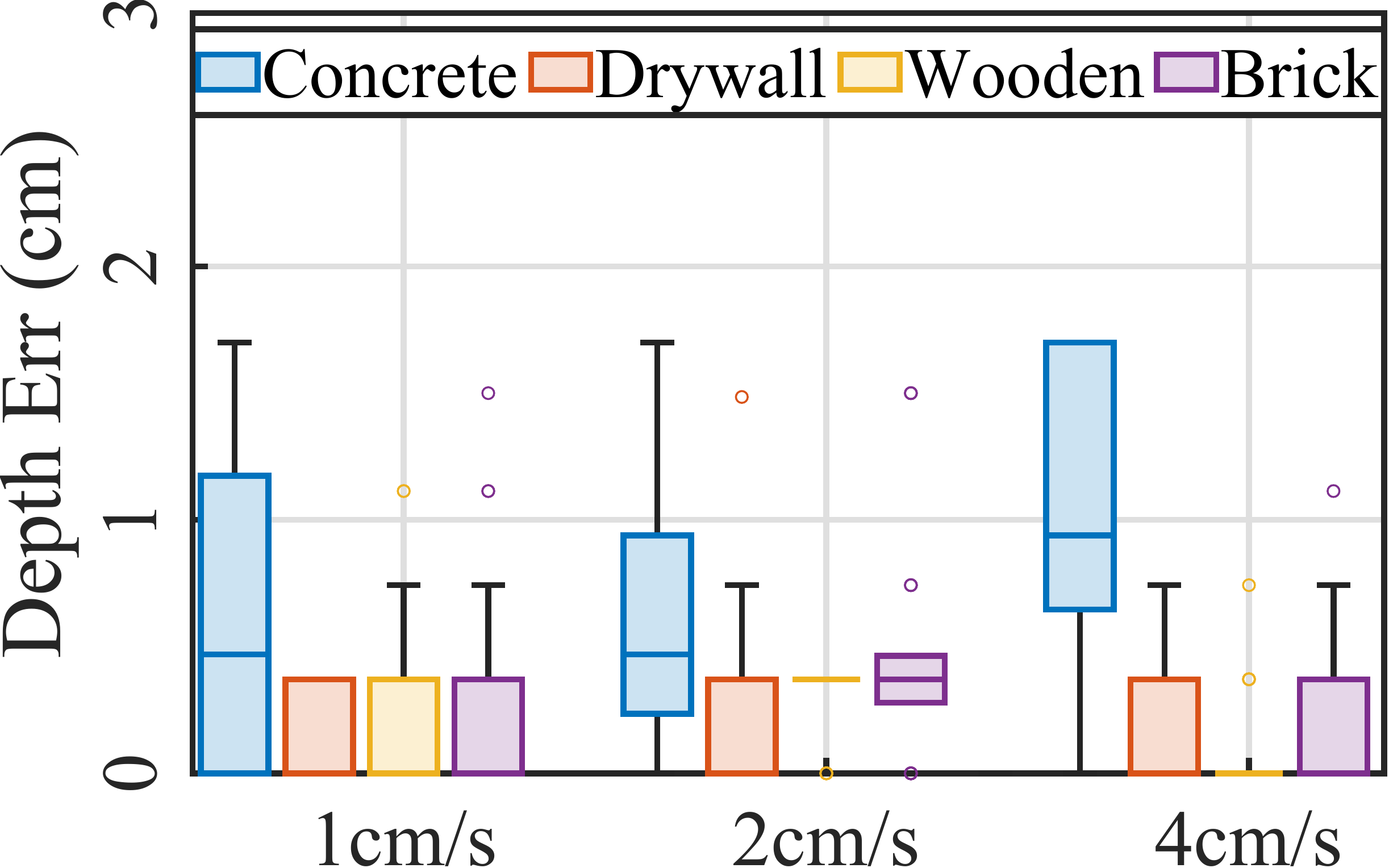}
			    \label{subfig:velocity_err}
			\end{minipage}
		}
		\subfloat[Material.]{
		    \begin{minipage}[b]{0.24\linewidth}
		        \centering
			    \includegraphics[width = 0.96\textwidth]{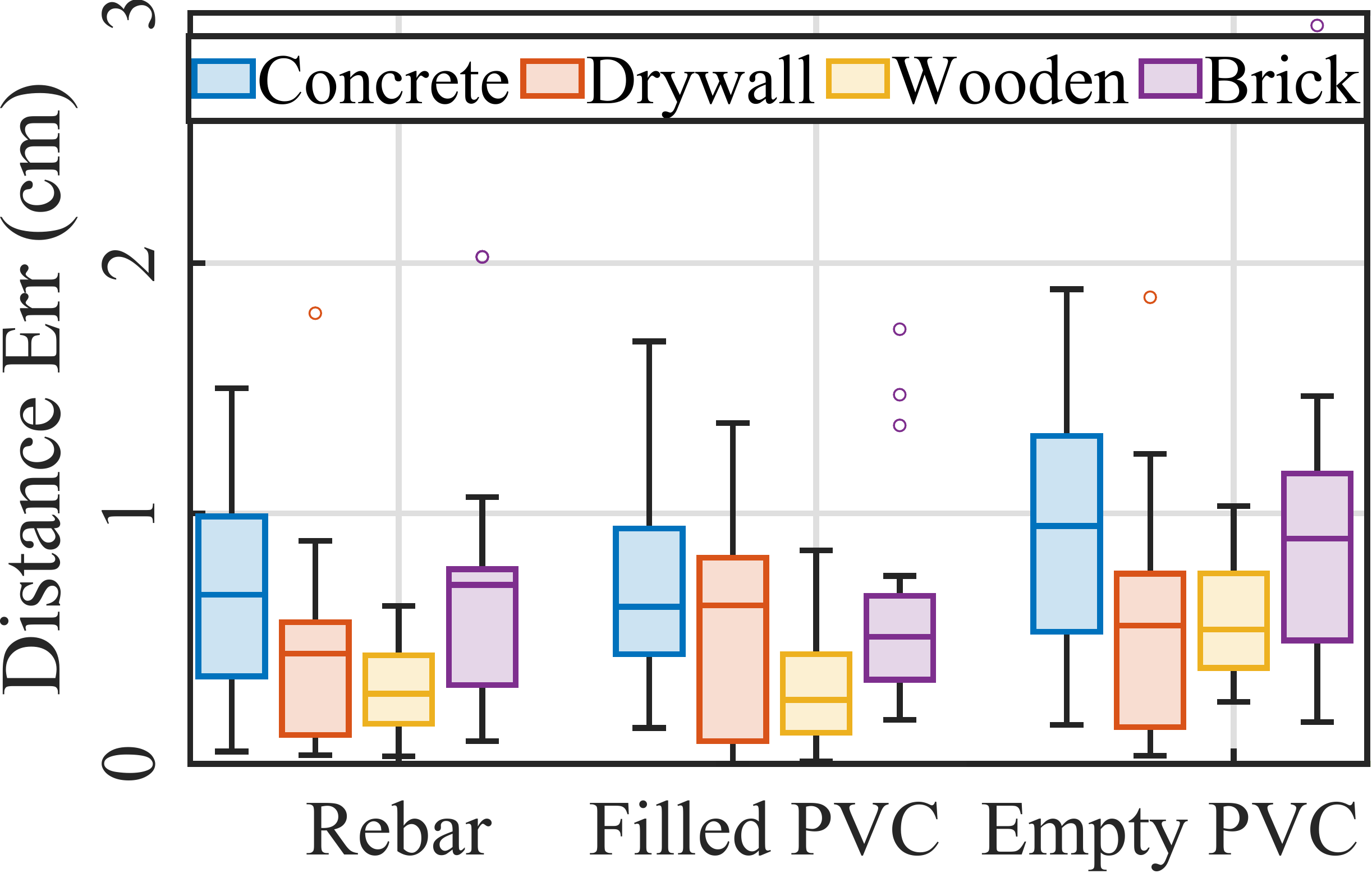}
			    \includegraphics[width = 0.96\textwidth]{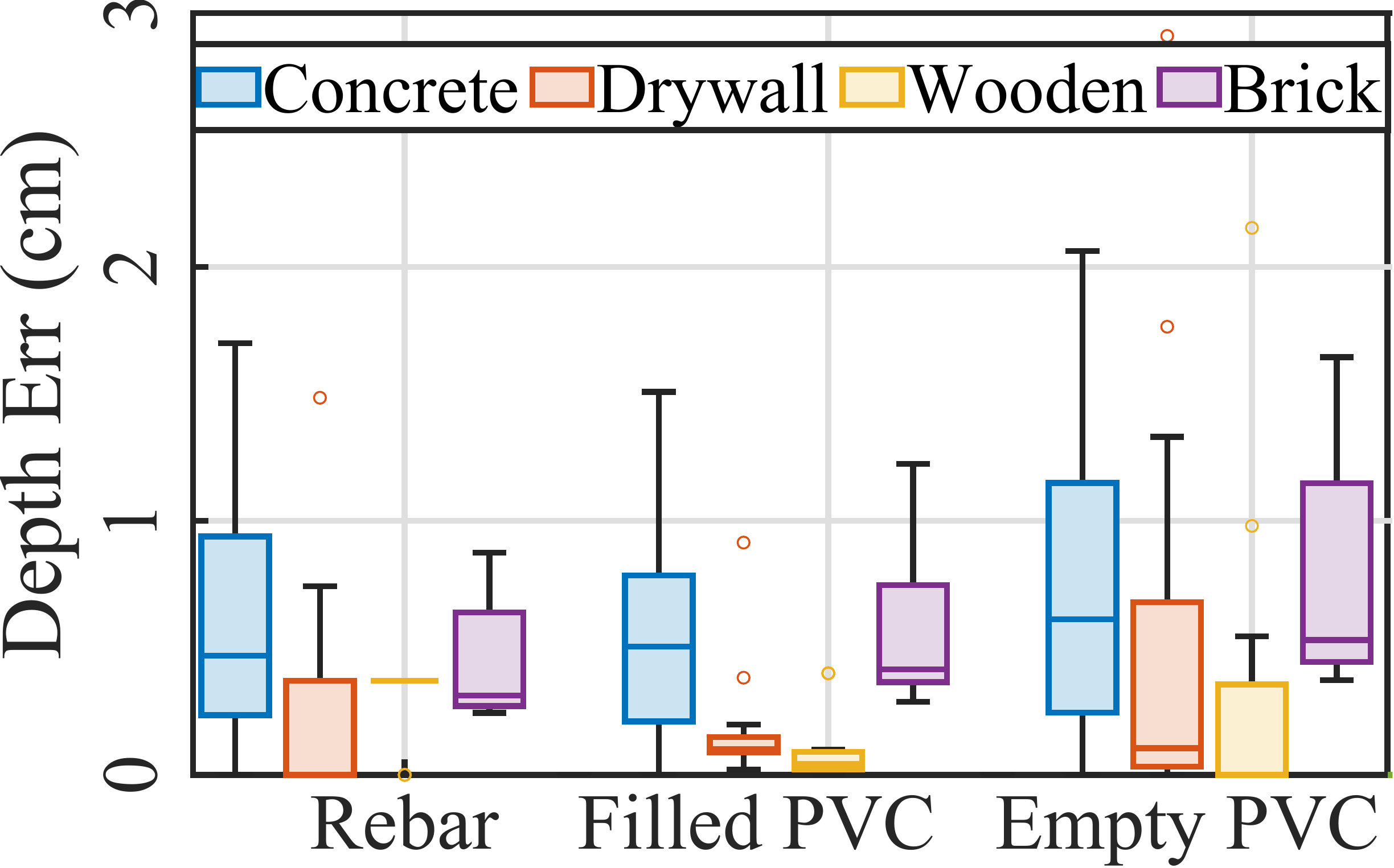}
			    \label{subfig:material_err}
			\end{minipage}
		}
		\caption{Relationship between localization errors and various environment and operating variables.}
		\label{fig:boxchart}
\end{figure*}

\subsubsection{Impact of Water Content}
We then evaluate the effect of water content on the imaging result. In many countries with rainy and humid climates, the water content in the wall can vary greatly and affect the performance of \sysname. This effect on imaging is twofold: i) water absorbs electromagnetic radiation in the RF spectrum ranging from 1~\!GHz to 1000~\!GHz~\cite{hasted1964microwave, wang2020soil}, and ii) water causes a drastic permittivity change of a wall~\cite{ogunsola2005shielding}, both leading to the failure of RMA. The evaluation results in Figure~\ref{subfig:moisture_err} show that as the water content in wall increases from 0~\!g/m\textsuperscript{3} to 100~\!g/m\textsuperscript{3}, the median horizontal distance error goes up for 0.3~\!cm on average, while the median depth error increases by around 0.2~\!cm. The gradual increase of error indicates that, as the water content increases, more RF signals are absorbed/blocked. It is worth noting that the increase in water content does not lead to a steep rise in error, confirming the robustness of I-Net in adapting to permittivity variations and validating our claim that I-Net is calibration-free.

\vspace{1.5ex}
\subsubsection{Impact of Depth}~\label{sssec:i_depth}
RF signals are greatly attenuated as they travel deep into walls, hence the depth of an in-wall structure is a key limiting factor of \sysname's imaging and localization capabilities. We now test whether I-Net can discover subsurface structures deep inside the wall. We put different materials at depths of 3~\!cm, 5~\!cm, and 10~\!cm, and evaluate the accuracies in both horizontal distance and depth, with the results shown in Figure~\ref{subfig:depth_err}. Take concrete wall as an example, the median horizontal distance errors are 0.4~\!cm, 0.7~\!cm, and 1.5~\!cm while the depth errors are 0.3~\!cm, 0.5~\!cm, and 1.2~\!cm, respectively. Other walls follow the same trend but appear less sensitive to depth. Apparently, \sysname\ is still able to deliver a decent accuracy as the depth increases.

\subsubsection{Impact of Hand Movement Speed}
As we discussed in Section~\ref{ssec:conventional}, traditional RMA algorithm is sensitive to the movement speed of the radar probe. In this section, we validate that the I-Net used in \sysname\ is not sensitive to this speed. We let a \sysname\ user move the probe on a wall with constant speeds of 1~\!cm/s, 2~\!cm/s, and 4~\!cm/s. The evaluation results are shown in Figure~\ref{subfig:velocity_err}. It can be observed that the median errors in horizontal distance and depth remain almost the same. This confirms that I-Net is robust to different hand movement speeds. We do observe that the error variances slightly grow as the speed increases, which may partially be attributed to the fact that the \sysname\ user cannot swipe the probe steadily on the wall at a higher speed. \rev{As we explained earlier, a distinct speed variation pattern happens to every trial regardless of how hard a user tries to maintain a stable speed. Fortunately, I-Net is shown to be robust to such minor variations.}

\subsubsection{Impact of Material}
The material of subsurface structures also plays an important role in the imaging and localization of I-Net. To investigate the effects of different materials, we choose three common ones: rebar, water-filled PVC pipe, and empty PVC pipe. In Figure~\ref{subfig:material_err}, we observe that, for rebar and water-filled PVC pipes, more than 75~\!\% of the horizontal distance and depth errors are smaller than 1~\!cm;
although empty PVC pipes incur slightly higher errors in percentile, all median errors in horizontal distance and depth are below 1~\!cm. We believe that the difference in reflectivity (metal and water are strong reflectors, but PVC plastics are weaker) may have caused the discrepancy in errors.

\subsubsection{I-Net Efficiency}
As the last step of evaluating I-Net, we investigate its runtime latency. In order to make our results compatible with most computers, we consider CPU-only inference, and we compare our algorithm with RMA on different devices. The PC workstation we use is equipped with Intel Xeon W-2133 CPU and 16~\!GB RAM, the laptop is equipped with AMD Ryzen 5 3500U CPU and 8~\!GB RAM, Raspberry Pi 3 is equipped with Arm Cortex-A53 CPU and 1~\!GB RAM, and Raspberry Pi 4 is equipped with ARM Cortex-A72 CPU and 2~\!GB RAM. Assuming an input 2-D signal matrix of size $200\times 200$, Table~\ref{tab:runtime} reports the inference times under the above four devices.
It appears that the runtime latency of I-Net is slightly higher than that of RMA, but the latency of I-Net on PC workstation and laptop can satisfy the requirement of real-time processing. We plan to further optimize the structure of I-Net, so as to reduce its complexity for embedded platforms (e.g., Raspberry Pi)  while maintaining the imaging performance. This should enable 
a standalone \sysname\ prototype with real-time capability.
\begin{table}[h]
\vspace{-1ex}
\caption{Runtime latency of RMA and I-Net.}
\vspace{-1ex}
\begin{tabular}{|l|c|c|c|c|}
\hline
      & PC Workstation  & Laptop & RPi 3  & RPi 4    \\ \hline
RMA   & 0.05~\!s  & 0.12~\!s & 2.97~\!s & 1.22~\!s  \\ \hline
I-Net & 0.08~\!s  & 0.15~\!s & 3.09~\!s & 1.43~\!s \\ \hline
\end{tabular}
\label{tab:runtime}
\vspace{-1ex}
\end{table}

\subsection{Material Identification}
In this section, we first discuss the overall material identification performance of \sysname's M-Net. Then we investigate M-Net's robustness to depth and water content changes. We also discuss the necessity of dual polarization for accurate material identification. \rev{We further compare M-Net to 2-D ResNet~\cite{he2016deep} and Multi-Layer Perceptron (MLP)~\cite{mlp}. ResNet is a popular deep learning network for classification tasks, and MLP is a common feed-forward network used in a recent proposal~\cite{zhang2019feasibility} for material identification.} At last, we verify the computation efficiency of M-Net. %

\subsubsection{Overall Performance}
We prepare four different in-wall structures: non-corroded rebar, corroded rebar, non-leaked PVC pipe, and leaked PVC pipe. 
We first use confusion matrices to characterize the identification performance under different walls in Figure~\ref{fig:conf_mat}, where we observe that concrete and brick walls have more adverse effects on material identification than wooden wall and drywall. It is also interesting to note the correlated adverse effect between concrete and rebar, as well as that between brick and PVC. By summarizing these data, we could also derive that the overall accuracy for correctly identifying material (i.e., steel or PVC) is 95.2\%, while that for further correctly diagnosing their health status is slightly lower at 91.8\%.
\setcounter{figure}{21}
\begin{figure}[h]
     \setlength\abovecaptionskip{6pt}
    \vspace{-2.5ex}
	   \captionsetup[subfigure]{justification=centering}
		\centering
		\subfloat[Concrete wall.]{
		  \begin{minipage}[b]{0.45\linewidth}
		        \centering
			    \includegraphics[width =0.9 \textwidth]{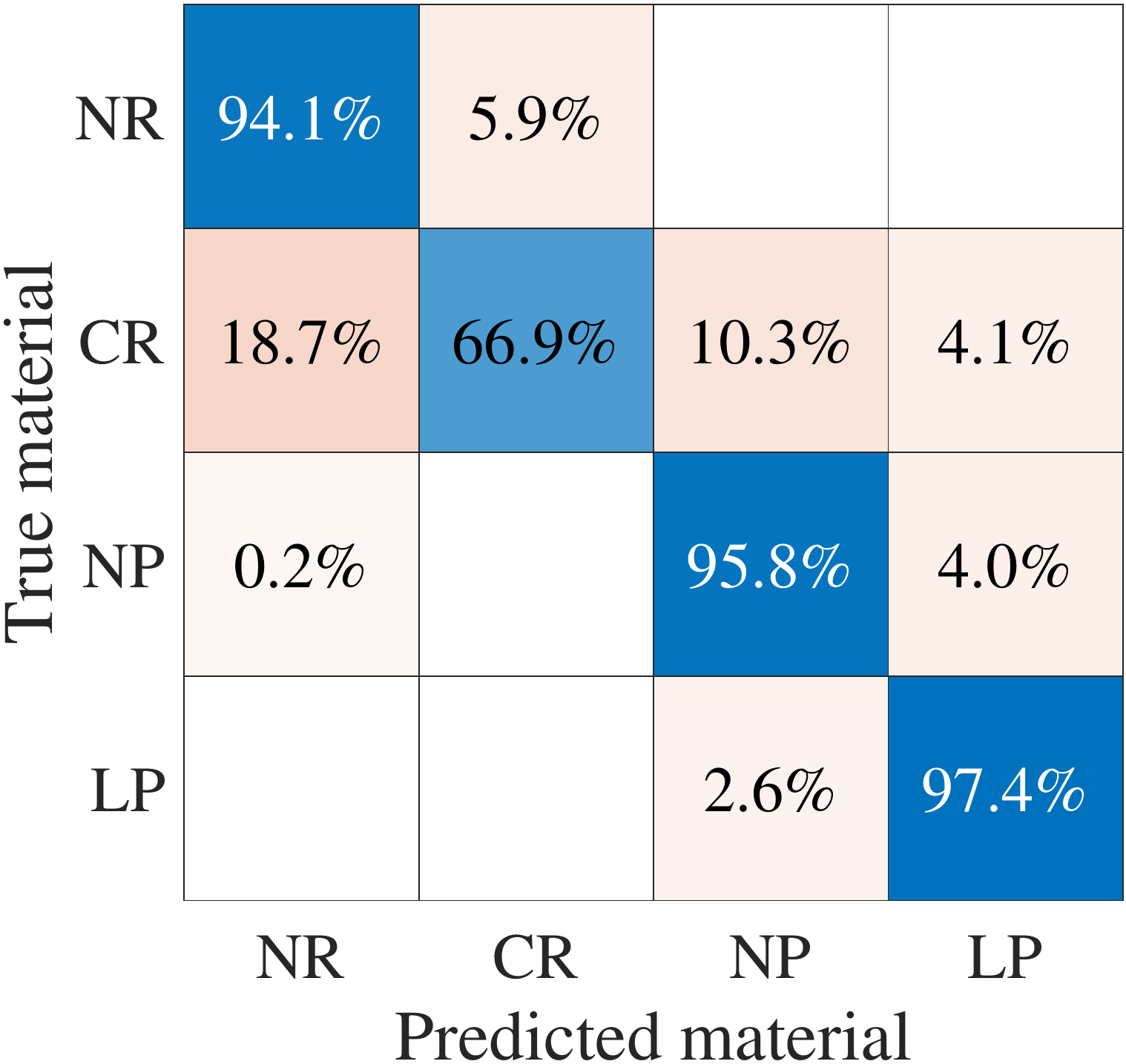}
			\end{minipage}
		}
		\subfloat[Drywall.]{
		    \begin{minipage}[b]{0.45\linewidth}
		        \centering
			    \includegraphics[width =0.9 \textwidth]{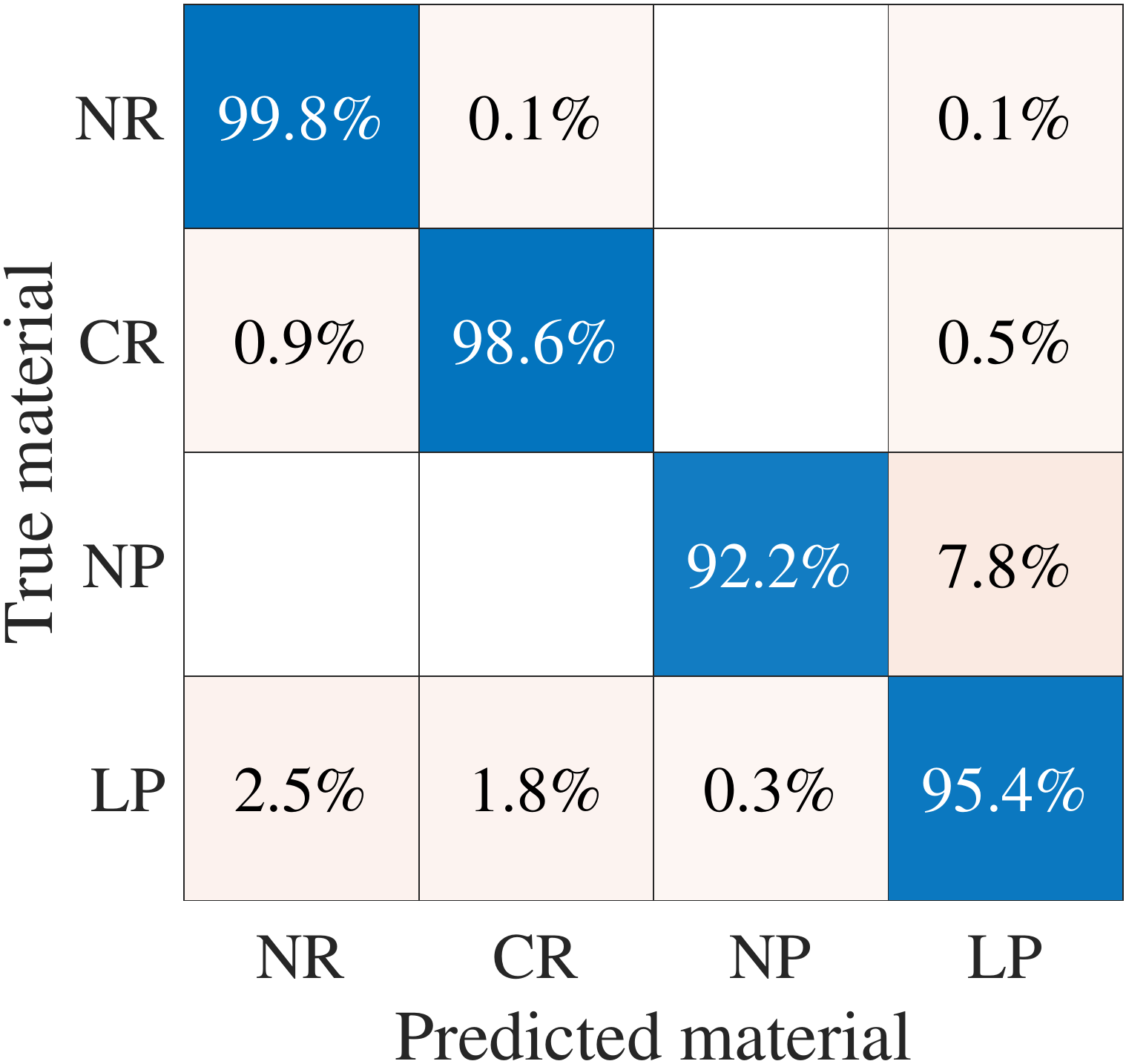}
			\end{minipage}
		}
		\\
		\subfloat[Wooden wall.]{
		  \begin{minipage}[b]{0.45\linewidth}
		        \centering
			    \includegraphics[width =0.9 \textwidth]{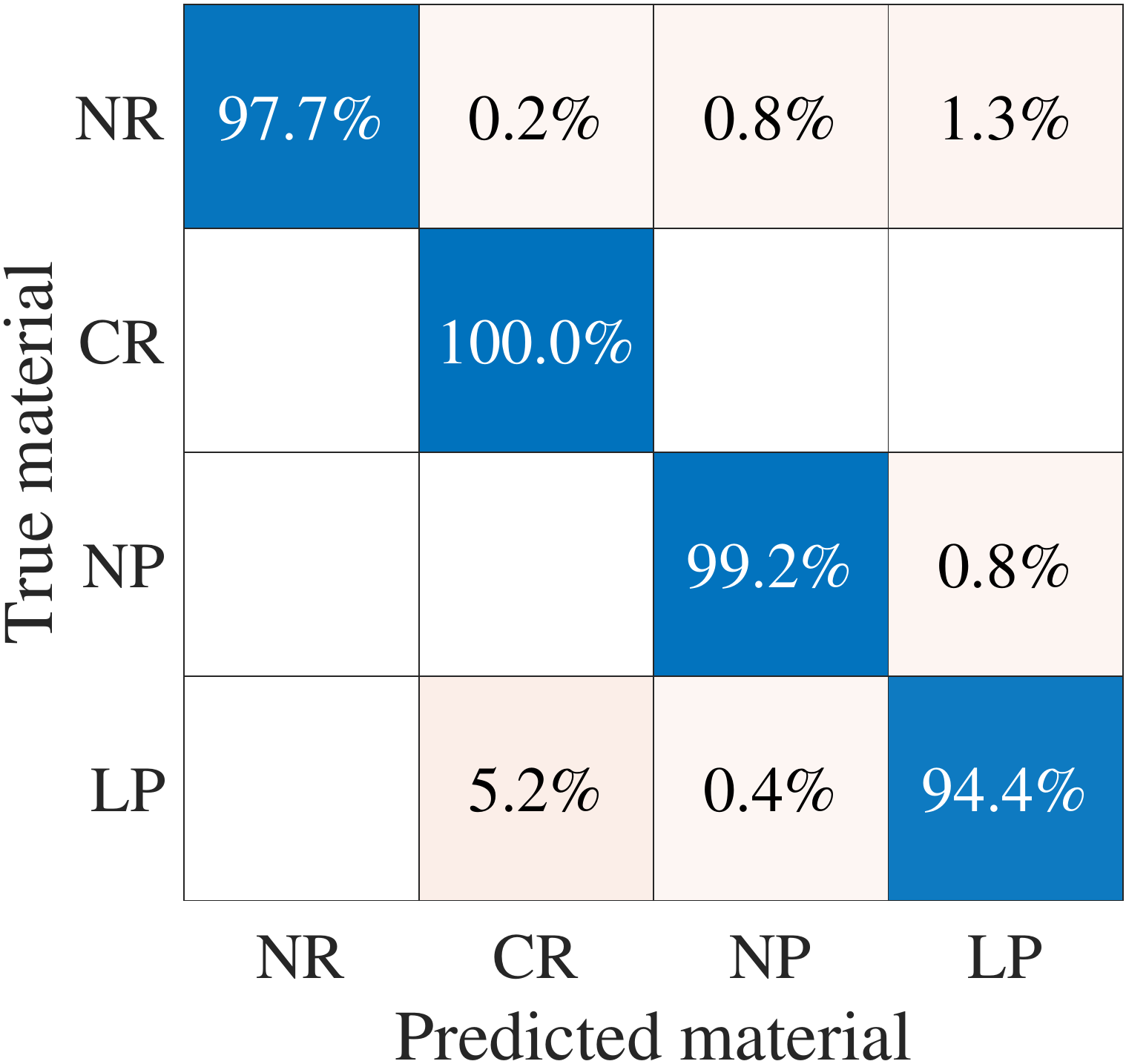}
			\end{minipage}
		}
		\subfloat[Brick wall.]{
		    \begin{minipage}[b]{0.45\linewidth}
		        \centering
			    \includegraphics[width =0.9 \textwidth]{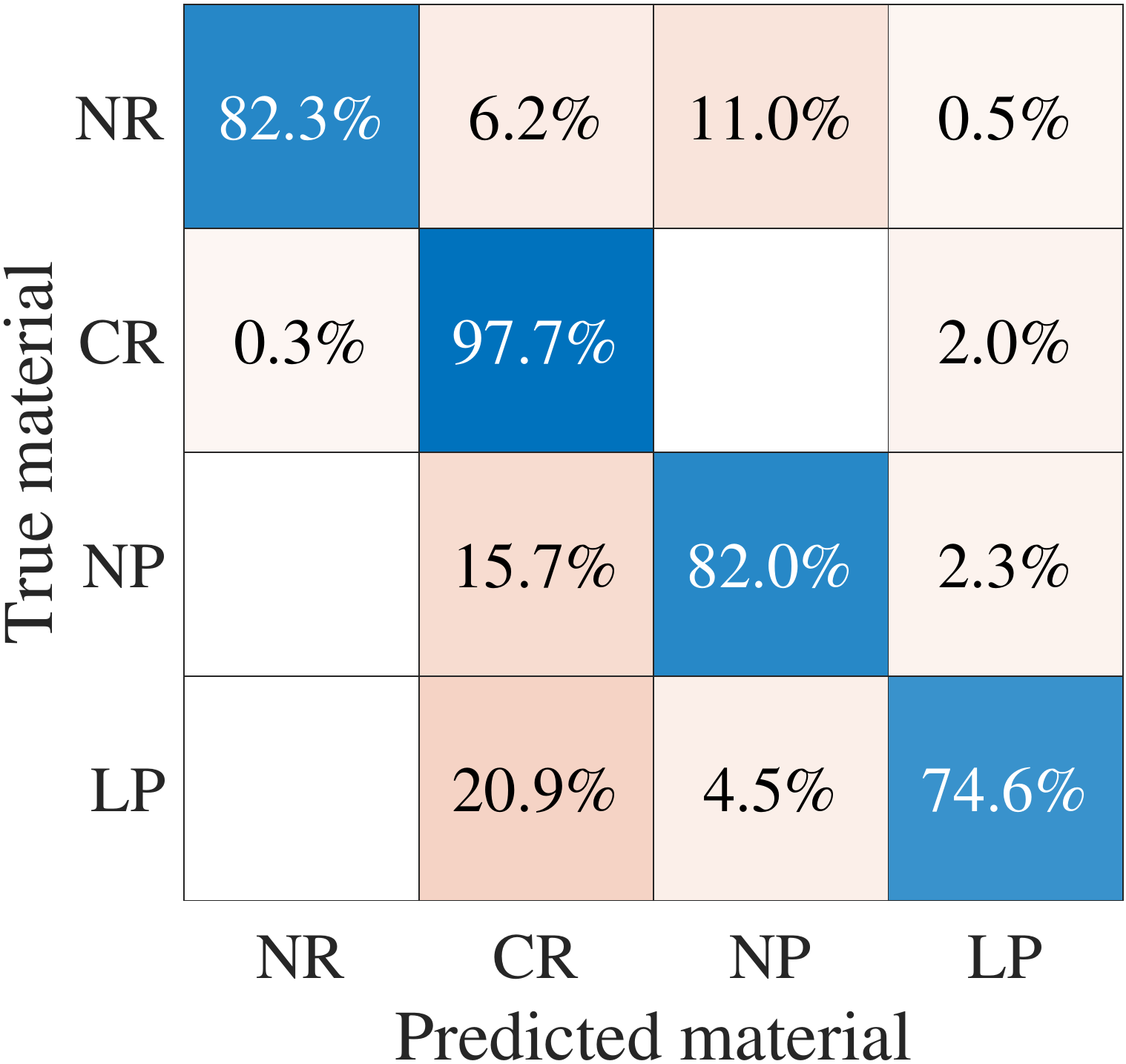}
			\end{minipage}
		}
		\caption{Confusion matrices of M-Net under different walls, where non-corroded rebar, corroded rebar, non-leaked PVC pipe, and leaked PVC pipe are abbreviated as NR, CR, NP, LP, respectively.}
		\label{fig:conf_mat}
		\vspace{-2ex}
\end{figure}

\subsubsection{Depth}
\begin{figure}[ht]
    \vspace{-2ex}
    \centering
	   \captionsetup[subfigure]{justification=centering}
		\centering
		\subfloat[Identification accuracy.]{
		    \begin{minipage}[b]{0.49\linewidth}
		        \centering
			    \includegraphics[width = \textwidth]{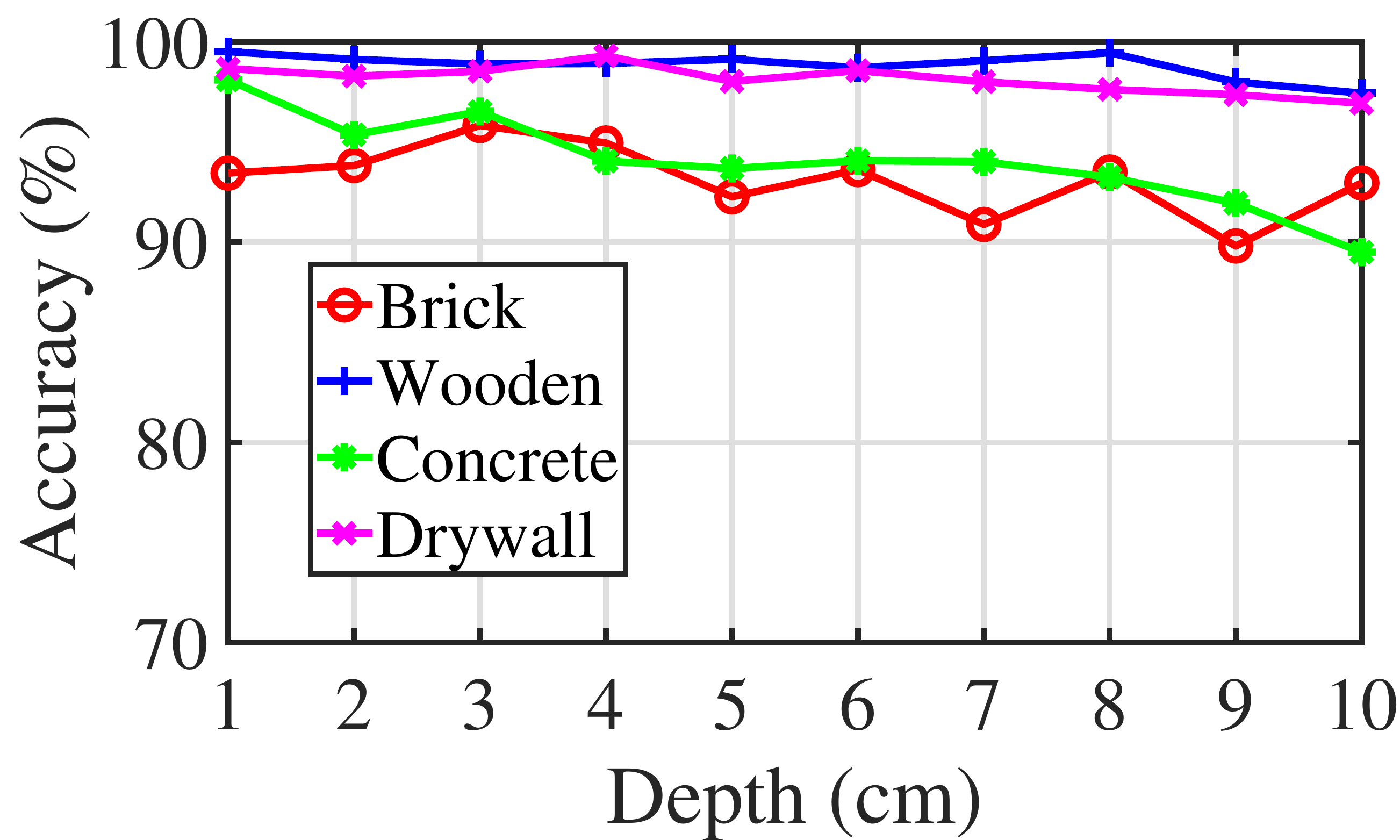}
			\end{minipage}
		}
		\subfloat[Diagnosis accuracy.]{
		  \begin{minipage}[b]{0.49\linewidth}
		        \centering
			    \includegraphics[width = \textwidth]{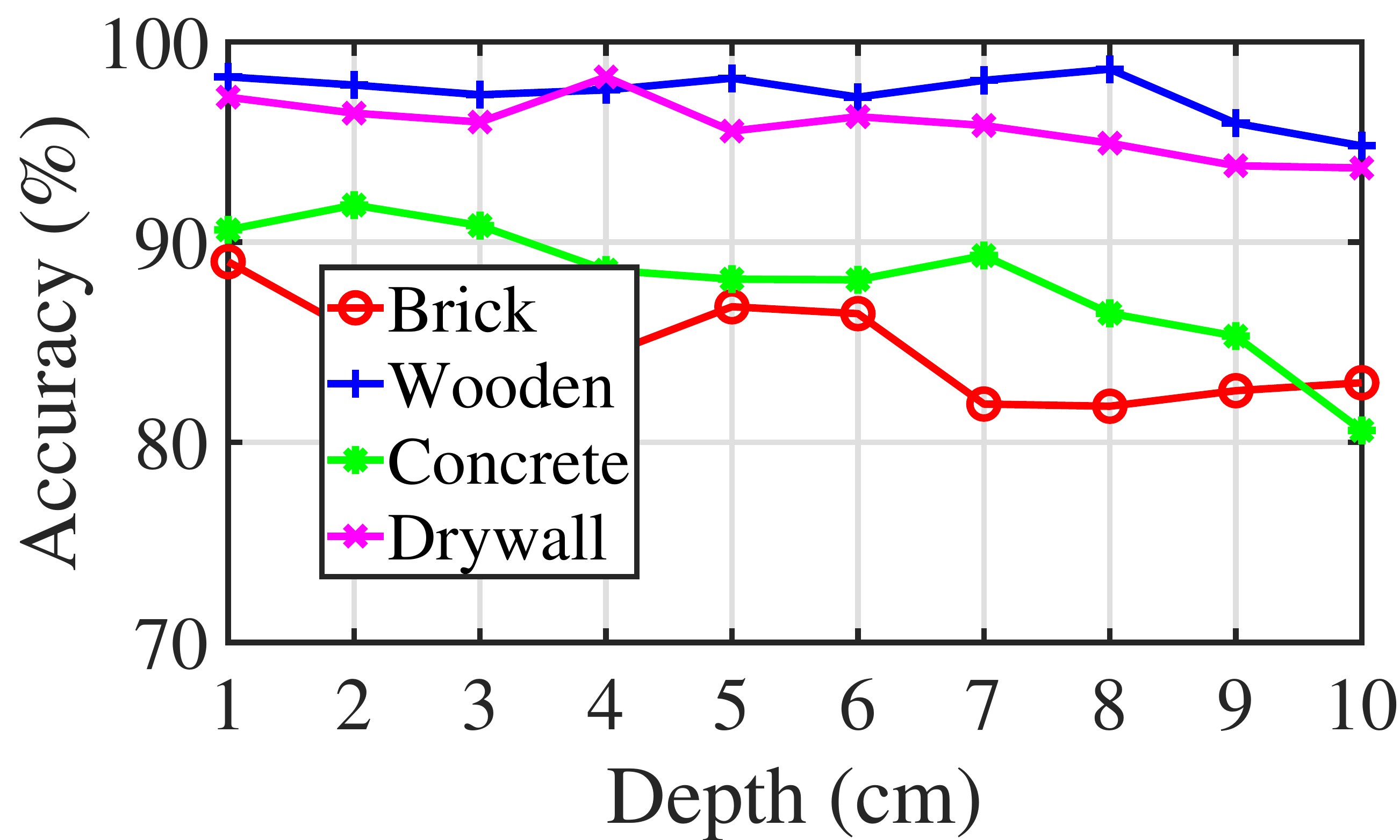}
			\end{minipage}
		}
		\vspace{-1ex}
		\caption{Impact of depth on material identification.}
		\vspace{-1ex}
		\label{fig:accuracy_depth}
	    \vspace{-1ex}
\end{figure}

As discussed in Section~\ref{sssec:i_depth}, attenuation of RF signals as they travel into walls makes it harder to detect in-wall structures. 
Therefore, we hereby study the effects of depth on material identification in this section, resulting in Figure~\ref{fig:accuracy_depth} to depict that the accuracy slightly decreases as the depth increases. 
Fortunately, while the identification accuracies are all over 90\%, the diagnosis accuracies remain above 80\%. The worst performance caused by brick walls may result from their impurity (e.g., cavities or stones inside), but these are also realities that have to be faced by \sysname. Overall, these results validate that our adversarial training strategy successfully removes the influence caused by varying wall depths and makes M-Net virtually immune to environment changes.

\subsubsection{Water Content}
\begin{figure}[ht]
    \vspace{-2ex}
    \centering
	   \captionsetup[subfigure]{justification=centering}
		\centering
		\subfloat[Identification accuracy.]{
		    \begin{minipage}[b]{0.49\linewidth}
		        \centering
			    \includegraphics[width = \textwidth]{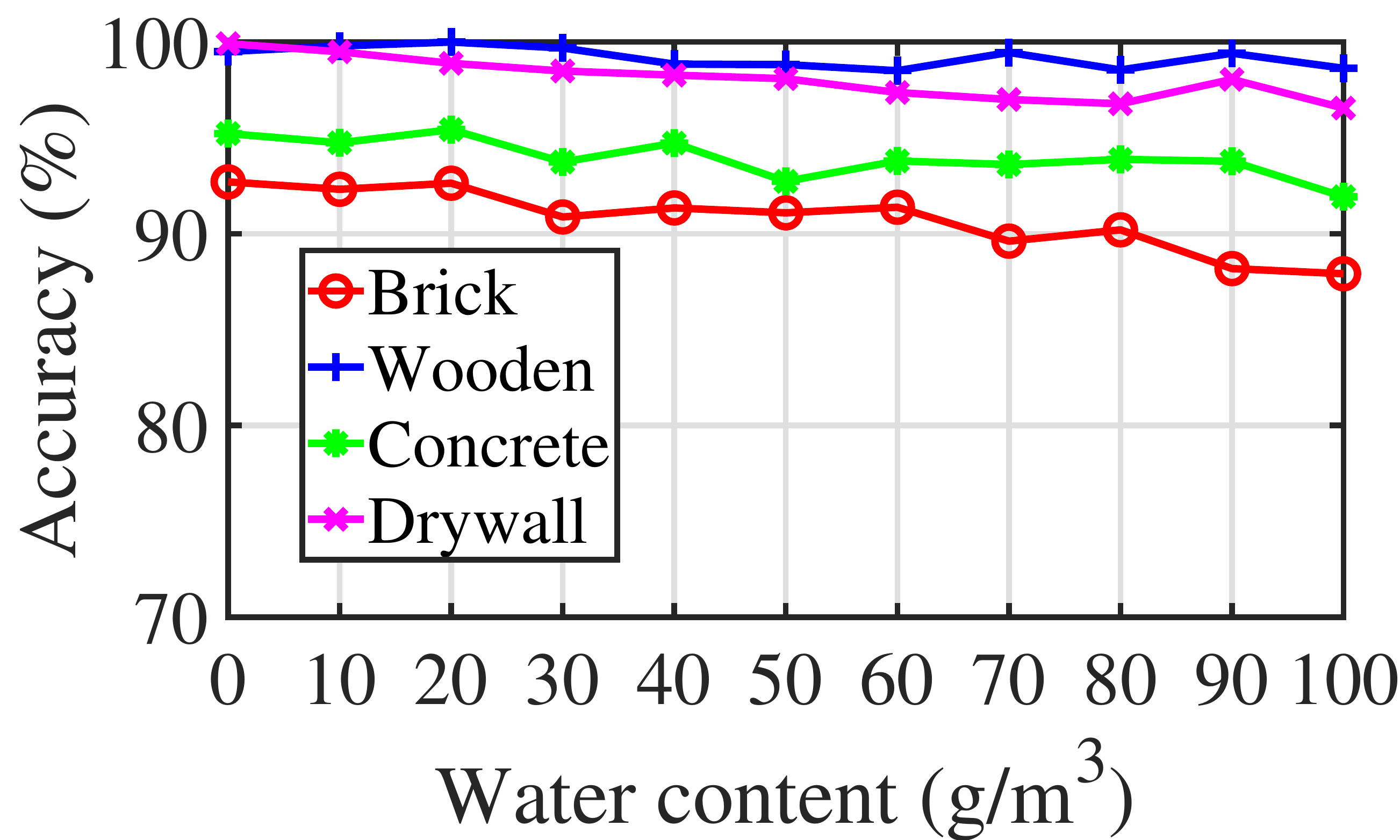}
			\end{minipage}
		}
		\subfloat[Diagnosis accuracy.]{
		  \begin{minipage}[b]{0.49\linewidth}
		        \centering
			    \includegraphics[width = \textwidth]{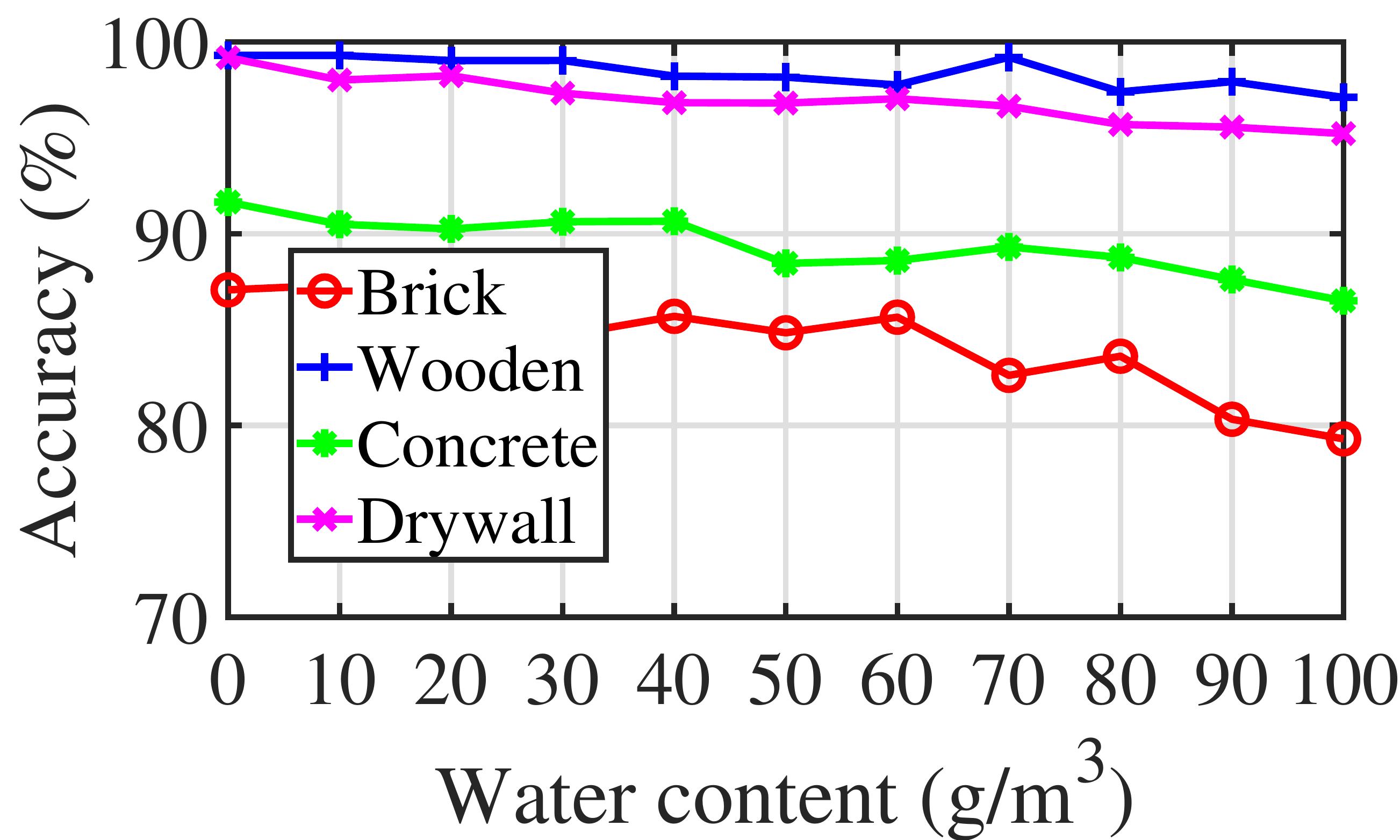}
			\end{minipage}
		}
		\vspace{-1ex}
		\caption{Impact of water on material identification.}
		\vspace{-1ex}
		\label{fig:accuracy_water}
	    \vspace{-1ex}
\end{figure}

We then test M-Net's robustness to various water contents in the wall. We gradually spray water onto the wall, from 0~\!g/m\textsuperscript{3} to 100~\!g/m\textsuperscript{3}, and employ M-Net to identify and diagnose in-wall materials. According to the results shown in Figure~\ref{fig:accuracy_water}, we observe that, though performance degrades due to water blocking and absorbing RF signals, the overall identification accuracies and diagnosis accuracies remain over 90\% and 80\%, respectively. These results again confirm the robustness of M-Net.

\subsubsection{Polarization Diversity} 
In Section~\ref{sssec:material_id}, we have discussed polarimetry exploiting the diversity of polarizations to identify materials. To prove the necessity of using two polarizations, we perform an ablation study by removing the cross-polarization. The overall identification and diagnosis accuracies are 95.2\% and 91.8\% for dual polarization, but 86.1\% and 70.5\% for single polarization. These large gaps validate the need for 
dual polarization.

\subsubsection{Other Neural Networks}
Table~\ref{tab:othernn} summarizes the material identification performance with different deep neural networks, where M-Net* is a variant of M-Net without environment discriminator. %
\begin{table}[b]
    \vspace{-1ex}
	\centering	
	\small
	\caption{Comparing M-Net with other networks.}
	\label{tab:othernn}
	\vspace{-1ex}
    \begin{tabular}{|l|c|c|c|c|c|}
    \hline
    & M-Net & M-Net* & 2-D ResNet & MLP~\cite{zhang2019feasibility} \\ \hline
    Identification acc (\%) & 95.2 & 92.9 & 91.4 & 89.8 \\ \hline
    Diagnosis acc (\%)  & 91.8 & 85.1 & 86.4 & 78.9 \\ \hline
    Latency on PC (s) & 0.032  & 0.031  & 0.057  & 0.013 \\ \hline 
    Latency on RPi 4 (s) & 0.96 & 0.96 & 1.63 & 0.51 \\ \hline
    \end{tabular}
\end{table}
To adapt 2-D ResNet for processing radar signal, we first transform the 1-D signal sequence to 2-D spectrogram containing both temporal and frequency information, then we employ the network to classify the spectrogram. Clearly, M-Net achieves the highest accuracy while maintaining a reasonable processing latency. We also observe that the runtime latency of M-Net on single-board Raspberry Pi can potentially satisfy real-time requirements.

\subsection{\rev{Case Studies}}
In this section, we extend the aforementioned evaluations
to more complex scenarios, including cracks in wall and multi-material wall. Due to 
the lack of sufficient specimens, we do not re-train SiWa. Instead, we examine if the structural imaging ability of I-Net trained on previous samples can be generalized to these complex scenarios.

\subsubsection{Cracks in Wall}
\begin{figure}[h]
    \setlength\abovecaptionskip{6pt}
    \vspace{-1ex}
	   \captionsetup[subfigure]{justification=centering}
		\centering
		\subfloat[Concrete block with cracks.]{
		  \begin{minipage}[b]{0.49\linewidth}
		        \centering
			    \includegraphics[width = .98\textwidth]{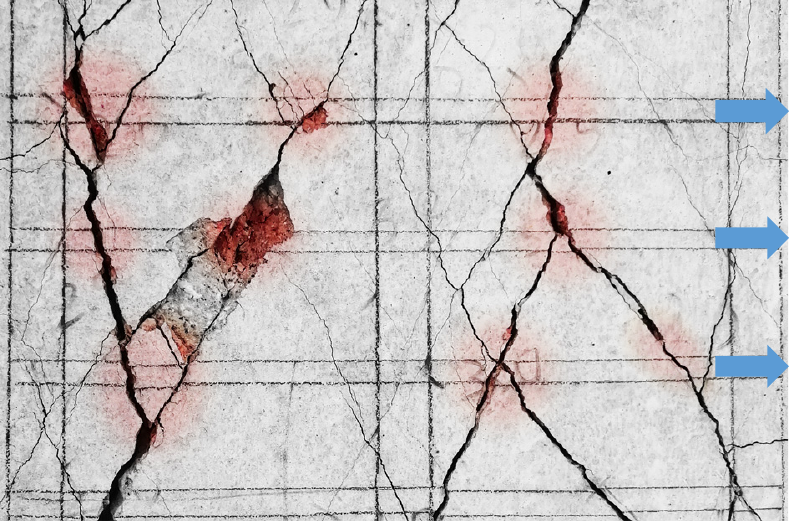}
			    \label{subfig:crack_photo}
			\end{minipage}
		}
		\subfloat[Imaging result.]{
		    \begin{minipage}[b]{0.49\linewidth}
		        \centering
			    \includegraphics[width = 0.98\textwidth]{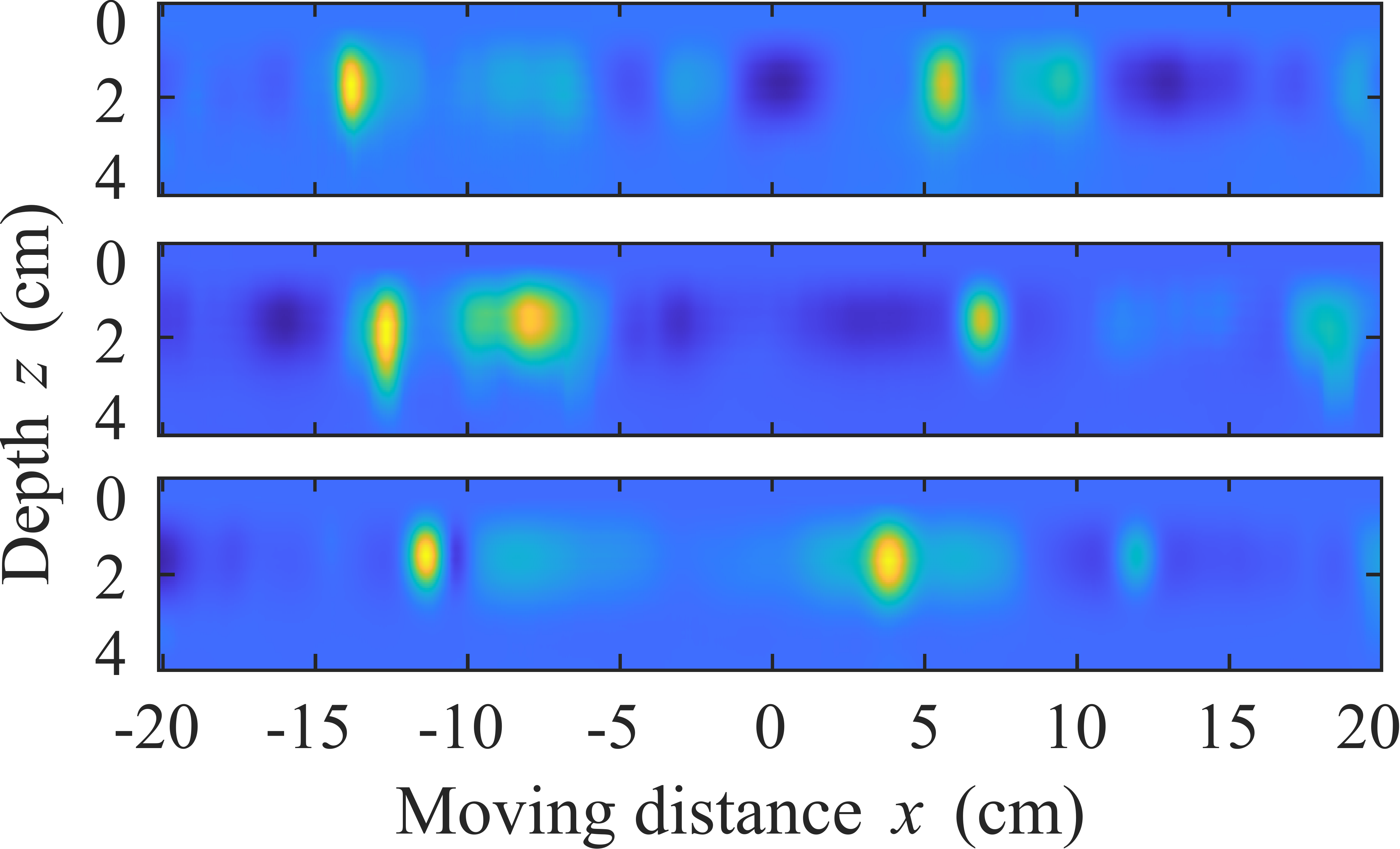}
			    \label{subfig:crack_result}
			\end{minipage}
		}
		\caption{Three scanning strips (with ground truth cracks marked) and their corresponding imaging results.}
		\label{fig:crack}
	    \vspace{-1ex}
\end{figure}

Wall crack detection~\cite{dung2019autonomous} is a critical goal of non-destructive inspection, and it is hence an objective of \sysname\ too. To verify \sysname's ability in crack detection, we prepare a cracked concrete block, as shown in Figure~\ref{subfig:crack_photo}, and we apply another %
cement layer on top of the cracked surface. This allows us to emulate in-wall cracks while still having full knowledge of the ground truth. Then we scan this augmented sample to test the imaging performance of \sysname\ in visualizing these cracks.

We scan three horizontal strips (marked along with underlying cracks in Figure~\ref{subfig:crack_photo}) on the newly applied cement layer; the resulting \sysname\ imaging outcomes are depicted in Figure~\ref{subfig:crack_result},
with each subfigure showing a 2-D cross-section image along the width $x$ (moving dimension) and depth $z$. In these images, one may evidently discern major cracks and dents (marked red in Figure~\ref{subfig:crack_photo}) as bright spots. 
One may also observe that small cracks only generate bright patches and hence may not be precisely identified. This phenomenon can be attributed to the relatively weak reflection and the lack of crack samples in the training dataset. 
Nonetheless, the current results clearly demonstrate that \sysname\ is sufficiently competent to identify major issues even without any training.

\subsubsection{Multi-material Wall}
\begin{figure}[h]
    \setlength\abovecaptionskip{6pt}
    \vspace{-2ex}
	   \captionsetup[subfigure]{justification=centering}
		\centering
		\subfloat[Multi-material wall.]{
		  \begin{minipage}[b]{0.47\linewidth}
		        \centering
			    \includegraphics[width = 0.96\textwidth]{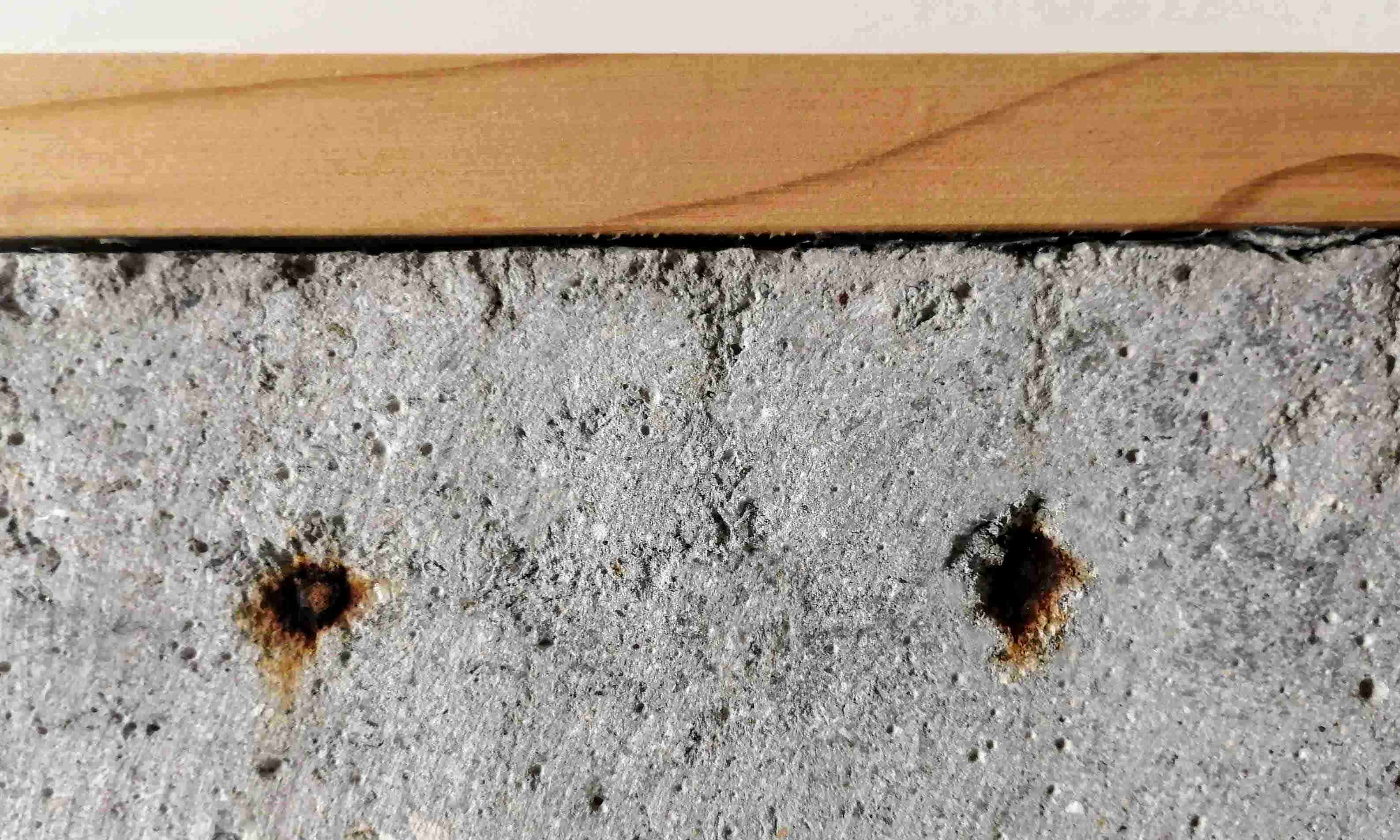}
			    \label{subfig:multiphoto}
			\end{minipage}
		}
		\subfloat[Imaging result.]{
		    \begin{minipage}[b]{0.47\linewidth}
		        \centering
			    \includegraphics[width = 0.96\textwidth]{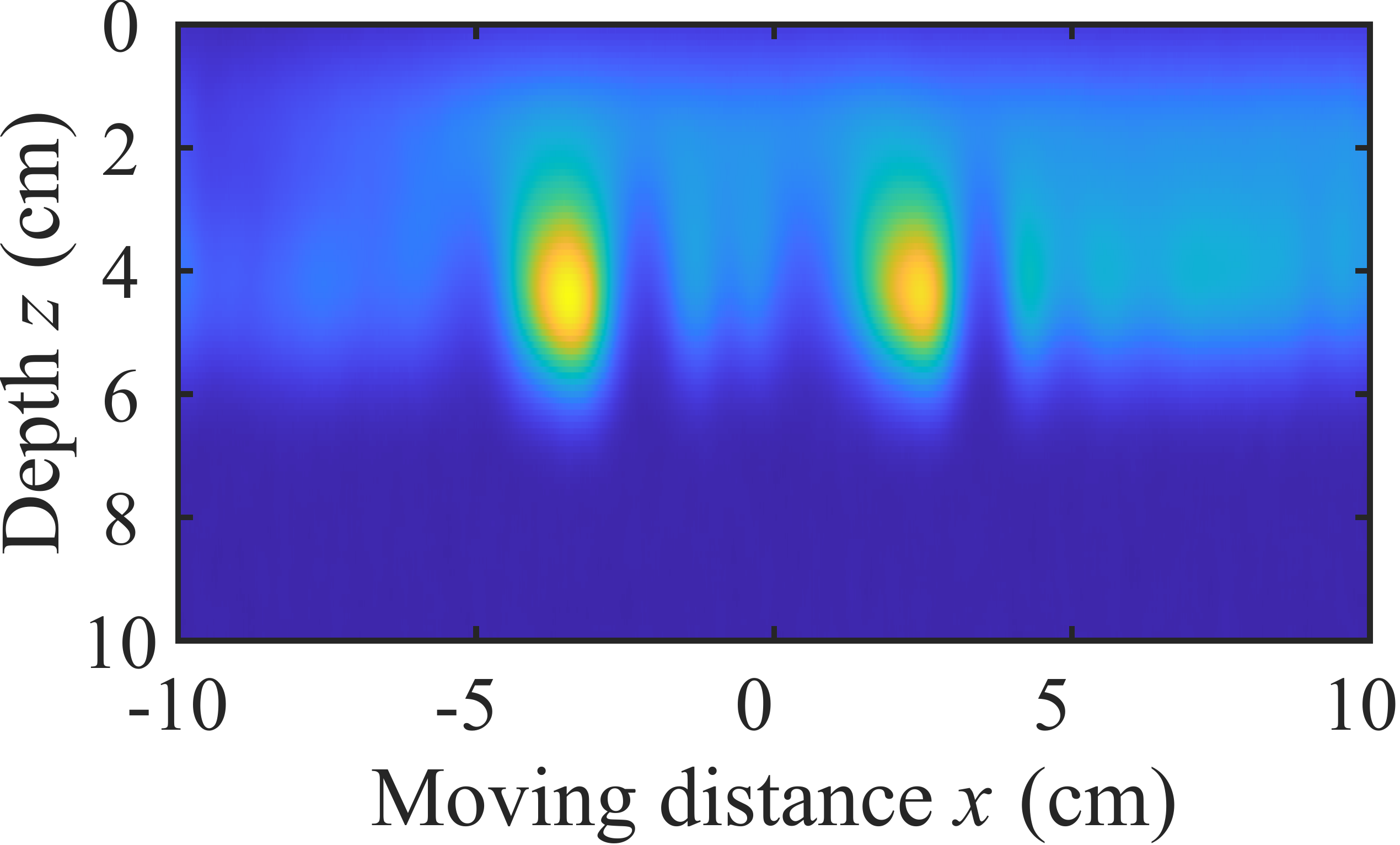}
			    \label{subfig:multiresult}
			\end{minipage}
		}
		\caption{Multi-material wall and imaging result.}
		\label{fig:multi}
\end{figure}

By far we have considered only homogeneous walls made of a single material, but multi-layer walls made of different materials can be adopted for interior finish; for example, wooden tiles are often used to cover concrete walls/floors for decoration purpose.
To prove that \sysname's in-wall detection ability can be generalized to multi-material walls, we cover a reinforced concrete block with 1.5~\!cm thick wooden tile, as shown in Figure~\ref{subfig:multiphoto}. We can see that the two rebars are approximately 4.5~\!cm deep and 6~\!cm apart in the moving distance $x$ direction. 

We then conduct \sysname\ scan on the wooden surface, and the \sysname\ imaging result is shown in Figure~\ref{subfig:multiresult}. One may readily observe the two rebars imaged with high localization accuracy in both directions of horizontal distance $x$ and depth $z$. One may also notice that the wood-concrete interface introduces a certain level of noise to the imaging result and causes the imaging quality to degrade slightly. This can be explained by the fact that the signals are partly reflected back at the material discontinuity, hence negatively impacting SNR. Nevertheless, the current result is sufficiently good to confirm that \sysname\ supports imaging into multi-material walls, and we believe better imaging quality can be achieved if a sufficient amount of multi-material wall samples become available.

\section{Limitations and Future Work} \label{sec:related}
As \sysname\ is a proof-of-concept prototype using IR-UWB radar and novel deep learning modules for seeing into walls, it may still leave a number of technical limitations for future exploration. %
First, we only report 2-D imaging in Section~\ref{sec:evaluation}, as achieving tomography-like 3-D imaging is just a straightforward extension by multiple parallel scans (which is a common practice). 
Therefore, we consider implementing an antenna array for more effective and efficient 3-D imaging as a future direction~\cite{octopus}. 
Second, as a follow-up of the previous point, we also need to consider using deep generative models~\cite{RFMag} to improve the robustness of the imaging ability (especially for 3-D imaging). Third, we have only tested four different objects in the material identification experiments, but we believe that M-Net can identify other materials as well.
We are discussing with local industries to put \sysname\ into use, so that we may gather more training and performance indicating data.
Last but not least, as the latency of I-Net and M-Net on single-board computers are only quasi-real-time, we are planning to use techniques such as model compression~\cite{yao2017deepiot, liu2018demand, bhattacharya2016sparsification} to speed them up.

\section{Conclusion} \label{sec:conclusion}
Seeing into walls without undermining structural integrity is a challenging yet important problem pertinent to building health diagnostics.
To this end, we have implemented \sysname, a portable and low-cost in-wall scanning prototype based on an IR-UWB radar, aiming to synthesize in-wall images and identify the material status. Employing carefully designed deep learning modules, \sysname\ performs structural imaging and material identification robustly in a self-adaptive manner. With extensive experiments on different walls and materials, we have demonstrated the promising performance of 
\sysname\ in accurate imaging and effective identification robust to various environments. 
We are on the way to identifying practical deployment scenarios for performing real-life tests on \sysname, so as to push it to the application market.

\section*{Acknowledgments}
We are grateful to the anonymous reviewers for their valuable and constructive comments. We further thank Energy Research Institute~@~NTU (ERI@N) and Interdisciplinary Graduate Programme (IGP) of NTU for supporting the PhD scholarship of Tianyue Zheng.

\newpage
\balance
\bibliographystyle{ACM-Reference-Format}
\bibliography{acmart}
\end{document}